\pdfoutput=1
\documentclass[12pt]{article}
\usepackage{cite}
\usepackage{dsfont}
\usepackage{graphicx}
\usepackage{latexsym}
\usepackage{mathrsfs}
\usepackage[overload]{textcase}

\font\grande=cmr9.5 scaled \magstep4
\font\medio=cmr9.5 scaled \magstep2
\outer\def\beginsection#1\par{\medbreak\bigskip
      \message{#1}\leftline{\bf#1}\nobreak\medskip
\vskip-\parskip
      \noindent}
\vfuzz=\maxdimen

\begin{document}
\bibliographystyle{unsrt}
\titlepage
\vspace{1cm}
\begin{center}
{\grande Hanbury-Brown Twiss effect, }\\
\vspace{0.5 cm}
{\grande  squeezed gravitons and the photon correlations}
\vspace{1.5 cm}\\
Massimo Giovannini\footnote{e-mail address: massimo.giovannini@cern.ch}\\
\vspace{1cm}
{{\sl Department of Physics, CERN, 1211 Geneva 23, Switzerland }}\\
\vspace{0.5cm}
{{\sl INFN, Section of Milan-Bicocca, 20126 Milan, Italy}}
\vspace*{1cm}
\end{center}
\vskip 0.3cm
\centerline{\medio  Abstract}
\vskip 0.5cm
Unlike the gravitational waves classically generated by moving macroscopic masses, the diffuse backgrounds of gravitons originate most likely from zero-point fluctuations of the gravitational field amplified by the evolution of the space-time curvature. The resulting entangled states lead to specific second-order correlation effects that could be eventually detected. For a quantitative analysis of this empirical expectation we scrutinize the interactions between the cosmic gravitons and the fundamental mode of a quantized electromagnetic field confined inside a closed optical resonator with perfectly-reflecting walls. We show that the Hanbury-Brown Twiss correlations of the photons are insensitive to the degrees of second-order coherence of the gravitons even barring for the exceedingly small couplings of the problem. Since the degree of second-order coherence of the photons does not reflect the correlation properties of the gravitons,  the statistical properties of the gravitons (and their super-Poissonian statistics) cannot be inferred, even in principle, from the intensity correlations of the cavity modes.
\noindent
\vspace{5mm}
\vfill
\newpage
\renewcommand{\theequation}{1.\arabic{equation}}
\setcounter{equation}{0}
\section{Introduction}
Since the experiments of Hanbury-Brown and Twiss \cite{QQ1,QQ2} (see also \cite{QQ3}), the degrees of coherence became an essential diagnostic in many quantum technologies \cite{QQ4,QQ5}. As originally suggested by Glauber and Sudarshan \cite{QQ6,QQ7}, the coherence of an optical source is unambiguously related to the minimization of the indetermination relation of the underlying quantum states of the photons. The mutual coupling between gravity and electromagnetism would naively suggest that the degrees of coherence of the photons must be eventually sensitive to the statistical properties of the gravitons but, according to a strictly empirical logic, this argument could be preliminarily discarded for two complementary reasons. The first one is that the coupling of photons to gravity waves is ridiculously small; the second motivation is that gravitons do not exist, at least in the same sense in which we could discuss photons. We find that, between these two objections, the latter is far more disputable than the former since the adiabatic paradigm of structure formation  \cite{ETA1,ETA2,ETA3} predicts that the large-scale curvature inhomogeneities are complemented by a diffuse background of gravitons. From a stricter observational viewpoint the consistency of the adiabatic paradigm with the temperature and polarization anisotropies of the microwave background \cite{SF1,SF2} has been subsequently confirmed by different sets of observations \cite{SF3,SF4,SF5,SF6,SF7} in the last twenty years. In this framework the phonons (associated with the inhomogeneities of the scalar curvature) are produced together with the gravitons  (corresponding to the tensor modes of the geometry). Therefore, while it is true that classical gravitational waves are typically produced by moving macroscopic masses, it is equally well established that gravitons originate from zero-point fluctuations of the gravitational field amplified by the evolution of the space-time curvature \cite{IOTA1,IOTA2}. Rather than indulging in various theoretical speculations on the quantized modes of the gravitational field it seems more productive to focus on the entangled multiparticle states of relic gravitons whose existence is predicted by the current success of the adiabatic paradigm. This is, in a nutshell, the perspective of this paper where we intend to investigate the connection between the quantum coherence of the photons and of the gravitons. 

To clarify the physical scales of the problem we remind that the shortest wavenumbers characterizing the diffuse backgrounds of cosmic gravitons are ${\mathcal O}(k_{p})$ where $k_{p} = 0.002\, \mathrm{Mpc}^{-1}$ is the pivot scale at which the scalar and the tensor power spectra are customarily assigned in the context of the concordance paradigm. The corresponding (comoving) frequency falls in the aHz range since $\nu_{p} = k_{p}/(2\pi)= 3.09 \, \mathrm{aHz}$ (where\footnote{The conventional prefixes of the International System of units are employed throughout (e.g. $1\, \mathrm{aHz} =10^{-18} \, \mathrm{Hz}$, $1 \, \mathrm{fHz} = 10^{-15} \mathrm{Hz}$ and so on and so forth). Since the value of the scale factor at the present time is normalized to $1$ the comoving and physical frequencies coincide today.}
$1\, \mathrm{aHz} = 10^{-18} \mathrm{Hz}$). The constraints on the aHz gravitons are customarily introduced as limits on the tensor-to-scalar-ratio\footnote{We remind that $r_{T} = {\mathcal A}_{T}/{\mathcal A}_{{\mathcal R}}$ where  ${\mathcal A}_{T}$ and ${\mathcal A}_{{\mathcal R}}$ denote, respectively,  the amplitudes of the tensor and of the scalar power spectra either at a conventional reference wavenumber $k_{p}$ or at the corresponding pivotal frequency $\nu_{p}$.} $r_{T}$.  These limits have been slowly but steadily improving in the last three decades: while the different releases of the WMAP collaboration did set upper limits $r_{T} < {\mathcal O}(0.1)$ \cite{SF1,SF2}, the recent determinations suggest even smaller values, i.e. $r_{T} < {\mathcal O}(0.06)$ or even $r_{T} < {\mathcal O}(0.03)$  \cite{SF3,SF4,SF5,SF6,SF7}.  For frequencies larger than $\nu_{p}$  the spectral energy density of the diffuse backgrounds is customarily measured in terms of  the spectral energy density in critical units (i.e. $\Omega_{gw}(\nu, \tau_{0})$ in what follows); this quantity depends upon the averaged multiplicity of the gravitons in a logarithmic interval of frequency $\nu$. In the nHz range the direct bounds on $\Omega_{gw}(\nu,\tau_{0})$ come from the Pulsar Timing Arrays \cite{NANO,PPTA,EPTA} 
and for even higher frequencies in the audio band (i.e. between few Hz and $10$ kHz) 
the operating wide-band interferometers \cite{LIGO1,LIGO2} set the highest frequency limits to date. Although we shall be using hereunder the natural system of units\footnote{We remind that, in this system, $\hbar = c= \kappa_{B} = 1$ where $\kappa_{B}$ is the Boltzmann constant. Furthermore the Planck mass and its reduced counterpart are given, respectively, by $M_{P} = 1.22 \times 10^{19} \, \mathrm{GeV}$ and by $\overline{M}_{P} = M_{P}/\sqrt{8 \pi}$. Finally, within the present notations, the Planck length is denoted by $\ell_{P} = 1/\overline{M}_{P}$.} $\Omega_{gw}(\nu, \tau_{0})$ is ${\mathcal O}(\hbar^2)$; this means, as correctly stressed in Ref. \cite{IOTA3}, that we are dealing here with an effect that vanishes in the limit $\hbar\to 0$. This observation clarifies the logic of the present analysis and also suggests, as stressed above, that relic gravitons are a unique laboratory for the exploration of quantum gravitational effects especially in the high frequency domain. One might also say that since $\Omega_{gw}(\nu,\tau_{0})$ scales as $\nu^4$ when the averaged multiplicity is (approximately) frequency-independent (as in the coherent case), single gravitons (or bunches of few gravitons) could be preferentially detected in the high frequency domain. A similar argument is in fact due to Dyson \cite{IOTA4} who suggested that only at high frequencies it will be eventually possible to detect single gravitons. More 
concrete estimates to be discussed later on suggest that the maximal frequency of cosmic gravitons in the adiabatic paradigm 
must not exceed the THz range.

Having established that gravitons exist for frequencies ranging between the aHz and the THz, it is natural to be curious about the specific diagnostics able to distinguish classical gravitational waves from a quantum state of gravitons. As anticipated above it has been pointed out long ago  that the quantumness of a source can be gauged from the degrees of coherence of the primary optical fields \cite{QQ6,QQ7}. Similarly the degrees of quantum coherence of the gravitons are a key diagnostic \cite{MG1} since they encode, as in the case of the photons, the statistical properties of the underlying quantum states. Furthermore the same is also true in the case of the {\em phonons} associated with the large scale curvature inhomogeneities \cite{MG2}. According to the ideas propounded in Refs. \cite{MG1,MG2} (see also, for instance,  \cite{MG3,MG4,MG5}) the degrees of coherence reveal testable features of the underlying quantum states. From a technical viewpoint these conclusion follows by examining the generalized Glauber correlators associated to tensor modes: this step leads to a proper definition of the degrees of first- and second-order coherence that are ultimately associated with the correlation of the intensities of a given quantum field. Owing to their first appearance in Refs. \cite{QQ3,QQ4} the intensity correlations are customarily referred to as Hanbury Brown-Twiss (HBT) correlations.  If the first- and second-order Glauber correlators are evaluated in different dynamical regimes (both in the polarized and unpolarized cases) the 
HBT correlations consistently display super-Poissonian statistics with quantitative features that depend on the initial quantum state and polarization averaging  \cite{MG3}. This result holds, in particular, 
in the case of the adiabatic paradigm \cite{ETA1,ETA2,ETA3} when a decelerated stage of expansion follows a conventional inflationary epoch. Finally, although HBT correlations can probe the quantum origin of relic gravitons within the Hubble radius, the resulting distributions universally preserve their super-Poissonian character regardless of the initial state.

The idea explored here is, in short, the following: we ought to consider the degrees of coherence 
 of the photons and see in what sense they are sensitive to the HBT correlations 
 of the gravitons. It is expected that the photons must be sensitive to the gravitons 
 and to their degrees of coherence. This physical expectation is scrutinized hereunder 
with the purpose of relating the degrees of quantum coherence of the photons and of the gravitons. 
Given the degrees of first- and second-order coherence of the 
gravitons, what is the statistics of the photons in a Fabry-P\'erot cavity? Is it the same 
of the one of the gravitons?  Is it determined by some other consideration? These are some of the questions 
addressed in this investigation whose general layout is the following. In section \ref{sec2} the 
Glauber correlators are introduced for the scalar, vectors and tensor cases. As already pointed out, this step is required for an accurate definition of the degrees of quantum coherence both in the case of the photons and of the gravitons.
In section \ref{sec3} we then consider the interaction of the relic gravitons with a Fabry-P\'erot cavity. 
In section \ref{sec4} we analyze the effective Hamiltonian of the problem. Section \ref{sec5} deals with the coupled dynamics of the system and section \ref{sec6} is devoted to the specific analysis of the HBT correlations of the cavity mode and of the gravitons. Finally section \ref{sec7} contains some concluding considerations.  Since it is essential to consider the couplings of the photons and of the gravitons in curved backgrounds some relevant details have been collected in the appendix \ref{APPA}. Although the problem  discussed here involves a multimode Hamiltonian it is also possible to simplify the problem, with some caveats, to the case of a three-mode Hamiltonian; this correspondence (together with some other considerations complementing the discussions of sections \ref{sec4} and \ref{sec5}) is presented in appendix \ref{APPB}.

\renewcommand{\theequation}{2.\arabic{equation}}
\setcounter{equation}{0}
\section{The quantum coherence of relic gravitons}
\label{sec2}
The Glauber-Sudarshan theory of optical coherence 
is formulated in terms of (quantum) vector fields but it implicitly applies to the scalar case when a single electromagnetic polarization is considered; this exclusive approach
complements the inclusive description where all the polarizations are simultaneously 
considered. When employed in the analysis of the curvature phonons and of the gravitons 
the Glauber correlators must be extended to the scalar and tensor cases. For we are discussing separately the degrees of second-order coherence of  the photons {\em and} of the gravitons it is preferable to start from the general forms 
of the Glauber correlators in the vector, tensor and scalar cases by swiftly recalling 
the original construction of \cite{QQ6,QQ7} and its extensions \cite{MG1,MG2}.

\subsection{Original Glauber correlators}
We consider first the solenoidal vector field operators $\widehat{A}_{i}(x)$ (where $x$ represents the space-time point) 
in the Minkowski background\footnote{The same construction holds in the case of a conformally flat metric of Friedmann-Robertson-Walker type that will be introduced later on. For simplicity in this section we stick to
the Minkowski case.}.  As in \cite{QQ6,QQ7} they consist of a positive $(+)$ and of a negative $(-)$ frequency part [i.e. $\widehat{A}_{i}(x) = \widehat{A}_{i}^{(+)}(x) + \widehat{A}_{i}^{(-)}(x)$]; since the fields are Hermitian 
we also have  $\widehat{A}_{i}^{(+)}(x)= \widehat{A}_{i}^{(-)\,\dagger}(x)$.
Exactly the same decomposition holds for the electric and for the magnetic fields, i.e. 
$\widehat{E}_{i}(x) = \widehat{E}_{i}^{(+)}(x) + \widehat{E}_{i}^{(-)}(x)$ and  $\widehat{B}_{i}(x) = \widehat{B}_{i}^{(+)}(x) + \widehat{B}_{i}^{(-)}(x)$. The description based on  the Coulomb gauge is more suitable in view of a Weyl rescaling of the flat space-time results\footnote{The Lorenz gauge condition is not invariant under Weyl rescaling; as a consequence the Coulomb is 
often preferable in curved backgrounds and especially in cosmological backgrounds \cite{LFC}.}.
If the state $|\mathrm{vac}\rangle $  minimizes the associated Hamiltonian, it is annihilated by $\widehat{A}_{i}^{(+)}(x)$ (i.e.  $\widehat{A}_{i}^{(+)}(x) |\mathrm{vac} \rangle=0$ and 
$\langle \mathrm{vac} |\, \widehat{A}_{i}^{(-)}(x) =0$). The explicit form of the {\em vector} Glauber correlator can then be written as:
\begin{eqnarray}
&& {\mathcal V}^{(n,m)}_{(i_{1}), \,.\,.\,(i_{n}), \, (i_{n+1}),\, .\,.\,, (i_{n +m}) }(x_{1}, \,.\,.\,x_{n}, \, x_{n+1},\, .\,.\,.\,, x_{n +m})  
\nonumber\\
&&= \mathrm{Tr}\biggl[ \widehat{\rho} \,\,\,\widehat{A}_{i_{1}}^{(-)}(x_{1})\,.\,.\,.\, \widehat{A}_{i_{n}}^{(-)}(x_{n})
\, \widehat{A}_{i_{n+1}}^{(+)}(x_{n+1})\,.\,.\,.\,\widehat{A}_{i_{n+m}}^{(+)}(x_{n+m})\biggr],
\label{vec1}
\end{eqnarray}
where $\widehat{\rho}$ indicates the density operator representing the (generally mixed) state of the field.  When 
$\widehat{A}_{i}(x)$ is replaced either by $\widehat{E}_{i}(x)$ or by $\widehat{B}_{i}(x)$ we obtain, respectively, the Glauber correlators valid in the electric and magnetic case.  In full analogy with Eq. (\ref{vec1}) we introduce next the field operators $\widehat{\mu}_{ij}(x) = \widehat{\mu}_{ij}^{(+)}(x) + \widehat{\mu}_{ij}^{(-)}(x)$ consisting, as in the vector situation of a positive and of a negative frequency part, with 
$\widehat{\mu}_{ij}^{(+)}(x)= \widehat{\mu}_{ij}^{(-)\,\dagger}(x)$. The field operator $\widehat{\mu}_{ij}(x)$ is both traceless and solenoidal i.e. $\partial_{i} \widehat{\mu}^{i\, j} = \widehat{\mu}_{i}^{\,\,i} =0$ and, in the tensor case,
the corresponding Glauber correlator becomes:
\begin{eqnarray}
&& {\mathcal T}^{(n,m)}_{(i_{1}\,\,j_{1}), \,.\,.\,.\,(i_{n}\,\,j_{n}), \, (i_{n+1}\,\,j_{n+1}),\, .\,.\, (i_{n +m}\,\,j_{n+m}) }(x_{1}, \,.\,.\,x_{n}, \, x_{n+1},\, .\,.\, x_{n +m}) 
\nonumber\\
&& = \mathrm{Tr}\biggl[ \widehat{\rho} \, \widehat{\mu}_{i_{1}\,\,j_{1}}^{(-)}(x_{1})\,.\,.\, \widehat{\mu}_{i_{n}\,\,j_{n}}^{(-)}(x_{n})
\, \widehat{\mu}_{(i_{n+1}\,\,j_{n+1})}^{(+)}(x_{n+1})\,.\,.\,\widehat{\mu}_{(i_{n+m}\,\,j_{n +m})}^{(+)}(x_{n+m})\biggr],
\label{tens1}
\end{eqnarray}
where, as before, $\widehat{\rho}$ indicates the density operator representing the (generally mixed) state of the field $\widehat{\mu}_{ij}$. In Eq. (\ref{tens1}) instead of the $(n+m)$ {\em vector} indices of Eq. (\ref{vec1}) \cite{QQ6,QQ7} there are $(n+m)$ pairs of {\em tensor} indices (i.e. $(i_{1}\, j_{1})\,...\,(i_{n}\, j_{n})\,...\, (i_{n+m}\, j_{n+m})$). Finally the {\em scalar} Glauber correlator is obtained by dropping all the vector and tensor indices:
\begin{eqnarray}
{\mathcal S}^{(n,m)}(x_{1}, \,.\,.\,.\,x_{n}, \, x_{n+1},\, .\,.\,.\,, x_{n +m})  = \mathrm{Tr}\biggl[ \widehat{\rho} \, \widehat{\phi}^{(-)}(x_{1})\,.\,.\,.\, \widehat{\phi}^{(-)}(x_{n})
\, \widehat{\phi}^{(+)}(x_{n+1})\,.\,.\,.\,\widehat{\phi}^{(+)}(x_{n+m})\biggr],
\label{scal1}
\end{eqnarray}
where $\widehat{\phi}(x) = \widehat{\phi}^{(+)}(x) + \widehat{\mu}^{(-)}(x)$, with 
$\widehat{\phi}^{(+)}(x)= \widehat{\phi}^{(-)\,\dagger}(x)$. 
As  in the vector case, the scalar and the tensor correlators of Eqs. (\ref{vec1})--(\ref{tens1}) are analyzed either within an inclusive perspective (where all the polarizations are included) or in an exclusive approach (where only one polarization 
is considered at a time). In quantum optics an exclusive perspective is adopted when one 
of the two polarizations of the photon is neglected \cite{loudon}. The single-polarization approximation is motivated by  various experiments dealing with a single polarization (for instance in a cavity).  The same logic selects a single polarization in the case of the gravitons. It is also common to treat the Mach-Zehnder and Hanbury Brown-Twiss interferometry in terms of  a single mode of the field. The single-mode experiments use plane parallel light beams whose transverse intensity profiles are not important for the measured quantities. In these situations it is often sufficient to consider the light beams as exciting a single mode of the field. 

\subsection{Delayed photoelectric coincidence measurements}
Following the tenets of the quantum theory of optical coherence an operator 
of the type 
\begin{equation}
\widehat{E}^{(-)}_{i_{1}}(x_{1})\,.\,.\,.\, \widehat{E}_{i_{n}}^{(-)}(x_{n})
\, \widehat{E}_{i_{1}}^{(+)}(x_{1})\,.\,.\,.\, \widehat{E}_{i_{n}}^{(+)}(x_{n})
\label{COIN1}
\end{equation}
is required for the description of the $n$-fold photoelectric (delayed) coincidence 
measurements of the electric field at the space-time points $x_{1}, \,.\,.\,.\, x_{n}$ with detectors 
sensitive to the polarizations $i_{1}, \,.\,.\,.\, i_{n}$ (see \cite{QQ3} and \cite{QQ6,QQ7}). 
If we now introduce the multiparticle states $|\{\,a\}\rangle$ and $|\{\,b\}\rangle$ (indicating, respectively, the states 
of the field {\em after}  and {\em before} the measurement), the matrix element corresponding 
to the absorption of photons with given polarizations at each detector (and at different times) is \cite{QQ6,QQ7}
\begin{equation}
\langle \{ a\}\, | \widehat{E}_{i_{1}}^{(+)}(x_{1})\,.\,.\,.\, \widehat{E}_{i_{n}}^{(+)}(x_{n}) | \{\,b\}\rangle.
\label{COIN2}
\end{equation}
The rate of the absorptions, summed over the final states, becomes then 
\begin{eqnarray}
\sum_{\{a\}} \biggl|\langle \{ a\}\, | \widehat{E}_{i_{1}}^{(+)}(x_{1})\,.\,.\,.\, \widehat{E}_{i_{n}}^{(+)}(x_{n})| \{ b\} \rangle \biggr|^2=  \langle \{ b\}| \widehat{E}_{i_{1}}^{(-)}(x_{1})\,.\,.\,.\, \widehat{E}_{i_{n}}^{(-)}(x_{n})
 \widehat{E}_{i_{1}}^{(+)}(x_{1})\,.\,.\,.\, \widehat{E}_{i_{n}}^{(+)}(x_{n}) |\{ b \}\rangle,
\label{COIN3}
\end{eqnarray}
where the right hand side follows from the left hand side thanks to the completeness relation 
applied to the final state. When the right-hand side of this equation is averaged over the ensemble of the initial states, the correlation function of Eq. (\ref{vec1}) reappears if the pairs of the space-time points and the vector indices are 
identified; more specifically we must have $x_{n + r} = x_{r}$,  $i_{n+r} = i_{r}$, $m=n$ and $r= 1, 2, \,.\,.\,., n$. The 
explicit form of the vector correlator becomes then:
\begin{eqnarray}
&& {\mathcal V}^{(n)}_{(i_{1}), \,.\,.\,.\,(i_{n})\,.\,.\,.\,(i_{2n})}(x_{1}, \,.\,.\,.\,x_{n}, \, x_{n+1},\, .\,.\,.\,, x_{2n}) 
\nonumber\\
&&= \mathrm{Tr}\biggl[ \widehat{\rho} \,\,\, \widehat{E}^{(-)}_{i_{1}}(x_{1})\,.\,.\,.\, \widehat{E}^{(-)}_{i_{n}}(x_{n})
\, \widehat{E}_{i_{n+1}}^{(+)}(x_{n+1})\,.\,.\,.\, \widehat{E}_{i_{2n}}^{(+)}(x_{2n})\biggr].
\label{COIN4}
\end{eqnarray}
The same argument leading to Eq. (\ref{COIN4}) applies in the case of the tensor correlation 
function 
\begin{eqnarray}
&&{\mathcal T}^{(n)}_{(i_{1}\,\,j_{1}), \,.\,.\,.\,(i_{n}\,\,j_{n})\,.\,.\,.\,(i_{2n}\,\,j_{2n})}(x_{1}, \,.\,.\,.\,x_{n}, \, x_{n+1},\, .\,.\,.\,, x_{2n}) 
\nonumber\\
&&= \mathrm{Tr}\biggl[ \widehat{\rho} \, \widehat{\mu}^{(-)}_{(i_{1}\,\,j_{1})}(x_{1})\,.\,.\,.\, \widehat{\mu}^{(-)}_{(i_{n}\,\,j_{n})}(x_{n})
\, \widehat{\mu}_{(i_{n+1}\,\,j_{n+1})}^{(+)}(x_{n+1})\,.\,.\,.\, \widehat{\mu}_{(i_{2n}\,\,j_{2n})}^{(+)}(x_{2n})\biggr].
\label{COIN5}
\end{eqnarray}
Finally, the analog of Eq. (\ref{COIN5}) in the scalar case is given by:
\begin{equation}
{\mathcal S}^{(n)}(x_{1}, \,.\,.\,.\,x_{n}, \, x_{n+1},\, .\,.\,.\,, x_{2n}) = \mathrm{Tr}\biggl[ \widehat{\rho} \, \widehat{\phi}^{(-)}(x_{1})\,.\,.\,.\, \widehat{\phi}^{(-)}(x_{n})
\, \widehat{\phi}(x_{n+1})\,.\,.\,.\, \widehat{\phi}(x_{2n})\biggr].
\label{COIN6}
\end{equation}
From the Glauber correlators introduced in Eqs. (\ref{COIN4}), (\ref{COIN5}) and (\ref{COIN6}) we shall now get the proper definition of the degrees of quantum coherence and of the intensity correlation functions arising in the analysis of the HBT effect. 

\subsection{Degrees of first-order coherence and Young interferometry}
From Eq. (\ref{COIN4}) in the case $n=1$ we obtain the explicit form of the first Glauber
correlation function associated with the degree of first-order coherence:
\begin{equation}
{\mathcal V}^{(1)}_{(i),\,(j)}(x_{1},\, x_{2}) = \langle  \widehat{E}^{(-)}_{i}(x_{1}) \, \widehat{E}^{(+)}_{j}(x_{2}) \rangle,
\label{HBT1}
\end{equation}
where, as usual, $ \langle .\,.\,. \rangle = \mathrm{Tr}[ \widehat{\rho}.\,.\,.]$. When not otherwise stated, the same shorthand notation shall be also adopted hereunder. The operator appearing inside the expectation value (\ref{HBT1}) defines the degree of first-order coherence: 
\begin{equation}
g^{(1)}_{\gamma}(x_{1}; x_{2}) = {\mathcal V}^{(1)}(x_{1},x_{2})/\sqrt{{\mathcal V}^{(1)}(x_{1}) \,{\mathcal V}^{(1)}(x_{2})}, \qquad {\mathcal V}^{(1)}(x_{1},x_{2}) = \langle  \widehat{E}^{(-)}_{i}(x_{1}) \, \widehat{E}^{(+)}_{i}(x_{2}) \rangle.
\label{HBT2}
\end{equation}
For simplicity in Eq. (\ref{HBT2})  the case of a single polarization has been considered.
The degree of first-order coherence introduced in Eq. (\ref{HBT2}) is in fact probed by Young-type (two-slit) correlation 
experiment; in this context when $g^{(1)}_{\gamma}(x_{1}, x_{2}) =1$ the field is 
said to be first-order coherent while it is only partially coherent when $0< g^{(1)}_{\gamma}(x_{1}, x_{2}) < 1$. The case $g^{(1)}_{\gamma}(x_{1}, x_{2}) \to 0$ corresponds to an incoherent field. It follows that Young interferometry does not provide information on the statistical properties of the quantum state of the
radiation field: various states with opposite physical properties (such as laser light and chaotic light) lead to comparable degrees of first-order coherence. The same logic leading to Eqs. (\ref{HBT1})--(\ref{HBT2}) can be applied to the 
gravitons; thus,  from Eq. (\ref{COIN5}) in the case $n =1 $ we obtain 
\begin{equation}
{\mathcal T}^{(1)}_{(i\,\,j),\,(k\,\,\ell)}(x_{1},\, x_{2}) = \langle  \widehat{\mu}^{(-)}_{i\,\,j}(x_{1}) \, \widehat{\mu}^{(+)}_{k\,\,\ell}(x_{2}) \rangle.
\label{HBT3}
\end{equation}
The operator whose expectation value appears inside Eq. (\ref{HBT3}) defines the degree of first-order coherence: 
\begin{equation}
g^{(1)}_{g}(x_{1}; x_{2}) = {\mathcal T}^{(1)}(x_{1},x_{2})/\sqrt{{\mathcal T}^{(1)}(x_{1}) \,{\mathcal T}^{(1)}(x_{2})}, \qquad {\mathcal T}^{(1)}(x_{1},x_{2}) = \langle  \widehat{\mu}^{(-)}_{i\,j}(x_{1}) \, \widehat{\mu}^{(+)}_{i\,j}(x_{2}) \rangle.
\label{HBT4}
\end{equation}
Again, in the case of a single tensor polarization the degree of first-order coherence keeps the same form but 
the tensor indices can be dropped:
\begin{equation}
g_{g}^{(1)}(x_{1}; x_{2}) = \frac{\langle \widehat{\mu}^{(-)}(x_{1})\,\, \widehat{\mu}^{(+)}(x_{2}) \rangle }{\sqrt{\langle \widehat{\mu}^{(-)}(x_{1})\,\, \widehat{\mu}^{(+)}(x_{2}) \rangle\,\,\,\langle \widehat{\mu}^{(-)}(x_{2})\,\, \widehat{\mu}^{(+)}(x_{2}) \rangle }}.
\label{HBT4a}
\end{equation}

\subsection{Degrees of second-order coherence and HBT interferometry}
The degree of second-order coherence for the photons follows from Eq. (\ref{COIN4}) when $n =2$:
\begin{equation}
{\mathcal V}^{(2)}_{(i_{1}), \,(i_{2})\,(i_{3}),\,(i_{4}) }(x_{1}, x_{2}, x_{3}, x_{4}) = \langle  \widehat{E}^{(-)}_{i_{1}}(x_{1}) \, \widehat{E}^{(-)}_{i_{2}}(x_{2})\widehat{E}^{(+)}_{i_{3}}(x_{3}) \widehat{E}^{(+)}_{i_{4}}(x_{4}) \rangle.
\label{HBT5}
\end{equation} 
Equation (\ref{HBT5}) describes the generalized intensity correlations provided the operator appearing inside 
the expectation value is Hermitian; this happens when the four coordinates coincide two by two (i.e.  $x_{2} = x_{3}$ and $ x_{4}=  x_{1}$) and similarly for the polarizations (i.e. $i_{2}=  i_{3}$ and $i_{1} =  i_{4}$).
From Eq. (\ref{HBT5}) the degree of second-order coherence for the photons 
reads 
\begin{equation}
g_{\gamma}^{(2)}(x_{1}; x_{2}) = \frac{{\mathcal V}^{(2)}(x_{1}, x_{2})}{{\mathcal V}^{(1)}(x_{1})\,{\mathcal V}^{(1)}(x_{2})}, \qquad {\mathcal V}^{(2)}(x_{1}, x_{2}) =
 \langle  \widehat{E}^{(-)}_{i}(x_{1}) \, \widehat{E}^{(-)}_{j}(x_{2})\widehat{E}^{(+)}_{j}(x_{2}) \widehat{E}^{(+)}_{i}(x_{1}) \rangle.
\label{HBT6}
\end{equation}
The applications of the HBT effect \cite{QQ1,QQ2} are 
 associated, either directly or indirectly, with the analysis of Eq. (\ref{HBT6}) and of its descendants. 
While the degree of first-order coherence is generally 
insensitive to the statistical properties of the underlying quantum state, 
the opposite is true in the case of $g_{\gamma}^{(2)}(x_{1}, x_{2}) $.
From Eq. (\ref{COIN5}) we obtain next the degree of second-order coherence 
for the gravitons always for $n = 2$
\begin{equation}
{\mathcal T}^{(2)}_{(i_{1}\,\,j_{1}), \,(i_{2}\,\,j_{2})\,(i_{3}\,\,j_{3}),\,(i_{4}\,\,j_{4}) }(x_{1}, x_{2}, x_{3}, x_{4}) = \langle  \hat{\mu}^{(-)}_{i_{1}\,\,j_{1}}(x_{1}) \, \hat{\mu}^{(-)}_{i_{2}\,\,j_{2}}(x_{2})\hat{\mu}^{(+)}_{i_{3}\,\,j_{3}}(x_{3}) \hat{\mu}^{(+)}_{i_{4}\,\,j_{4}}(x_{4}) \rangle.
\label{HBT7}
\end{equation} 
Again, since the intensity must be Hermitian, the standard HBT correlators 
follow from Eqs.  (\ref{HBT7}) by requiring 
$(i_{2}\,\,j_{2}) =  (i_{3}\,\,j_{3})$ and $(i_{1}\,\,j_{1}) =  (i_{4}\,\,j_{4})$ when $x_{2} \to x_{3}$ and $x_{4} \to x_{1}$; thus the degree of second-order coherence in the case of the gravitons is:
\begin{equation}
g_{g}^{(2)}(x_{1}; x_{2}) = \frac{{\mathcal T}^{(2)}(x_{1}, x_{2})}{{\mathcal T}^{(1)}(x_{1})\,{\mathcal T}^{(1)}(x_{2})},\quad {\mathcal T}^{(2)}(x_{1}, x_{2}) = 
 \langle  \hat{\mu}^{(-)}_{i\,\,j}(x_{1}) \, \hat{\mu}^{(-)}_{k\,\,\ell}(x_{2})\hat{\mu}^{(+)}_{k\,\ell}(x_{2}) \hat{\mu}^{(+)}_{i\,j}(x_{1}) \rangle.
 \label{HBT8}
 \end{equation}
The degrees of second-order coherence obtained so far can be simplified in various limits. If the 
quantum fields are defined in a cavity the dependence simplifies between the numerator and the denominator (as we are going to see) and the degree of second-order coherence of the photons becomes:
\begin{equation}
g^{(2)}_{\gamma}(t; t + \Delta t )=  \frac{\langle  \widehat{E}^{(-)}_{i}(t) \, \widehat{E}^{(-)}_{j}(t + \Delta t)\widehat{E}^{(+)}_{j}(t + \Delta t) \widehat{E}^{(+)}_{i}(t) \rangle}{\langle \widehat{E}^{(-)}_{i}(t)\widehat{E}^{(+)}_{i}(t)\rangle\,\,\langle \widehat{E}^{(-)}_{i}(t+ \Delta t)\widehat{E}^{(+)}_{i}(t + \Delta t)\rangle}.
\label{HBT9}
\end{equation}
 Since many experiments
use plane parallel light beams (whose transverse intensity profiles are not important for the
measured quantities), it is often preferable to consider the light beams as exciting a single mode of the field. 
In the single mode approximation (for instance inside a cavity) one can then focus on that 
single optical mode which i close to the laser frequency; the degree of second-order 
coherence becomes, in this case, 
\begin{equation}
g^{(2)}_{c}(t; t + \Delta t )=  \frac{\langle  \widehat{c}^{\dagger}(t) \,\, \widehat{c}^{\dagger}(t + \Delta t)\,\,\widehat{c}(t + \Delta t) \,\,\widehat{c}(t) \rangle}{\langle \widehat{c}^{\dagger}(t)\,\,\widehat{c}(t)\rangle\,\,\, \langle \widehat{c}^{\dagger}(t + \Delta t)\,\,\widehat{c}(t + \Delta t)\rangle},
\label{HBT10}
\end{equation}
where $\widehat{c}$ and $\widehat{c}^{\dagger}$ are the standard creation and annihilation operators for a single mode of the cavity field. If the photons and the gravitons interact it is plausible 
to expect that the degrees of quantum coherence of the photons [i.e. $g_{\gamma}^{(1)}(t; t + \Delta t)$ and 
$g_{\gamma}^{(2)}(t; t + \Delta t)$] could bear the mark of the degrees of quantum coherence of the gravitons [i.e. $g_{g}^{(1)}(t; t + \Delta t)$ and $g_{g}^{(2)}(t; t + \Delta t)$]. The coupling between relic gravitons and the cavity modes of the photons 
is specifically discussed in the next section.

\renewcommand{\theequation}{3.\arabic{equation}}
\setcounter{equation}{0}
\section{Relic gravitons in a cavity}
\label{sec3}
The analysis of the HBT correlations depend upon the interactions between 
 the relic gravitons and the modes of the electromagnetic field confined inside an optical resonator with perfectly-reflecting walls. 
 Here we are going to consider the simplest situation where the gauge fields are minimally coupled. Although a shortcut  would be 
 to scrutinize the problem directly in flat space-time, such a choice would hamper any accurate analysis since the interactions with the space-time curvature are the ultimate source of the correlation exhibited later on by the gravitons. It is therefore mandatory to investigate the interactions of gravitons and photons with the space-time curvature. In this respect, instead of a Minkowski background it is preferable to choose a spatially flat metric of Friedmann-Robertson-Walker type since this choice is compatible with the current version of the adiabatic paradigm\footnote{We remind that the conformally flat backgrounds are also the ones favoured, through the years, by the observational determinations of the cosmological parameters \cite{SF1,SF2,SF3,SF4,SF5,SF6,SF7}. } \cite{ETA1,ETA2,ETA3}. In this concrete situation the gravitons with opposite three-momenta are produced while the evolution  of the cavity mode can be phrased in terms of the rescaled electric and magnetic fields. The background metric can then be written in a conformally flat parametrization $\overline{g}_{\mu\nu}(\tau) = a^2(\tau) \, \eta_{\mu\nu}$ where $a(\tau)$ is the scale factor, 
$\tau$ is the conformal time coordinate and $\eta_{\mu\nu}$ the Minkowski metric with 
signature $(1, \, -1,\, -1, \,-1)$. The space-time metric will then consist of $\overline{g}_{\mu\nu}(\tau)$ 
supplemented by the corresponding fluctuations (see Eqs. (\ref{AP1a})--(\ref{AP1b}) and discussions therein). The details of this analysis have been reported, for the sake of accuracy, in appendix \ref{APPA} [see in particular Eqs.  (\ref{APG4}) and (\ref{APG6})]. In this section we are going to focus on of the physical implications and swiftly account of the technical results that can be followed through in the appendix.

\subsection{The general form of the action}
By perturbing the full action to second-order in the amplitude of the tensor modes of the geometry (see appendix \ref{APPA}) we obtain the system that describes the coupled evolution of the relic gravitons and of the quantum field inside the Fabry-Perot cavity with perfectly reflecting walls. In short the quantum field of the cavity will be referred to as the cavity mode. 
Bearing in mind this terminology the total action takes the form $S = S_{g} + S_{\gamma} + S_{\gamma\,g}$ where 
$S_{g}$ denotes the action of the relic gravitons, $S_{\gamma}$ indicates the contribution of the cavity mode and $S_{\gamma\,g}$ describes the mutual interaction between the two previous terms. The full action can be written as\footnote{The Latin (lowercase) indices are spatial; the Greek (lowercase) indices run instead 
from $0$ to $3$; see, in this respect, appendix \ref{APPA} and discussion therein.}
\begin{eqnarray}
S&=&  \frac{1}{8 \ell_{P}^2} \int d^{4} x \,\, a^2(\tau)\, \, \biggl[ 
\partial_{\tau} h_{i\,j} \,\partial_{\tau} h^{\,\,i\,j} - \partial_{k} h_{i\,j} \,\partial^{k} h^{\,\,i\,j}\biggr]
\nonumber\\
&+& \frac{1}{2} \int d^{4} x\, (E^2 - B^2) -  \frac{1}{4}\int d^{4} x  (E_{i} E_{j} + B_{i} B_{j}) \, \,h^{i\, j} 
\nonumber\\
&-& \frac{1}{4}\int d^{4} x\,(E^{i} E^{j} + B^{i} B^{j})\,\, h_{i\, j}.
\label{AC1}
\end{eqnarray}
Equation (\ref{AC1}) has been written in terms of the Planck length $\ell_{P} = \sqrt{8 \pi G}$ and assumes the natural system of units $\hbar= c= 1$. The first line of Eq. (\ref{AC1}) corresponds to $S_{g}$ and this is the 
action originally derived by Ford and Parker in conformally flat background geometries \cite{IOTA2}
(see also \cite{ACF2} and \cite{ACF3}). The second term is the free action of the cavity mode, i.e. $S_{\gamma}$; the last two terms of Eq. (\ref{AC1}) give the interaction term $S_{\gamma\,g}$. We remind that tensor amplitude $h_{i\,j}$ appearing in Eq. (\ref{AC2}) is solenoidal and traceless, i.e. $\partial_{i} h^{\,\,i}_{j} = h_{i}^{\,\,\,i} =0$ (see also Eqs. (\ref{AP1a})--(\ref{AP1b}) and discussion thereafter). The vectors $\vec{E}$ and $\vec{B}$ are the comoving electric and magnetic fields and their relation to the physical fields\footnote{In terms of the physical fields $\vec{e}(\vec{x}, \tau)$ and $\vec{b}(\vec{x},\tau)$ the components of field strength are  $F^{i\,0} = e^{i}(\vec{x},\tau)/a^2(\tau)$ and $F^{i\,j} = - \epsilon^{i\,j\,k} b_{k}/a^2(\tau)$.} 
is  given by $\vec{E}(\vec{x},\tau) = a^2(\tau) \,\,\vec{e}(\vec{x}, \tau)$ and $\vec{B}(\vec{x},\tau) = a^2(\tau) \,\,\vec{b}(\vec{x},\tau)$; as usual in Eq. (\ref{AC1}) the notations $E^2= \vec{E}\cdot \vec{E}$ and $E^2= \vec{B}\cdot \vec{B}$ have been employed. Finally the contributions  
$S_{\gamma}$ and $S_{\gamma\, g}$ (second and third lines in Eq. (\ref{AC1})) can also be expressed by using the vector potential  in the Coulomb gauge, i.e. $\vec{E} = - \partial_{\tau} \vec{A}$ and $\vec{B} = \vec{\nabla} \times \vec{A}$:
\begin{eqnarray}
S_{\gamma} + S_{\gamma\,g} &=& \frac{1}{2} \int d^{4}x \biggl[ \bigl(\partial_{\tau} A_{m}\bigr) \bigl(\partial_{\tau} A^{m}\bigr) - \bigl(\partial_{k} A_{m}\bigr) \bigl(\partial^{k} A^{m}\bigr) \biggr]
\nonumber\\
&-& \frac{1}{4} \int d^{4} x \biggl[\partial_{\tau} A_{i} \, \partial_{\tau} A_{j}+ (\vec{\nabla}\times \vec{A})_{i} (\vec{\nabla}\times \vec{A})_{j}\biggr] h^{i\, j}
\nonumber\\
&-& \frac{1}{4} \int d^{4} x \biggl[\partial_{\tau} A^{i} \, \partial_{\tau} A^{j}+ (\vec{\nabla}\times \vec{A})^{i} (\vec{\nabla}\times \vec{A})^{j}\biggr] h_{i\, j}.
\label{AC2}
\end{eqnarray}
The explicit expressions for $(\vec{\nabla}\times \vec{A})_{i} (\vec{\nabla}\times \vec{A})_{j}$ can be 
found in appendix \ref{APPA} (see, in particular, Eq. (\ref{APG7})); as explained 
there the integrands of the second and third  terms of Eq. (\ref{AC2}) can also be written as
$\Pi^{i,\,j} \, h_{i\,j}$ and as $\Pi_{i,\,j} \, h^{i\,j}$ where $\Pi_{i}^{\,\,j}$ is the anisotropic stress. Indeed from the total energy-momentum tensor of the cavity modes 
\begin{equation}
T_{\mu}^{\,\,\nu} = - F_{\mu\alpha} \, F^{\nu\,\alpha} + \delta_{\mu}^{\nu}\,\,F_{\alpha\beta} \, F^{\alpha\beta}/4,
\label{AC3}
\end{equation}
the terms with purely spatial components (i.e. $T_{i}^{\,\,j}$) can be written as $T_{i}^{\,\,\,j} = \Pi_{i}^{\,\,\,j}/a^4 - (p_{E}/a^4 + p_{B}/a^4) \delta_{i}^{\,\,j}$ where $p_{E} = E^2/6$ and $p_{B}= B^2/6$ while
\begin{equation} 
\Pi_{i}^{\,\,\,j} = E_{i}\, E^{j} + B_{i}\, B^{j} - (B^2 + E^2) \delta_{i}^{j}/3.
\label{AC4}
\end{equation}
Thanks to Eq. (\ref{AC4}) the interaction terms of Eq. (\ref{AC1}) can also be expressed as
\begin{eqnarray}
S_{\gamma g} &=& - \frac{1}{4} \int d^{4} x \, \sqrt{- \overline{g}} \,\,T_{i\, j} \,\, h^{i\,j} - 
\frac{1}{4} \int d^{4} x \, \sqrt{- \overline{g}}\,\, T^{i\, j} \,\, h_{i\,j} 
\nonumber\\
&=& -\frac{1}{4}\int d^{4} x  (E_{i} E_{j} + B_{i} B_{j}) \, \,h^{i\, j}  - \frac{1}{4}\int d^{4} x\,(E^{i} E^{j} + B^{i} B^{j})\,\, h_{i\, j},
\label{AC5}
\end{eqnarray}
where $\overline{g}$ indicates the determinant of the background metric. Although not strictly necessary we like to distinguish the terms $h^{i\,j}$ from their covariant analogs; this implies that, in the variation, $h_{i\,j}$ and $h^{i\, j}$ are considered independent.
The second line of Eq. (\ref{AC5}) clarifies that all the terms proportional to the Kroeneker deltas appearing in $\Pi_{i\,j}$ vanish when contracted with the (traceless) tensor 
amplitude $h^{i\,j}$. When we consider a single polarization (e.g. $h= h^{x \, x}$, where $x$ denotes the first Cartesian axis) it would 
be tempting to write the interaction term as 
\begin{equation}
S_{g\,\gamma} = \frac{1}{6} \int d^{4} x (B^2 + E^2) \,\,h_{x}^{\, x},
\label{AC5a}
\end{equation}
having assumed an appropriate geometry of the cavity. This choice is however inconsistent with the traceless nature of the fluctuation (i.e. $h_{i}^{\,\,i}=0$), as we shall discuss at the end of this section. Finally the action (\ref{AC1}) (and its associated Lagrangian density ${\mathcal L}_{g}(\vec{x},\tau)$)
can be expressed in a different way  by introducing the rescaled tensor amplitudes $\mu_{i j} = a\, h_{i\,j}$ and $\mu^{i\,j} = a \, h^{i\,j}$. The result of this transformation appears in Eq. 
(\ref{APG5}) and ${\mathcal L}_{g}(\vec{x},\tau)$ becomes\footnote{In Eq. (\ref{AC6}) we employ the usual notations, i.e. ${\mathcal H} = a^{\prime}/a$ where the prime denotes a derivation with respect to the cosmic time coordinate $\tau$; we remind, in this respect, that ${\mathcal H} = a \, H$ and $H$ is the standard Hubble rate (see also appendix \ref{APPA}).}
\begin{equation}
{\mathcal L}_{g}(\vec{x},\tau) = \frac{1}{8 \ell_{P}^2} \biggl\{(\partial_{\tau} \mu_{i\, j}) (\partial_{\tau} \mu^{i\,j}) + {\mathcal H}^2 \mu_{i\,j} \mu^{i\,j} - {\mathcal H} \biggl[\mu_{i\,j} (\partial_{\tau} \mu^{i\,j}) + (\partial_{\tau} \mu_{i\, j}) \, \mu^{i\,j}\biggr]\biggr\},
\label{AC6}
\end{equation}
where, by definition, $S_{g}= \int d\, \tau L_{g}(\tau)$ and $L_{g}(\tau) = \int d^{3} x\, {\mathcal L}_{g}(\vec{x},\tau)$.

\subsection{The canonical Hamiltonian}
From Eq. (\ref{AC6}) the canonical momenta can be written as $\pi_{i\,j}= (\partial_{\tau} \mu_{i\,j} - {\mathcal H} \mu_{i\,j})/(8 \ell_{P}^2)$ and $\pi^{i\,j}= (\partial_{\tau} \mu^{i\,j} - {\mathcal H} \mu^{i\,j})/(8 \ell_{P}^2)$. Thus, the gravitational part of the canonical Hamiltonian is:
\begin{eqnarray}
H_{g}(\tau) &=& \int \, d^{3} \biggl[\pi_{i\,j} \partial_{\tau} \mu^{i\,j} + \pi^{i\,j} \partial_{\tau} \mu_{i\,j} - {\mathcal L}_{g}(\vec{x},\tau) \biggr]
\nonumber\\
&=& \int d^{3} x \biggl[ 8 \ell_{P}^2 \pi_{i\,j} \,\,\pi^{i\,j} + 
{\mathcal H} \bigl(\mu_{i\,j} \,\,\pi^{i\,j} + \mu^{i\, j} \,\,\pi_{i\,j} )+ 
\frac{1}{8\ell_{P}^2} \partial_{k}\mu_{i\,j} \partial^{k} \mu^{i\,j}\biggr].
\label{AC7}
\end{eqnarray}
With the same procedure we obtain the canonical Hamiltonian of the gauge part of the action
after recalling that the conjugate  momentum is, in this case, $\Pi_{m} = \partial_{\tau} A_{m}$; the results 
for $H_{\gamma}(\tau)$ and $H_{\gamma\, g}(\tau)$ are:
\begin{eqnarray}
H_{\gamma}(\tau) &=&  \frac{1}{2} \int d^{3} x \biggl[ \Pi_{m}^2 + (\partial_{k} A_{m})^2 \biggr],
\nonumber\\
H_{\gamma \,g}(\tau) &=& -\frac{1}{4\, a} \int d^{3} x \biggl[ \Pi_{i}\, \Pi_{j} + \epsilon_{m n i} \epsilon_{a b j} 
(\partial^{m} A^{n} )\,(\partial^{a} A^{b})\biggr] \mu^{i\,j} 
\nonumber\\
  &-& \frac{1}{4\, a} \int d^{3} x \biggl[\Pi^{i}\, \Pi^{j} + \epsilon^{m n i} \epsilon^{a b j} 
(\partial_{m} A_{n} )\,(\partial_{a} A_{b}) \biggr] \,\mu_{i\,j}. 
\label{AC8}
\end{eqnarray}
The total Hamiltonian of the problem will then be given by the sum of the three different 
contributions:
\begin{eqnarray}
H(\tau) &=& \int d^{3} x \biggl[ 8 \ell_{P}^2 \pi_{i\,j} \pi^{i\,j} + 
{\mathcal H} \bigl(\mu_{i\,j} \pi^{i\,j} + \mu^{i\, j} \pi_{i\,j} )+ 
\frac{1}{8\ell_{P}^2} \partial_{k}\mu_{i\,j} \partial^{k} \mu^{i\,j}\biggr]
\nonumber\\
&+& \frac{1}{2} \int d^{3} x \biggl[ \Pi_{m}^2 + (\partial_{k} A_{m})^2 \biggr]
+ \frac{1}{2\, a} \int d^{3} x \biggl[ \Pi_{i}\, \Pi^{j} + \epsilon_{m n i} \epsilon^{a b j} 
(\partial^{m} A^{n} )\,(\partial_{a} A_{b})\biggr] \mu^{i}_{\,\,j}.
\label{HAM3}
\end{eqnarray}

\subsection{Polarizations and cavity interactions}
The Hamiltonian (\ref{HAM3}) describes the interaction of the photons 
and the gravitons but in this paragraph we consider the general restrictions 
imposed by the cavity modes of the electromagnetic field. We recall that if the propagation of the tensor disturbance is directed along a certain direction $\hat{k}$ the two tensor polarizations can be written as:
\begin{eqnarray}
e_{i\,j}^{\oplus}(\hat{k}) = \hat{m}_{i} \hat{m}_{j} - \hat{n}_{i} \hat{n}_{j}, \quad 
e_{i\,j}^{\otimes}(\hat{k}) = \hat{m}_{i} \hat{n}_{j} + \hat{n}_{i} \hat{m}_{j}, 
\label{POL1}
\end{eqnarray}
where $\hat{m}$, $\hat{n}$ and $\hat{k}$ three mutually orthogonal unit vectors 
obeying $\hat{m} \times \hat{n} = \hat{k}$. According to Eq. (\ref{POL1}) both polarizations 
are traceless and transverse with respect to the propagation direction $\hat{k}$. For simplicity we now consider a plane wave propagating along $\hat{k}$ and write the tensor amplitude as 
\begin{equation}
h_{i\, j}(\vec{x}, \tau) = \sum_{\alpha= \oplus,\otimes} h_{\alpha}(\vec{x},\tau) \, 
e_{i\,j}^{\alpha}(\hat{k}).
\label{POL2}
\end{equation}
Using Eqs. (\ref{POL1})--(\ref{POL2}), the interaction Hamiltonian can be written as 
\begin{eqnarray}
H_{\gamma\, g}(\tau) &=& \frac{1}{4} \int d^{3}\, x (E_{i} E_{j} + B_{i} B_{j}) h^{i\, j} + \frac{1}{4} \int d^{3}\, x (E^{i} E^{j} + B^{i} B^{j}) h_{i\, j} 
\nonumber\\
&=& \frac{1}{2} \int d^{3}\, x \biggl\{ h_{\oplus} \bigl[ (\vec{E}\cdot\hat{m})^2 +(\vec{B}\cdot\hat{m})^2 - (\vec{E}\cdot\hat{n})^2  -(\vec{B}\cdot\hat{n})^2 \bigr] 
\nonumber\\
&+& 2 h_{\otimes} \bigl[ (\vec{E}\cdot\hat{m}) (\vec{E}\cdot\hat{n})+ 
(\vec{B}\cdot\hat{m}) (\vec{B}\cdot\hat{n})\bigr]\biggr\}.
\label{POL3}
\end{eqnarray}
\begin{figure}[!ht]
\centering
\includegraphics[height=8cm]{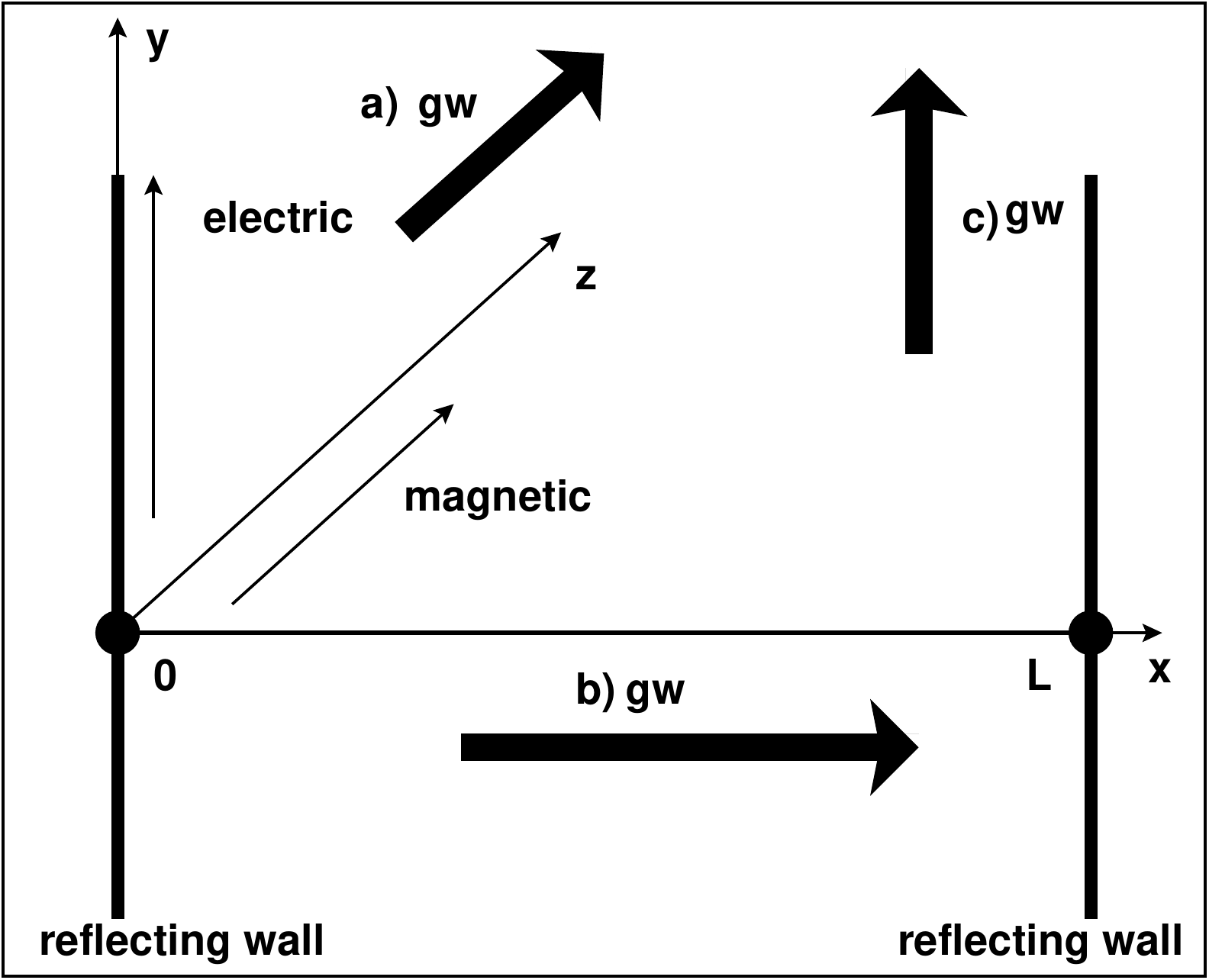}
\caption[a]{We illustrate the geometry of a cavity where the walls are located along the $\hat{x}$ direction. If $\hat{k} = \hat{x}$ the gravitational wave propagates long the cavity axis. In the other cases the gravitational wave propagates orthogonally to the axis of the cavity. }
\label{Figure1}      
\end{figure}
We now consider a cavity with perfectly reflecting walls of length $L$ and directed 
along $\hat{x}$ (i.e. the direction of the first Cartesian axis). The transverse dimensions have length 
$L_{\perp}$ so that the volume of the cavity is $V= L \, L_{\perp}^2$. In Fig. \ref{Figure1} we schematically 
illustrate the cavity oriented along $\hat{x}$ and we distinguish three physical situations 
[denoted by $a)$, $b)$ and $c)$] that are specifically described in what follows.
\begin{itemize}
\item{} If we suppose that $\hat{k} = \hat{z}$ [corresponding to the case $a)$ in Fig. \ref{Figure1}], in Eq. (\ref{POL3}) $\hat{m} = \hat{x}$ and $\hat{n} = \hat{y}$ since, by construction, 
$\hat{x} \times \hat{y} = \hat{z} = \hat{k}$. In case the walls are located on the $\hat{x}$ 
axis (as in Fig. \ref{Figure1}) then the vector potential and the electric field are polarized 
along $\hat{y}$ (i.e. $\vec{E} = E_{y} \hat{y}$) while $\vec{B} = B_{z} \hat{z}$. Thus the only term surviving in Eq. (\ref{POL3}) is $ - h_{\oplus} E_{y}^2$. 
\item{} If $\hat{k}$ is instead oriented along the same direction of the cavity, i.e. $\hat{k} = \hat{x}$ this 
corresponds to the case $b)$ of Fig. \ref{Figure1}. We must have that $\hat{m} = \hat{y}$ and $\hat{n} =\hat{z}$ since $\hat{y} \times \hat{z} = \hat{x} = \hat{k}$.
But because $\vec{E} = E_{y} \hat{y}$ the term surviving in Eq. (\ref{POL3}) becomes
$h_{\oplus}(\vec{k},\tau) ( E_{y}^2 - B_{z}^2)$. 
\item{} Finally, when $\hat{k} = \hat{y}$ (see case $c)$ in Fig. \ref{Figure1}) we have that 
$\hat{m} = \hat{z}$ and $\hat{n} = \hat{x}$; the only term surviving in Eq. (\ref{POL3}) is given by
$B_{z}^2 h_{\oplus}$. 
\end{itemize}
In Fig. \ref{Figure1} we illustrated the basic geometry of the ideal cavity that will be discussed in the 
forthcoming sections. The construction suggested in Fig. \ref{Figure1} can be slightly 
modified: we could for instance identify the unit vector $\hat{k}$ with a generic radial direction 
$\hat{r} = (\cos{\varphi} \sin{\vartheta}, \sin{\varphi} \sin{\vartheta}, \cos{\vartheta})$. In this case 
$\hat{m} = \hat{\vartheta}= (\cos{\varphi} \cos{\vartheta}, \sin{\varphi} \cos{\vartheta}, - \sin{\vartheta})$
and $\hat{n} = \hat{\vartheta}= (- \sin{\varphi}, \cos{\varphi}, 0)$. If the direction of the cavity is now fixed 
along a specific Cartesian direction, the same arguments discussed above can be repeated. 
The difference between the interaction terms along the various directions is only apparent 
since, as it will be shown, in the quantum Hamiltonian what matters are the leading terms in the rotating-wave 
approximation. Finally, these considerations developed here demonstrate that the interaction term of Eq. (\ref{AC5a}) does not properly account of the polarization structure of the problem. It is however true that 
also that interaction term is of the same order of the others in the rotating-wave 
approximation (see the related discussion in section \ref{sec4}).

\renewcommand{\theequation}{4.\arabic{equation}}
\setcounter{equation}{0}
\section{Relic gravitons and the cavity modes}
\label{sec4}
\subsection{General considerations}
The light beams are often treated as exciting a single
mode of the field and this common approach is customarily 
employed in the quantum optical analysis of the HBT correlations. 
The plane parallel light beams have thus transverse
intensity profiles that are not essential for the measurements. 
We are now going to discuss the interaction between gravitons and 
the quantized excitations of the  electromagnetic field
confined inside a closed optical resonator with perfectly reflecting walls.
The field excitations are limited to a discrete set of spatial modes determined by the boundary conditions at the cavity walls and while higher-order modes always physically exist,  the fundamental mode dominates the  coupling strength.

\subsection{Cavity Hamiltonian}
As illustrated in Fig. \ref{Figure1} we consider the radiation field
confined to a one-dimensional cavity along the $x$-axis with perfectly conducting
walls at $x = 0$ and $x= L$. Thanks to Eq. (\ref{HAM3}), the concrete 
Hamiltonian of the problem consists of the cavity part supplemented, respectively, by the graviton contribution and by the 
interaction terms. For we are considering a cavity directed along $\hat{m} = \hat{x}$ the vector potential is given by
\begin{equation}
A_{y}(x,\tau) = \sum_{n = 1}^{\infty} {\mathcal A}_{n}(\tau) u_{n}(x), \qquad
u_{n}(x) = \sqrt{\frac{2}{L\, S}} \sin{k_{n} x },
\label{HAM3a}
\end{equation}
where $L$ is the length of the cavity, $S = L_{\perp}^2$ is its transverse section and $k_{n} = n \, \pi/L$;
by definition  $V_{c} = L\, S= L\, L_{\perp}^2$ is the volume of the cavity
With the help of Eq. (\ref{HAM3a}) the classical Hamiltonian of the fundamental mode  
(conventionally denoted hereunder by $\omega_{c} = \pi/L$) becomes 
\begin{eqnarray}
H_{c}(\tau) =  \frac{1}{2} \int d^{3} x \biggl[ \Pi_{y}^2 + (\partial_{x} A_{y})^2\biggr]= \frac{{\mathcal A}^{\prime\,2}}{2\pi} \int_{0}^{\pi} [1 - \cos{2 \alpha}] \, d\alpha 
+ \frac{\omega_{c}^2 \, {\mathcal A}^{2}}{2\pi}  \int_{0}^{\pi} [1 + \cos{2 \alpha}] \, d\alpha,
\label{HAM4}
\end{eqnarray}
where the stenographic notation 
${\mathcal A}^{\prime} = \partial_{\tau} {\mathcal A}$ has been employed. The classical fields can now be promoted to the status of quantum mechanical operators obeying 
canonical commutation relations at equal times:
\begin{equation} 
{\mathcal A}(\tau) \to \widehat{q}_{c}(\tau), \quad {\mathcal A}^{\prime}(\tau) \to \widehat{p}_{c}(\tau), 
\quad [ \widehat{q}_{c}(\tau), \widehat{p}_{c}(\tau) ] = i.
\label{HAM4a}
\end{equation}
From the indicated integrations of Eq. (\ref{HAM4}) and after the subtraction of the zero-point energy 
the quantum Hamiltonian $\widehat{H}_{c}(\tau)$ becomes
\begin{equation}
\widehat{H}_{c}(\tau) = \omega_{c} \widehat{c}^{\dagger} \, \widehat{c}, \quad \widehat{q}_{c} = (\widehat{c} + \widehat{c}^{\dagger})/\sqrt{2 \omega_{c}}, \quad  \widehat{p}_{c} = - i\, \sqrt{\omega_{c}/2} ( \widehat{c} - \widehat{c}^{\dagger}),
\label{HAM5}
\end{equation}
where $[\widehat{c}, \widehat{c}^{\dagger}] =1$. In quantum optical applications the electric field of the cavity is sometimes associated with the canonical field while the magnetic field with the canonical momentum (see e.g. \cite{loudon} and also \cite{loudon2}). Here we use instead, as canonical variables, 
the vector potential and its time derivative as it is customary when quantizing the electromagnetic field in the Coulomb gauge \cite{BDF} which is invariant under Weyl rescaling of the four-dimensional metric \cite{LFC}. In our geometry (and in the Coulomb gauge) the canonical momentum corresponds to $\Pi_{y} =  \partial_{\tau} A_{y} = - E_{y}$ and this is why we associated $\widehat{p}_{c}$ to ${\mathcal A}^{\prime}(\tau)$. 

\subsection{Graviton Hamiltonian}
To derive the Hamiltonian of the gravitons we first express 
the appropriately normalized field operators $\widehat{\mu}_{i\,j}(\vec{x}, \tau)$ and $\widehat{\pi}_{i\,j}(\vec{x}, \tau)$ in Fourier space:
\begin{eqnarray}
\widehat{\mu}_{i\,j}(\vec{x}, \tau) &=& \frac{\sqrt{2} \ell_{P}}{(2 \pi)^{3/2}} \, 
\sum_{\alpha} \int d^{3} k\,\, e^{(\alpha)}_{i\,j}(\hat{k})\,\widehat{\mu}_{\vec{k}, \alpha} \,\, e^{ - i \vec{k}\cdot\vec{x}}, 
\label{HAMG1}\\
\widehat{\pi}_{i\,j}(\vec{x}, \tau) &=& \frac{1}{4 \sqrt{2} \ell_{P} \,(2 \pi)^{3/2}} \,\,
\sum_{\alpha} \int d^{3} k\,\, e^{(\alpha)}_{i\,j}(\hat{k})\,\widehat{\pi}_{\vec{k}, \alpha} \,\, e^{ - i \vec{k}\cdot\vec{x}},
\label{HAMG2}
\end{eqnarray}
where $\alpha = \oplus, \,\otimes$ since the sum runs over the two tensor 
polarizations introduced in Eq. (\ref{POL1}).
Since the quantum fields and the canonical momenta are both Hermitian, the corresponding operators in Fourier space obey $\widehat{\mu}_{\vec{k}, \alpha}^{\dagger} = \widehat{\mu}_{-\vec{k}, \alpha}$ and $\widehat{\pi}_{\vec{k}, \alpha}^{\dagger} = \widehat{\pi}_{-\vec{k}, \alpha}$. It is also convenient to adopt a decomposition that is explicitly Hermitian (but formally equivalent to the ones of Eqs. (\ref{HAMG1})--(\ref{HAMG2})):
\begin{equation}
\widehat{\mu}_{i\,j}(\vec{x}, \tau) = \frac{\ell_{P}}{\sqrt{2} (2 \pi)^{3/2}} \, 
\sum_{\alpha} \int d^{3} k\,\, e^{(\alpha)}_{i\,j}(\hat{k})\,\,\biggl[\widehat{\mu}_{\vec{k}, \alpha} \, e^{ - i \vec{k}\cdot\vec{x}} + \widehat{\mu}_{\vec{k}, \alpha}^{\dagger} \, e^{  i \vec{k}\cdot\vec{x}}\biggr],
\label{HAMG1a}
\end{equation}
and similarly for $\widehat{\pi}_{i\,j}(\vec{x}, \tau)$; in the overall normalization there is a factor $2$ between Eq. (\ref{HAMG1}) and (\ref{HAMG1a}) accounting for the presence of a further term 
in Fourier space. In terms of the creation and annihilation operators we have that 
\begin{equation}
\widehat{\mu}_{\vec{k}, \alpha} = (\widehat{a}_{\vec{k},\alpha} + \widehat{a}_{-\vec{k},\alpha}^{\dagger})/\sqrt{2 k}, \qquad \pi_{\vec{k}, \alpha} = 
- i\,\sqrt{k/2} (\widehat{a}_{\vec{k},\alpha} - \widehat{a}_{-\vec{k}, \alpha}^{\dagger}).
\label{HAMG3}
\end{equation}
Since $[\widehat{a}_{\vec{k}, \alpha}, \widehat{a}_{\vec{p}, \beta}^{\dagger}] = \delta_{\alpha\beta} \,\delta^{(3)}(\vec{k} - \vec{p})$ we also 
have that  $[\widehat{\mu}_{\vec{k}, \alpha}, \,  \pi_{\vec{p}, \beta} ] = i \, \delta_{\alpha\beta} \, \delta^{(3)}(\vec{k} + \vec{p})$ while $[\widehat{\mu}_{\vec{k}, \alpha}, \,  \pi_{\vec{p}, \beta}^{\dagger} ] = i \, \delta_{\alpha\beta} \, \delta^{(3)}(\vec{k} - \vec{p})$; it then follows from Eq. (\ref{AC7}) that 
the Hamiltonian of the relic gravitons is given by 
\begin{eqnarray}
\widehat{H}_{g}(\tau) = \frac{1}{2} \sum_{\alpha} \, \int d^{3} k \, \biggl[ 
k( \widehat{a}_{\vec{k}, \alpha}^{\dagger} \,  \widehat{a}_{\vec{k}, \alpha} + 
 \widehat{a}_{-\vec{k}, \alpha}\,  \widehat{a}_{-\vec{k}, \alpha}^{\dagger}) 
 + i {\mathcal H} \,\,(\widehat{a}_{\vec{k}, \alpha}^{\dagger} \,  \widehat{a}_{-\vec{k}, \alpha}^{\dagger} - \widehat{a}_{\vec{k}, \alpha}\,  \widehat{a}_{-\vec{k}, \alpha})\biggr],
\label{HAMG4}
\end{eqnarray}
where, as before, ${\mathcal H} = a^{\prime}/a$ indicates the coupling to the space-time curvature.

\subsection{Interaction Hamiltonian and rotating wave approximation}
Thanks to Eq. (\ref{POL3}), the interaction between the cavity and the gravitons can be written as 
\begin{eqnarray}
\widehat{H}_{g\,c}(\tau) &=& \frac{\sqrt{2}\ell_{P}}{4 a (2\pi)^{3/2}}  \int d^{3} x \int d^{3} k \biggl\{ {\mathcal C}_{\oplus}(\vec{x},\hat{m},\hat{n})\biggl[ \widehat{\mu}_{\vec{k},\oplus} e^{ - i \vec{k}\cdot\vec{x}} + \widehat{\mu}_{\vec{k},\oplus}^{\dagger} \, e^{ i \vec{k}\cdot\vec{x}} \biggr] 
\nonumber\\
&+& {\mathcal C}_{\otimes}(\vec{x},\hat{m},\hat{n}) \biggl[\widehat{\mu}_{\vec{k},\otimes} e^{ - i \vec{k}\cdot\vec{x}} + \widehat{\mu}_{\vec{k},\otimes}^{\dagger} \, e^{ i \vec{k}\cdot\vec{x}} \biggr]\biggr\},
\label{HAMG5}
\end{eqnarray}
where, for convenience, we introduced the two  terms
${\mathcal C}_{\oplus}(\vec{x},\hat{m},\hat{n})$ and ${\mathcal C}_{\otimes}(\vec{x},\hat{m},\hat{n})$
that depend upon the geometry of the cavity and upon the tensor polarizations:
\begin{eqnarray}
{\mathcal C}_{\oplus}(\vec{x},\hat{m},\hat{n}) &=& (\vec{E}\cdot\hat{m})^2 + (\vec{B}\cdot\hat{m})^2 - (\vec{E}\cdot\hat{n})^2 - (\vec{B}\cdot\hat{n})^2,
\nonumber\\
{\mathcal C}_{\otimes}(\vec{x},\hat{m},\hat{n}) &=& 2 [ (\vec{E}\cdot\hat{m}) (\vec{E}\cdot\hat{n}) +(\vec{B}\cdot\hat{m}) (\vec{B}\cdot\hat{n})].
\label{HAMG5a}
\end{eqnarray}

\subsubsection{Parallel and orthogonal propagation}
The interaction Hamiltonian seems to depend upon the propagation 
of the gravitons: it is maximal when $\hat{k}$ lies in the plane orthogonal to the axis 
of the cavity and it is minimal when $\hat{k}$ is aligned 
with the direction of the cavity. To clarify this point we first analyze two simple cases and then introduce the general parametrization. 
If $\hat{k} = \hat{z}$ (with $\hat{m} = \hat{x}$ and $\hat{n} = \hat{y}$) we obtain $\vec{E}\cdot\hat{m} = \vec{B}\cdot\hat{m} = \vec{B}\cdot\hat{n} =0$ and the only term that survives in Eq. (\ref{HAMG5a}) is ${\mathcal C}_{\oplus}(\vec{x},\hat{m},\hat{n}) = -(\vec{E}\cdot\hat{n})^2 = -(\vec{E}\cdot\hat{y})^2$. In this case the grvitons propagate orthogonally to the axis of the cavity and the explicit form of the interaction Hamiltonian is:
\begin{equation}
 \widehat{H}_{g\,c}(\tau) = - \frac{\sqrt{2}\ell_{P}}{4 a (2\pi)^{3/2}}  \int d^{3} x \int d^{3} k \, (\vec{E}\cdot\hat{n})^2 \biggl[ \widehat{\mu}_{\vec{k},\oplus} e^{ - i \vec{k}\cdot\vec{x}} + \widehat{\mu}_{\vec{k},\oplus}^{\dagger} \, e^{ i \vec{k}\cdot\vec{x}} \biggr].
 \label{HAMG6}
 \end{equation}
We now recall that $(\vec{E}\cdot\hat{n})^2 = E_{y}^2 = (\partial_{\tau}A_{y})^2$. Thus 
$(\vec{E}\cdot\hat{n})^2 = 2 \, \widehat{p}_{c}^2 \sin^2{\omega_{c} x}/(L \,L_{\perp}^2)$
(where $\widehat{p}_{c} = \widehat{{\mathcal A}}^{\prime}$); as before, $\omega_{c}$ is the fundamental mode of the cavity. Once this result is inserted into Eq. (\ref{HAMG6})  we obtain 
\begin{eqnarray}
\widehat{H}_{g\,c}(\tau) &=& - \frac{\sqrt{2}\, \ell_{P}\, \widehat{p}_{c}^2}{2 a (2\pi)^{3/2} L L_{\perp}^2} \int d^{3} k\,  \int_{0}^{L} d x\, \int_{0}^{L_{\perp}} d y \, \int_{0}^{L_{\perp}} d z \sin^2{\omega_{c} x}
\biggl[ \widehat{\mu}_{\vec{k},\oplus} e^{ - i k z} + \widehat{\mu}_{\vec{k},\oplus}^{\dagger} \, e^{ i k z} \biggr],
\label{HAMG6b}\\
&=& - \frac{\sqrt{2}\, \ell_{P}\, \widehat{p}_{c}^2}{4 a (2\pi)^{3/2}} \int d^{3} k\,  
\biggl[ \widehat{\mu}_{\vec{k},\oplus} W(k L_{\perp}) + \widehat{\mu}_{\vec{k},\oplus}^{\dagger} W^{\ast}(k L_{\perp})  \biggr],
\label{HAMG6c}
\end{eqnarray}
where the complex function 
\begin{equation}
W(k L_{\perp})= ( 1 - e^{- i k L_{\perp}})/(i k L_{\perp}),  \qquad W^{\ast}(k L_{\perp})= (e^{i k L_{\perp}} - 1)/(i k L_{\perp}),
\label{FUNC}
\end{equation}
has been introduced and will appear frequently in the following discussions. For the long-wavelength modes we have 
$k L_{\perp} \ll 1$ and $W(k L_{\perp}) \to 1$; in the opposite limit $|W(k L_{\perp})|^2 \to 0$ when $| k L_{\perp} | > 1$. The interaction Hamiltonian can the be written as:
\begin{eqnarray}
\widehat{H}_{g\,c}(\tau) = - \frac{\sqrt{2}\, \ell_{P}\, \widehat{p}_{c}^2}{4 a (2\pi)^{3/2}} \int \frac{d^{3} k}{\sqrt{2 k}}\,  
\bigl[ \widehat{a}_{\vec{k}, \oplus} + \widehat{a}_{- \vec{k},\oplus}^{\dagger} + 
 \widehat{a}_{-\vec{k}, \oplus} + \widehat{a}_{ \vec{k},\oplus}^{\dagger}\bigr].
\label{HAMG6d}
\end{eqnarray}
The structure of the interaction Hamiltonian reflects the two-mode structure 
of Eq. (\ref{HAMG4}): since gravitons are created in pairs of opposite three-momenta
they interact accordingly with the cavity mode. Let us then analyze the case $\hat{k} = \hat{y}$ (with $\hat{m} = \hat{z}$ and $\hat{n} = \hat{x})$; also in this situation the gravitons propagate orthogonally to the 
cavity and if we repeat the same procedure described above we do obtain 
a slightly different interaction term, namely:
\begin{eqnarray}
\widehat{H}_{g\, c}(\tau) &=& \frac{\sqrt{2}\, \ell_{P}\, \widehat{q}_{c}^2}{4 a (2\pi)^{3/2}} \int d^{3} k\,  
\biggl[ \widehat{\mu}_{\vec{k},\oplus} W(k L_{\perp}) + \widehat{\mu}_{\vec{k},\oplus}^{\dagger} W^{\ast}(k L_{\perp})  \biggr],
\nonumber\\
&\to& \frac{\sqrt{2}\, \ell_{P}\, \widehat{q}_{c}^2}{4 a (2\pi)^{3/2}} \int \frac{d^{3} k}{\sqrt{2 k}}\,  
\biggl[ \widehat{a}_{\vec{k}, \oplus} + \widehat{a}_{- \vec{k},\oplus}^{\dagger} + 
 \widehat{a}_{-\vec{k}, \oplus} + \widehat{a}_{ \vec{k},\oplus}^{\dagger}\biggr],
\label{HAMG6f}
\end{eqnarray}
where the second expression holds in the same limit of Eq. (\ref{HAMG6d}),  $W(k L_{\perp}) 
=W^{\ast}(k L_{\perp}) \to 1$ for $|k \, L_{\perp} | \ll 1$. The two previous examples refer to the situation where the gravitons 
propagate orthogonally to the cavity: so we finally consider a propagation 
along the direction of the cavity (i.e. $\hat{k} = \hat{x}$) when $\hat{m} = \hat{y}$ 
and $\hat{n} = \hat{z}$. In this geometry ${\mathcal C}_{\oplus}(\vec{x},\hat{m},\hat{n}) = 
E_{y}^2 - B_{z}^2$ and the spatial integration along $x$ implies that the interaction 
Hamiltonian is:
\begin{eqnarray}
\widehat{H}_{g\, c}(\tau) &=& \frac{\ell_{P} \,\, (\widehat{p}_{c}^2 - \omega_{c}^2 \widehat{q}_{c}^2)}{4 \, \sqrt{2} \, a (2\pi)^{3/2}} \int \, d^{3} k\,\, \biggl\{ \widehat{\mu}_{\vec{k},\oplus}\biggl[ 2 \,W^{\ast}(k \, L) -W^{\ast}[ k\, L(1 + 2\omega_{c}/k)] - 
W^{\ast}[k\, L( 1 - 2 \omega_{c}/k)] \biggr]
\nonumber\\
&+& \widehat{\mu}_{\vec{k},\oplus}^{\dagger}\biggl[ 2 \,W(k \, L) -  W[ k\, L(1 + 2\omega_{c}/k)] - 
W[k\, L( 1 - 2 \omega_{c}/k)]\biggr]\biggr\}.
\label{HAMG6g}
\end{eqnarray}
The contributions containing $2\omega_{c}/k = 2\pi/(k\, L)$ imply that 
$W[ k\, L ( 1 \pm 2 \omega_{c}/k)] = W( k\, L \pm 2 \pi)$. Thus for $|k\, L| <1$ 
we have 
\begin{eqnarray}
\widehat{H}_{g\, c}(\tau) &=& \frac{\ell_{P} \,\, (\widehat{p}_{c}^2 - \omega_{c}^2 \widehat{q}_{c}^2)}{2 \, \sqrt{2} \, a (2\pi)^{3/2}} \int \, d^{3} k\,\, \biggl[ \widehat{\mu}_{\vec{k},\oplus} + \widehat{\mu}_{\vec{k},\oplus}^{\dagger} \biggr]
\nonumber\\
&\to& \frac{\sqrt{2}\, \ell_{P}\, (\widehat{p}_{c}^2 - \omega_{c}^2 \widehat{q}_{c}^2)}{4 a (2\pi)^{3/2}} \int \frac{d^{3} k}{\sqrt{2 k}}\,  
\biggl[ \widehat{a}_{\vec{k}, \oplus} + \widehat{a}_{- \vec{k},\oplus}^{\dagger} + 
 \widehat{a}_{-\vec{k}, \oplus} + \widehat{a}_{ \vec{k},\oplus}^{\dagger}\biggr].
\label{HAMG6h}
\end{eqnarray}
In the rotating wave approximation this result implies that the interaction Hamiltonian 
gets suppressed when the gravitons propagate exactly along the direction 
of the cavity. This is why it is useful to introduce a general parametrization 
where the momentum of the gravitons lies in the plane orthogonal to the 
cavity spanned by $\hat{y}$ and $\hat{z}$; in this case
\begin{equation}
\hat{k} = \cos{\gamma} \,\hat{y} + \sin{\gamma} \, \hat{z}, \qquad \hat{m} = \hat{x}, \qquad 
\hat{n} = \sin{\gamma} \, \hat{y} - \cos{\gamma} \, \hat{z},
\label{HAMG6i}
\end{equation}
where it can be readily verified that $\hat{m} \times \hat{n} = \hat{k}$. 
The case $\hat{k} = \hat{z}$ corresponds to $\gamma \to \pi/2$ while the case $\hat{k} 
= \hat{y} $ is realized\footnote{ Although for $\gamma \to 0$ 
$\hat{m} = \hat{x}$ and $\hat{n} = - \hat{z}$ the results coincide with the one obtained above when $\hat{m} = \hat{z}$ and $\hat{n} = \hat{x}$ due to the cyclic properties 
of the external products demanding that, in both cases, $\hat{m} \times\hat{n} = \hat{k}$.} for $\gamma\to 0$. We shall now consider the interaction Hamiltonian 
for the range $0< \gamma < \pi/2$; by repeating the analysis also in this case we first 
deduce that ${\mathcal C}_{\oplus}(\hat{m},\hat{n}, \hat{k}) = - \sin^2{\gamma}\,\, E_{y}^2 - \cos^2{\gamma} \, B_{z}^2$. The interaction Hamiltonian 
becomes then
\begin{eqnarray}
\widehat{H}_{g\, c}(\tau) &=& \frac{\ell_{P} (\widehat{p}_{c}^2 \sin^2{\gamma} 
+ \omega_{c}^2 \, \widehat{q}_{c}^2 \cos^2{\gamma})}{2 \, \sqrt{2} \, a (2\pi)^{3/2}} \,\,  \int \, d^{3} k\,\, \biggl[ \widehat{\mu}_{\vec{k},\oplus} \,W^{\ast}(k \, L_{\perp} \cos{\gamma})\,W^{\ast}(k \, L_{\perp} \sin{\gamma}) 
\nonumber\\
&+& \widehat{\mu}_{\vec{k},\oplus}^{\dagger} \,W(k \, L_{\perp} \cos{\gamma})\,W(k \, L_{\perp} \sin{\gamma})\biggr]
\nonumber\\
&\to& \frac{\ell_{P} (\widehat{p}_{c}^2 \sin^2{\gamma} 
+ \omega_{c}^2 \, \widehat{q}_{c}^2 \cos^2{\gamma})}{2 \, \sqrt{2} \, a (2\pi)^{3/2}} \int \frac{d^{3} k}{\sqrt{2 k}}\,  
\biggl[ \widehat{a}_{\vec{k}, \oplus} + \widehat{a}_{- \vec{k},\oplus}^{\dagger} + 
 \widehat{a}_{-\vec{k}, \oplus} + \widehat{a}_{ \vec{k},\oplus}^{\dagger}\biggr],
\label{HAMG6l}
\end{eqnarray}
where the final result holds, as before, in the limit $|k L_{\perp}| < 1$.

\subsubsection{Rotating wave approximation}
In Eq. (\ref{HAMG6d}) the modes of the gravitons are coupled through $\widehat{p}_{c}^2$
which actually contains various terms that rotate at different rates.
Since the optical mode is driven, in practice, by a resonant
pump field, it is natural to separate the fast time dependence of the pump which is rotating at $\omega_{c}$. In this corotating frame we find few terms oscillating at $2 \omega_{c}$ and since these effects occur on  timescales much shorter than the interaction dynamics, they can be ignored under the rotating wave approximation (RWA). This observation appears in analysis of different physical systems ranging from the Jaynes-Cummings Hamiltonian \cite{JC1,JC2,JC3} to the optomechanical applications
\cite{OPTO1,OPTO2}. The RWA is also commonly enforced in the description of nuclear magnetic resonance \cite{NMR}. Denoting for instance with $\omega_{A}$ and $\omega$ the frequencies of the atom and of the photon, in the Jaynes-Cummings Hamiltonian the approximate dependences of the products of operators go, respectively, as $(\omega_{A} - \omega)$  and as $(\omega_{A} + \omega)$: it is clear that when $\omega_{A} \simeq \omega$ the latter terms rotate much faster than the former and are therefore neglected. The same strategy used in the Jaynes-Cummings 
model to discard the anti-rotating contributions (i.e. $\omega+ \omega_{A} \sim 2 \omega_{A}$) can be also employed in the present context. Let us then consider, for this purpose, the general form of the Hamiltonian 
(\ref{HAMG6l}) and recall that $\widehat{q}_{c} = (\widehat{c} + \widehat{c}^{\dagger})/\sqrt{2 \omega_{c}}$ while $\widehat{p}_{c} = - i \sqrt{\omega_{c}/2} ( \widehat{c} - \widehat{c}^{\dagger})$. In the free-field case $\widehat{c}$ evolves as $e^{- i \omega_{c} \tau}$ and therefore, in the RWA all the contributions going 
as $\widehat{c}^2$ and $\widehat{c}^{\dagger\,\, 2}$ can be neglected in comparison 
with $\widehat{c}^{\dagger} \widehat{c}$. This means that in the RWA approximation 
the interaction Hamiltonian does not depend on the direction of propagation of the gravitons and it is given by 
\begin{equation}
\widehat{H}_{g\,c}(\tau) =  - \frac{\ell_{P}\,\,\widehat{c}^{\dagger} \, \widehat{c} }{2 \, \sqrt{2} \, a (2\pi)^{3/2}} \int \frac{d^{3} k}{\sqrt{2 k}}\,  
\biggl[ \widehat{a}_{\vec{k}, \oplus} + \widehat{a}_{- \vec{k},\oplus}^{\dagger} + 
 \widehat{a}_{-\vec{k}, \oplus} + \widehat{a}_{ \vec{k},\oplus}^{\dagger}\biggr],
\label{HAMF}
\end{equation}
The same result can be obtained by dealing directly with the evolution 
of $\widehat{p}_{c}$ and $\widehat{q}_{c}$ and by introducing a rotating 
frame where $\widehat{q}_{c}(\tau) \to \cos{(\omega_{c} \tau)} \,\, \widehat{q}_{c}(\tau) + \sin{(\omega_{c} \tau)} \,\,\widehat{p}_{c}(\tau)/\omega_{c}$ and 
$\widehat{p}_{c}(\tau) \to - \omega_{c} \, \sin{\omega_{c}\tau} \widehat{q}_{c}(\tau) + \cos{\omega_{c} \tau} \widehat{p}_{c}(\tau)$. The terms $\cos{(2 \omega_{c} \tau)}$ and $\sin{(2 \omega_{c} \tau)}$ appearing both in $\widehat{q}_{c}^2$ and $\widehat{p}_{c}^2$ rotate 
much faster and can then be neglected. 

\renewcommand{\theequation}{5.\arabic{equation}}
\setcounter{equation}{0}
\section{Dynamical evolution of the system}
\label{sec5}
\begin{figure}[!ht]
\centering
\includegraphics[height=8cm]{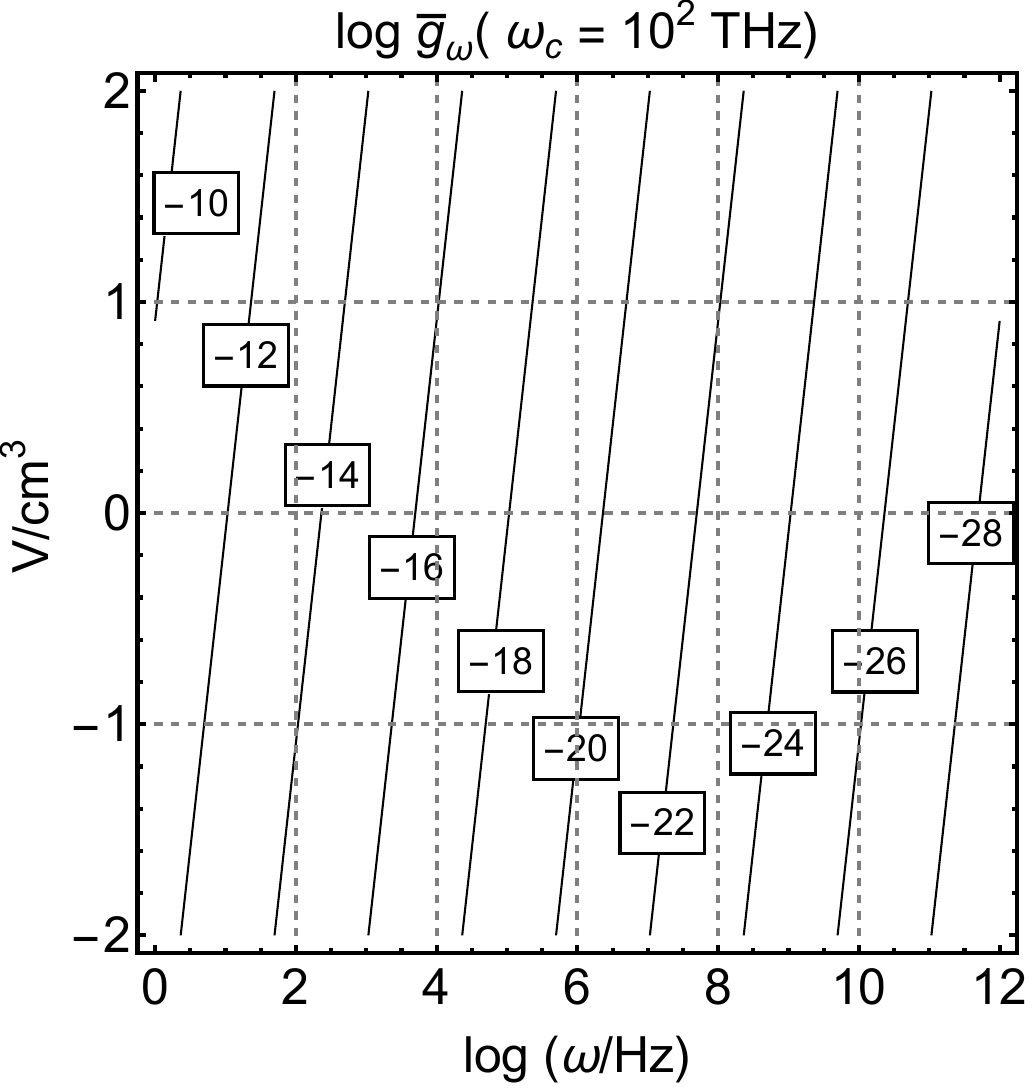}
\includegraphics[height=8cm]{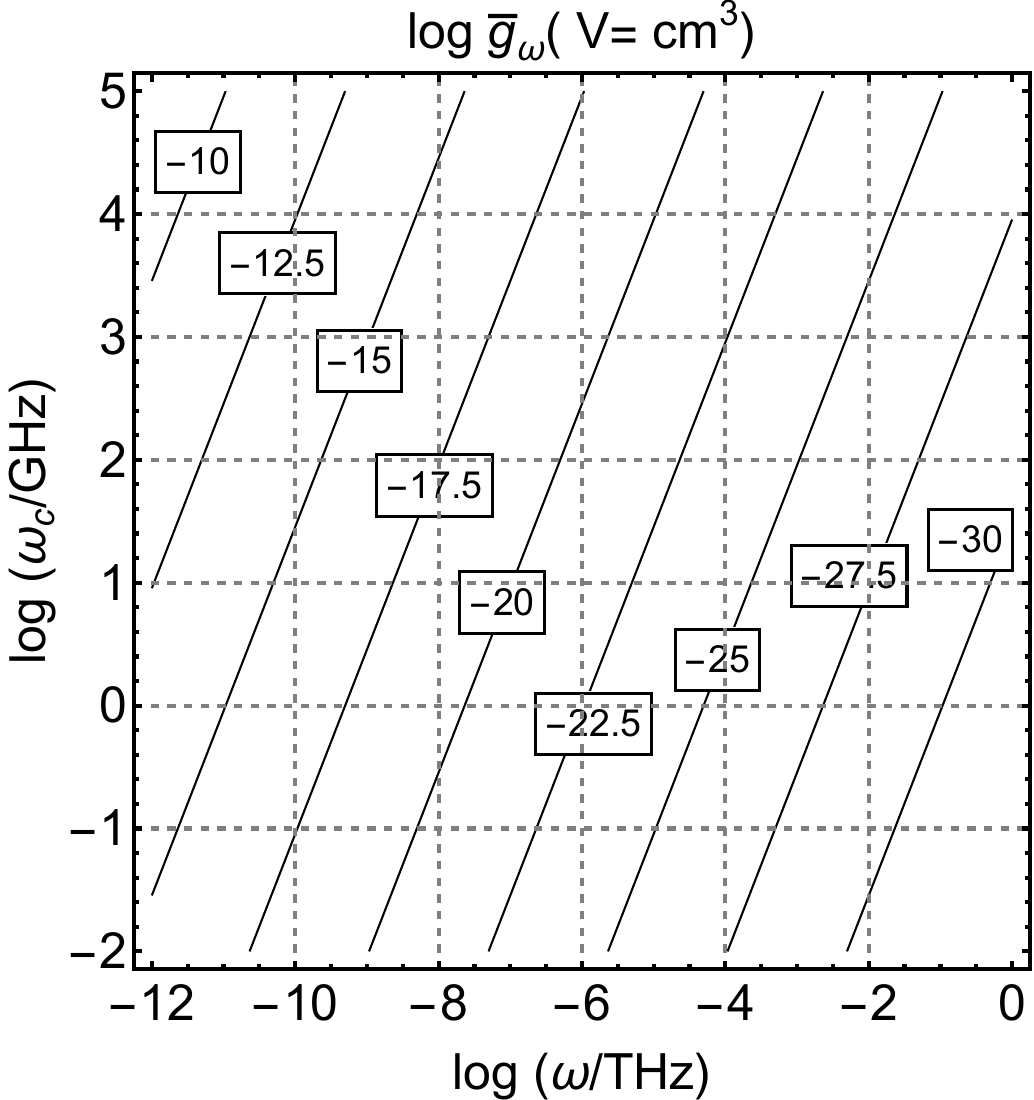}
\caption[a]{The common logarithm of the effective coupling is illustrated as a function of the frequencies of the gravitons and of the photons.}
\label{Figure2}      
\end{figure}

As we saw from Eqs. (\ref{HAMG6f})--(\ref{HAMG6g}) and (\ref{HAMG6l}) the interaction 
between gravitons and cavity modes does not select a 
component of the tensor quantum field but a specific tensor polarization.  In the continuous mode description the full expression of the Hamiltonian becomes then:
\begin{eqnarray}
\widehat{H}(\tau) &=&  \omega_{c} \, \widehat{c}^{\dagger} \, \widehat{c} + \frac{1}{2} \sum_{\alpha} \, \int d^{3} k \, \biggl[ 
\,\,( \widehat{a}_{\vec{k}, \alpha}^{\dagger} \,  \widehat{a}_{\vec{k}, \alpha} + 
 \widehat{a}_{-\vec{k}, \alpha}\,  \widehat{a}_{-\vec{k}, \alpha}^{\dagger}) 
 + i {\mathcal H} \,\,( \widehat{a}_{\vec{k}, \alpha}^{\dagger} \,  \widehat{a}_{-\vec{k}, \alpha}^{\dagger} - \widehat{a}_{\vec{k}, \alpha}\,  \widehat{a}_{-\vec{k}, \alpha})\biggr]
\nonumber\\
&-& \frac{\lambda}{2} \, \widehat{c}^{\dagger}\,\widehat{c} \,\, 
\int \frac{d^{3} k}{\sqrt{2 k}} \biggl[  \widehat{a}_{\vec{k}, \beta} +  \widehat{a}_{-\vec{k}, \beta}^{\dagger} +  \widehat{a}_{-\vec{k}, \beta} +  \widehat{a}_{\vec{k}, \beta}^{\dagger} \biggr],
\label{HG1}
\end{eqnarray}
where $\beta$ generically denotes a {\em fixed} tensor polarization; this happens since the interaction term in a cavity selects a single polarization that depends on the orientation of the cavity axis. Although in the forthcoming analysis Eq. (\ref{HG1}) will be privileged, to clarify 
the analogies (and the differences) between Eq. (\ref{HG1}) 
and some other quantum mechanical problem, it is possible to simplify the field content. The first simplification is to pass from the continuous mode description to the discrete mode representation; this step follows immediately by recalling the following dictionary:
\begin{equation}
\int d^{3} k \to \sum_{\vec{k}} (2\pi)^3/V, \quad \widehat{a}_{\vec{k}, \alpha} \to \widehat{a}_{\vec{k}, \alpha} \, \sqrt{V}/(2\pi)^{3/2}, \quad \delta^{(3)}(\vec{k}- \vec{p}) \to\delta_{\vec{k}\, \vec{p}} \, V/(2\pi)^{3},
\label{DIC1}
\end{equation}
where $\delta_{\vec{k}\, \vec{p}}$ now denotes the conventional Kroeneker delta functions over the three-momenta. In Eq. (\ref{DIC1}) $V$ indicates the normalization volume: since the gravitons constitute a diffuse background $V$ and $V_{c}$ might not always coincide. For an accurate comparison between the two approaches we must also consider that, in the discrete 
mode representation, the creation and annihilation operators are dimensionless.
With these caveats Eq. (\ref{HG1}) can be finally rewritten as: 
 \begin{eqnarray}
\widehat{H}(\tau) &=&  \omega_{c} \, \widehat{c}^{\dagger} \, \widehat{c} + \frac{1}{2}\sum_{\alpha} \, \sum_{\vec{k}} \, k  \biggl[ 
k( \widehat{a}_{\vec{k}, \alpha}^{\dagger} \,  \widehat{a}_{\vec{k}, \alpha} + 
 \widehat{a}_{-\vec{k}, \alpha}\,  \widehat{a}_{-\vec{k}, \alpha}^{\dagger}) 
 + i {\mathcal H} ( \widehat{a}_{\vec{k}, \alpha}^{\dagger} \,  \widehat{a}_{-\vec{k}, \alpha}^{\dagger} - \widehat{a}_{\vec{k}, \alpha}\,  \widehat{a}_{-\vec{k}, \alpha})\biggr]
\nonumber\\
&-& \widehat{c}^{\dagger}\,\widehat{c} \,\, \sum_{\vec{k}} \, g_{k}\, \biggl[  \widehat{a}_{\vec{k}, \beta} +  \widehat{a}_{-\vec{k}, \beta}^{\dagger} +  \widehat{a}_{-\vec{k}, \beta} +  \widehat{a}_{\vec{k}, \beta}^{\dagger} \biggr].
\label{HG1a}
\end{eqnarray}
As a result of the discrete mode description the coupling 
between the cavity mode and the gravitons inherits a volume dependence; therefore now $g_{k}$ is given by
\begin{equation}
g_{k} = \frac{\lambda}{2 \sqrt{V}} \equiv \frac{\omega_{c} \, \ell_{P}}{4 a(\tau) \sqrt{V} \sqrt{\omega}}.
\label{HG1b}
\end{equation}
We recall that, in the present units, $\omega= k = 2 \pi\, \nu$ so that
the coupling $g_{k}$ has dimensions of a wavenumber (or of a frequency); we can also 
introduce the dimensionless analog of Eq. (\ref{HG1b}), namely 
\begin{equation}
\overline{g}_{\omega} = \frac{g_{k}}{\omega} = \frac{\ell_{P} \, \omega_{c}}{ 4 \, a(\tau) \sqrt{V \, \omega^3}} = {\mathcal O}\biggl(10^{-13}\biggr) \biggl(\frac{\omega_{c}}{10^{2}\, \mathrm{THz}}\biggr) \biggl(\frac{\omega}{\mathrm{kHz}}\biggr)^{-3/2} \, \biggl(\frac{V}{\mathrm{cm}^3}\biggr)^{-1/2}.
\label{HG1c}
\end{equation}
As stressed above in the introductory section the frequencies of the gravitons range between ${\mathcal O}(\mathrm{aHz})$ and the THz. In Fig. \ref{Figure2} we illustrate the common logarithm of $\overline{g}_{\omega}$ as a function 
of $\omega$, $\omega_{c}$ and $V$.
The couplings illustrated in Fig. \ref{Figure2} are objectively very small but the problem analyzed here is not 
to find a way of increasing the sensitivity. What matters here is to see if the correlation properties of the 
gravitons could be inferred from the ones of the photons in a cavity.
\subsection{Three-modes Hamiltonian}
If the field content is artificially reduced Eq. (\ref{HAMG1}) can be mapped on to the following 
quantum mechanical Hamiltonian 
\begin{equation}
\widehat{H}(\tau) = \omega_{c} \,\,\widehat{c}^{\dagger} \,\widehat{c} + 
\omega \, (\widehat{a}^{\dagger} \,\widehat{a} + \widehat{b} \,\widehat{b}^{\dagger}) - g_{\omega}\, \widehat{c}^{\dagger} \widehat{c} \,\,(\widehat{a} + \widehat{b}^{\dagger} + \widehat{a}^{\dagger} + \widehat{b})  + 
i {\mathcal H} (\widehat{a}^{\dagger} \, \widehat{b}^{\dagger} - \widehat{a} \, \widehat{b}),
\label{HAMF2}
\end{equation}
that couples together three different modes. As before in Eq. (\ref{HAMF2}) $\widehat{c}$ may describe the fundamental mode of the cavity whereas $\widehat{a}$ and $\widehat{b}$ are associated with the the pairs of gravitons produced with opposite three-momenta (see Eq. (\ref{HG1a})). In case 
 one of the two modes associated with the gravitons vanishes, the remaining terms of Eq. (\ref{HAMF2}) 
lead to an Hamiltonian that assumes the conventional optomechanical from. For instance if $\widehat{a} \to 0$ 
in Eq. (\ref{HAMF2}) we obtain a result that coincides with the optomechanical Hamiltonian \cite{OPTO1} (see also \cite{OPTO2} and discussions therein):
\begin{equation}
\widehat{H}(\tau) = \omega_{c} \,\,\widehat{c}^{\dagger} \, \widehat{c} + \omega \,\widehat{b}^{\dagger} \,\widehat{b} 
-g_{\omega} \,\widehat{c}^{\dagger} \, \widehat{c} \,\, (\widehat{b} + \widehat{b}^{\dagger}),
\label{HAMF3}
\end{equation}
where, however, the mode $\widehat{b}$ corresponds to the mechanical oscillator (i.e. the position 
of a putative movable mirror).  It is relevant to stress here that, in spite of its simplicity, 
Eq. (\ref{HAMF3}) does not directly describe the physics 
of our problem since the terms associated with the production of gravitons 
with opposite three-momenta are missing. Furthermore, as already alluded to, in the optomechanical context $\omega_{c}$ denotes the frequency in the free spectral range 
whereas $\omega$ is related with the oscillation frequency of the mechanical oscillator \cite{OPTO1,OPTO2}. Although formally similar to the optomechanical situations, the meaning of Eqs. (\ref{HG1a}) and (\ref{HAMF3}) is physically different. We finally remark that when $\widehat{c} \to 0$, Eq. (\ref{HAMF2}) turns into the Hamiltonian that describes the quantum theory of parametric 
amplification \cite{PARM1,PARM2,PARM3}:
\begin{equation}
\widehat{H}(\tau) = 
\omega \, (\widehat{a}^{\dagger} \,\widehat{a} + \widehat{b} \,\widehat{b}^{\dagger})  + 
i {\mathcal H} \,(\widehat{a}^{\dagger} \, \widehat{b}^{\dagger} - \widehat{a} \, \widehat{b}).
\label{HAMF4}
\end{equation}
Equation (\ref{HAMF4}) is closely related to the formalism of two-mode squeezed states \cite{TWS1,TWS2,TWS3}. Its field theory counterpart describes the production of relic gravitons \cite{RG1}. 
Some quantum Hamiltonians potentially similar to the one of Eq. (\ref{HAMF2}) appeared 
in Refs. \cite{SQCAV1,SQCAV3} where the authors showed that two mechanical oscillators in an optomechanical cavity can be driven into a highly pure entangled two-mode squeezed state using only a single engineered reservoir, greatly simplifying experimental realization of quantum entanglement and continuous-variable teleportation: the role of the mechanical oscillators is played, in our context by the gravitons but also the way the dynamics is discussed is quite different. Finally Ref. \cite{SQCAV2} proposes a dissipation-engineering scheme that produces unlimited steady-state squeezing of a bosonic mode by coupling it to an appropriately designed reservoir.

\subsection{The multimode evolution}
The degrees of first- and second-order 
coherence involve the average of the field operators over the quantum states of the 
gravitons {\em and} of the cavity mode. In the Heisenberg description 
we need to determine the evolution of the various field operators throughout the different 
dynamical ranges of the problem. From Eq. (\ref{HG1a}) the evolution of $\widehat{a}_{\vec{p},\gamma}$ and of $\widehat{a}_{-\vec{p},\gamma}^{\dagger}$ reads\footnote{As already stressed the coupling to the photons selects a single tensor polarization and this is the reason of the terms containing $\delta_{\gamma\beta}$.}
\begin{eqnarray}
\partial_{\tau} \widehat{a}_{\vec{p},\gamma} &=& - i \, p \,\widehat{a}_{\vec{p},\gamma}+ {\mathcal H} \,\,\widehat{a}_{-\vec{p},\gamma}^{\dagger}
+i\, p\, \lambda_{p} \widehat{c}^{\dagger} \, \widehat{c} \,\,\delta_{\gamma\beta},
\label{HG2}\\
\partial_{\tau} \widehat{a}^{\dagger}_{-\vec{p},\gamma}&=&  i \,p \,\widehat{a}^{\dagger}_{-\vec{p},\gamma}+ {\mathcal H} \,\,\widehat{a}_{\vec{p},\gamma}
- i \,p\, \lambda_{p}\, \widehat{c}^{\dagger} \, \widehat{c}\,\, \delta_{\gamma\beta}.
\label{HG3}
\end{eqnarray}
 In Eqs. (\ref{HG2})--(\ref{HG3}) the momentum-dependent coupling 
$\lambda_{p} = \lambda/[a(\tau) \,p \, \sqrt{2 p}]$ has been introduced 
for the sake of convenience. The coupling $\lambda_{p}$ also scales with 
$a(\tau)$ but we actually need only to evaluate this dependence for typical times that are
comparable with the present epoch when $a_{0} = {\mathcal O}(1)$. For earlier times the coupling might 
get larger but the cavity mode is present at the current time and not before.
Equations (\ref{HG2})--(\ref{HG3}) describe two complementary processes that are separated in time: the production of gravitons 
(sensitive to ${\mathcal H}$) and the interaction with the cavity mode (controlled by $\lambda_{p}$). Once Eqs. (\ref{HG2})--(\ref{HG3}) are solved, we can compute the combination $[ \widehat{a}_{\vec{k}, \beta}  + \widehat{a}_{-\vec{k}, \beta}^{\dagger} + \widehat{a}_{-\vec{k}, \beta}  + \widehat{a}_{\vec{k}, \beta}^{\dagger}]$ entering  the evolution of the cavity mode
\begin{equation}
\partial_{\tau} \widehat{c} = - i \omega_{c} \widehat{c} + \frac{i}{2}\,\int d^{3} k\, k\, \lambda_{k}\, [ \widehat{a}_{\vec{k}, \beta}  + \widehat{a}_{-\vec{k}, \beta}^{\dagger} + \widehat{a}_{-\vec{k}, \beta}  + \widehat{a}_{\vec{k}, \beta}^{\dagger}] \,\, \widehat{c},
\label{HG4}
\end{equation}
as it follows from Eq. (\ref{HG1a}). The production of gravitons is determined 
by ${\mathcal H} = a\, H$ which is sensitive both to the inflationary evolution and to the 
subsequent (decelerated) stage of expansion. 
\begin{figure}[!ht]
\centering
\includegraphics[height=8cm]{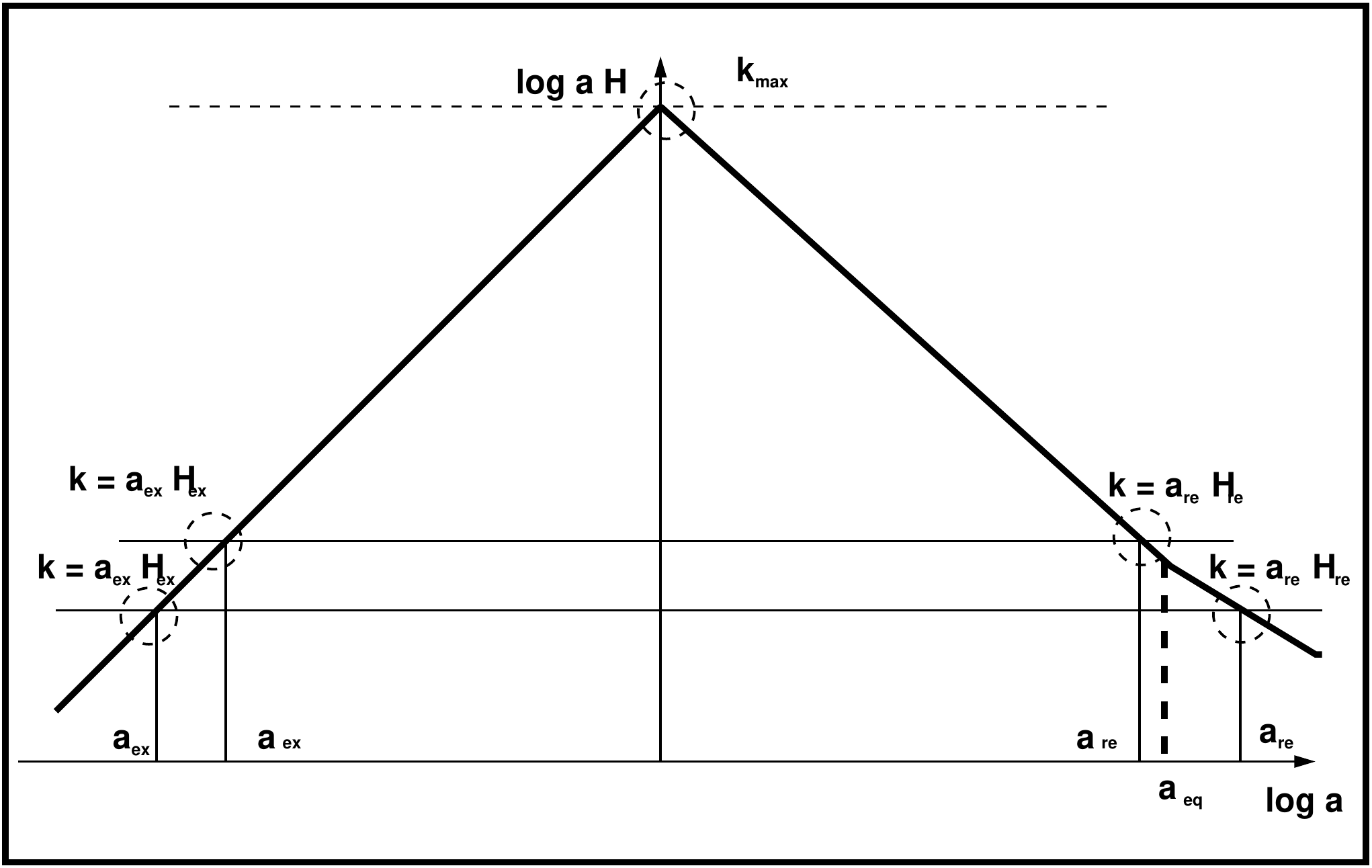}
\caption[a]{We approximately illustrate here the effective evolution of the comoving Hubble radius ${\mathcal H}$ as a function of the scale factor $a$. Common logarithms are employed on both axes. During the inflationary stage ${\mathcal H} \propto a$ while 
during the radiation and matter stages ${\mathcal H} \propto a^{-1}$ and ${\mathcal H} \propto a^{-1/2}$, respectively. What matters for the evolution of the field 
operators are the exit and reentry times corresponding to the moments 
where a given wavenumber crosses the comoving horizon. The reentry and the exit approximately coincide close to the maximal wavenumber of the spectrum $k_{max}$.}
\label{Figure3}      
\end{figure}
The explicit form of ${\mathcal H}$ has a typical shape illustrated in the cartoon of Fig. \ref{Figure3} that corresponds to the simplest situation compatible with the concordance scenario where a conventional inflationary stage 
is followed by a radiation epoch and, eventually, by a matter-dominated phase. 
Equations (\ref{HG2})--(\ref{HG3}) can be recast in a more tractable form by first considering the operators: 
\begin{eqnarray}
&& \partial_{\tau} (\widehat{a}_{\vec{p},\gamma} +  \widehat{a}^{\dagger}_{-\vec{p},\gamma}) = - i\, p\,( \widehat{a}_{\vec{p},\gamma} -  \widehat{a}^{\dagger}_{-\vec{p},\gamma}) + {\mathcal H} (\widehat{a}_{\vec{p},\gamma} +  \widehat{a}^{\dagger}_{-\vec{p},\gamma}),
\label{HG5}\\
&& \partial_{\tau} (\widehat{a}_{\vec{p},\gamma} -  \widehat{a}^{\dagger}_{-\vec{p},\gamma}) = - i\, p\, (\widehat{a}_{\vec{p},\gamma} -  \widehat{a}^{\dagger}_{-\vec{p},\gamma}) - {\mathcal H} (\widehat{a}_{\vec{p},\gamma} -  \widehat{a}^{\dagger}_{-\vec{p},\gamma})
+ 2 \,i \,p\, \lambda_{p}\,  \widehat{c}^{\dagger} \, \widehat{c} \,\,\delta_{\gamma\beta}.
\label{HG6}
\end{eqnarray}
After expressing  $( \widehat{a}_{\vec{p},\gamma} -  \widehat{a}^{\dagger}_{-\vec{p},\gamma})$ from Eq. (\ref{HG5}) the obtained result can be inserted into Eq. (\ref{HG6}) and the decoupled 
expression for $( \widehat{a}_{\vec{p},\gamma} +  \widehat{a}^{\dagger}_{-\vec{p},\gamma})$
becomes:
\begin{equation}
(\widehat{a}_{\vec{p},\gamma} +  \widehat{a}^{\dagger}_{-\vec{p},\gamma})^{\prime\prime} + \bigl[ p^2 - ({\mathcal H}^2 +{\mathcal H}^{\prime})\bigr] \, (\widehat{a}_{\vec{p},\gamma} +  \widehat{a}^{\dagger}_{-\vec{p},\gamma}) = 2 \, p^2\, \lambda_{p} \, \widehat{c}^{\dagger} \widehat{c} \,\,\delta_{\beta\gamma},
\label{HG7}
\end{equation}
 where the prime now denotes, as before, the derivation with respect to $\tau$. After 
 obtaining $(\widehat{a}_{\vec{p},\gamma} +  \widehat{a}^{\dagger}_{-\vec{p},\gamma})$ 
 from the solution of Eq. (\ref{HG7}) we can use Eq. (\ref{HG5}) to derive the explicit expression 
 for $( \widehat{a}_{\vec{p},\gamma} -  \widehat{a}^{\dagger}_{-\vec{p},\gamma})$. This is 
 however not essential if we need the evolution of $\widehat{c}(\tau)$ since the momentum integral appearing inside Eq. (\ref{HG4}) just follows from the solution of Eq. (\ref{HG7}) (and of its Hermitian conjugate). 

\subsection{The gravitational modes}
In Eq. (\ref{HG7}) there are two complementary regimes that are defined by the relative 
weight of the two terms appearing inside the squared bracket, i.e. $p^2$ and $({\mathcal H}^2 + {\mathcal H}^{\prime}) $. Between these two terms the former is somehow simpler than the latter that can be written in different ways;  for the present purposes, the most useful form is given by 
\begin{equation}
{\mathcal H}^2 + {\mathcal H}^{\prime} = a^2 \, H^2 (2 + \dot{H}/H^2) \equiv {\mathcal H}^2 
[2 - \epsilon(\tau)],
\label{HG7a}
\end{equation}
where, according to the standard notations, $H= \dot{a}/a$ and the overdot 
indicates a derivation with respect to the cosmic time coordinate $t$; in Eq. (\ref{HG7a}) 
$\epsilon(\tau) = - \dot{H}/H^2$ is the standard  slow-roll parameter. During 
the inflationary stage (see the leftmost region of Fig. \ref{Figure3}) $\epsilon \ll 1$ and a more explicit estimate follows by enforcing the so-called consistency relations 
stipulating that $n_{T} \simeq - 2 \epsilon \simeq - r_{T}/8$ where $r_{T} < {\mathcal O}(0.03)$ indicates, as usual, the tensor to scalar ratio. During the decelerated stage of expansion we have instead that $\epsilon = {\mathcal O}(1)$ (see the rightmost region of Fig. \ref{Figure3}). In both cases when 
a given wavenumber crosses the Hubble radius the comoving wavenumber becomes 
comparable with $a\, H$, i.e.
$k \simeq a \, H$; in  Fig. \ref{Figure3} we illustrated the crossing times for three different wavenumbers and the 
dashed circles correspond to the regions where $ k = {\mathcal O}(a\, H)$. From the viewpoint of our discussion 
this also means that when $p^2 \ll a^2 \, H^2 $ the source term in Eq. (\ref{HG7}) can be neglected since the operators evolve in a stage where the coupling to the photons of the cavity does not play a role. In this regime we have that 
\begin{equation}
(\widehat{a}_{\vec{p},\gamma} +  \widehat{a}^{\dagger}_{-\vec{p},\gamma}) = 
a(\tau) \widehat{A}_{\vec{p},\gamma} + a(\tau) \widehat{B}_{\vec{p},\gamma} \int^{\tau} \,\, d\tau^{\prime}/a^2(\tau^{\prime}).
\label{SH1}
\end{equation}
Equation (\ref{SH1}) holds in the range $\tau_{ex}< \tau < \tau_{re}$ where $\tau_{ex} \simeq 1/(a_{ex}\, H_{ex})$ and $\tau_{re}= 1/(a_{re} \, H_{re})$ are, respectively, the exit and reentry times illustrated in Fig. \ref{Figure3}. Moreover, two auxiliary operators $ \widehat{A}_{\vec{p},\gamma}$ and $\widehat{B}_{\vec{p},\gamma}$  are fixed by the initial condition at $\tau_{ex}$. Although more complicated cases can be envisaged, we now deal with the simplest 
physical situation where the initial state of the gravitons is annihilated by $\widehat{b}_{\vec{p},\, \gamma}$ and by $\widehat{b}_{-\vec{p},\, \gamma}$ in the limit 
$\tau \to - \infty$.  In this case the general solution of Eq. (\ref{HG7}) for $\tau > \tau_{re}$ becomes:
\begin{eqnarray}
(\widehat{a}_{\vec{p},\gamma} +  \widehat{a}^{\dagger}_{-\vec{p},\gamma}) &=& [ d_{+}(p) \, 
\widehat{b}_{\vec{p},\gamma} + d_{-}^{\ast}(p)\, \widehat{b}_{-\vec{p},\gamma}^{\dagger} ] \,\, e^{- i p \Delta \tau}
\nonumber\\
&+& [ d_{+}^{\ast}(p) \, 
\widehat{b}_{-\vec{p},\gamma}^{\dagger} + d_{-}(p)\, \widehat{b}_{\vec{p},\gamma}] \,\,e^{i p \Delta \tau}
+ 2 \,\lambda_{p} \,\,\widehat{c}^{\dagger}\, \widehat{c}\,\, [ 1 - \cos{p \Delta \tau} ]\,\,\delta_{\gamma\beta},
\label{SH2}
\end{eqnarray}
where, by definition, $\Delta \tau = (\tau - \tau_{re})$. In Eq. (\ref{SH2})
the coefficients $d_{\pm}(p)$ are determined by matching the solution for $\tau > \tau_{re}$ with the solutions valid when $\tau_{ex}< \tau< \tau_{re}$; a similar analysis can be found in Ref. \cite{WWW} in the case of the mode functions and in the absence of the coupling to the cavity mode.
The source term of Eq. (\ref{HG7}) is then integrated even without 
specific knowledge of the evolution of $\widehat{c}(\tau)$ (to be analyzed in the following paragraph) since the total Hamiltonian of Eq. (\ref{HG1a}) commutes with the number operator of the photons, i.e. 
$[\widehat{H}, \widehat{n}_{c}] =0$. The result of Eq. (\ref{SH2}) is valid for $\tau \geq \tau_{re}$ corresponding to the 
rightmost part of Fig. \ref{Figure3} when the wavelengths of the gravitons 
are shorter than the Hubble radius, i.e. $k > a \, H$. The result for $d_{\pm}(p) $ becomes:
\begin{equation}
d_{\pm}(p) = e^{- i p(\tau_{ex} - \tau_{re})}\biggl[ M_{p}(\tau_{ex}, \tau_{re}) \pm i N_{p}(\tau_{ex}, \tau_{re})\biggr]/2,
\label{SH3}
\end{equation}
where $M_{p}(\tau_{ex}, \tau_{re})$ and $N_{p}(\tau_{ex}, \tau_{re})$ are two complex 
functions defined as: 
\begin{eqnarray}
M_{p}(\tau_{ex},\tau_{re}) &=& \biggl(\frac{a_{re}}{a_{ex}}\biggr) \, J_{p}(\tau_{ex}, \tau_{re}),
\label{WKB6}\\
N_{p}(\tau_{ex}, \tau_{re}) &=& \biggl(\frac{{\mathcal H}_{re}}{p} \biggr) \,  \biggl(\frac{a_{re}}{a_{ex}}\biggr) \, J_{p}(\tau_{ex}, \tau_{re})
- \biggl(\frac{a_{ex}}{a_{re}}\biggr) \biggl(\frac{{\mathcal H}_{ex} + i p}{p}\biggr),
\label{WKB7}\\
J_{p}(\tau_{ex}, \tau_{re}) &=& 1 - ({\mathcal H}_{ex} + i p) \int_{\tau_{ex}}^{\tau_{re}} [a_{ex}^2/a^2(\tau^{\prime})] \,\, d \tau^{\prime}.
\label{WKB8}
\end{eqnarray}
In the previous equations $a_{re} = a(\tau_{re})$, $a_{ex} = a(\tau_{ex})$ while the other quantities, with obvious notations, are ${\mathcal H}_{re} = {\mathcal H}(\tau_{re})=  a_{re} H_{re}$ and  ${\mathcal H}_{ex} = {\mathcal H}(\tau_{ex})= a_{ex} H_{ex}$. Although Eqs. (\ref{SH3}) and (\ref{WKB6})--(\ref{WKB8}) 
are valid in general terms, the physical situation described by the cartoon of Fig. 
\ref{Figure3} implies that the scale factor always expands and, in this situation,
$a_{re}/a_{ex}$ is always much larger than $1$. It is then possible to deduce 
various approximate expressions which always have to comply with 
the constraints set by unitarity, i.e. $|d_{+}(p)|^2 - |d_{-}(p)|^2 =1$.

Having deduced the general forms of the field operators for $\tau \geq \tau_{re}$, from  Eq. (\ref{SH2}) the Hermitian combination appearing at the right hand side of Eq. (\ref{HG4}) can be constructed and it is:
\begin{eqnarray}
\widehat{a}_{\vec{p},\gamma} +  \widehat{a}^{\dagger}_{-\vec{p},\gamma} + 
\widehat{a}_{-\vec{p},\gamma} +  \widehat{a}^{\dagger}_{\vec{p},\gamma} = \widehat{{\mathcal P}}_{\vec{p},\gamma} \, \, e^{- i\, p\, \Delta\tau} + \widehat{{\mathcal P}}_{\vec{p},\gamma}^{\dagger} \, \, e^{i\, p\, \Delta\tau}.
\label{SH4}
\end{eqnarray}
For conciseness in Eq. (\ref{SH4}) the auxiliary operator $\widehat{{\mathcal P}}_{\vec{p},\gamma}$ has been introduced together with its Hermitian conjugate
\begin{eqnarray}
\widehat{{\mathcal P}}_{\vec{p},\gamma} = d_{+}(p)\,\, (\widehat{b}_{\vec{p},\gamma} + \widehat{b}_{-\vec{p},\gamma}) + d_{-}^{\ast}(p)\,\, (\widehat{b}_{- \vec{p},\gamma}^{\dagger} + \widehat{b}_{\vec{p},\gamma}^{\dagger})
= \widehat{\Sigma}^{\dagger}[z] (\widehat{b}_{\vec{p},\gamma} + \widehat{b}_{-\vec{p},\gamma}) \widehat{\Sigma}[z],
\label{SH6}
\end{eqnarray}
where now $\widehat{\Sigma}[z]$ is the multimode 
generalization of the squeezing operator \cite{TWS1,TWS2,TWS3} and it is defined as:
\begin{equation}
\widehat{\Sigma}[z] = \exp{[\widehat{\Lambda}(z)]}, \qquad 
\widehat{\Lambda}(z) = \frac{1}{2} \int d^{3} k\,\,\sum_{\alpha} \biggl[ 
z_{k,\alpha}^{\ast} \, \widehat{b}_{\vec{p},\,\alpha}\, \widehat{b}_{-\vec{p},\, \alpha} - z_{k,\,\alpha} 
\widehat{b}_{\vec{p},\,\alpha}^{\dagger}\, \widehat{b}_{-\vec{p},\, \alpha}^{\dagger} \biggr].
\label{SH7}
\end{equation}
implying $\widehat{\Sigma}^{\dagger}= \widehat{\Sigma}^{-1}$. In Eqs. (\ref{SH6})--(\ref{SH7}) we can choose the usual parametrization for the squeezing parameters, namely $z_{k,\, \alpha} = e^{i \theta_{k,\alpha}} \, r_{k,\alpha}$; thanks to the commutation relations in the continuous mode description (i.e.  $[\widehat{b}_{\vec{k},\alpha}, \, \widehat{b}_{\vec{p},\beta}^{\dagger}] = \delta_{\alpha\beta}\, \delta^{(3)}(\vec{k} - \vec{p})$)
we can then obtain
\begin{eqnarray}
&& \Sigma^{\dagger}\, \,\widehat{b}_{\vec{p},\, \beta}  \, \,\Sigma = \cosh{r_{p,\beta}} \, \widehat{b}_{\vec{p},\, \beta}- e^{i \theta_{p,\beta}} \, \sinh{r_{p,\beta}}\, 
\widehat{b}_{-\vec{p},\, \beta}^{\dagger},
\label{SH8}\\
&& \Sigma^{\dagger} \, \, \widehat{b}_{-\vec{p},\,\beta}^{\dagger} \, \,\Sigma = \cosh{r_{p,\,\beta}} \, \widehat{b}_{-\vec{p},\,\beta}^{\dagger} - e^{-i \theta_{p,\,\beta}} \, \sinh{r_{p,\,\beta}}\, 
\widehat{b}_{\vec{p},\,\beta}^{\dagger}.
\label{SH9}
\end{eqnarray}
Equations (\ref{SH8})--(\ref{SH9}) can also be written for a single tensor polarization: in this case the sum over $\alpha$ appearing in Eq. (\ref{SH7}) becomes pleonastic and the polarization indices disappear. For a single polarization the connection between $d_{\pm}(p)$ of Eq. (\ref{SH3}) and the variables of Eqs. (\ref{SH8})--(\ref{SH9}) is given by 
 $d_{+}(p) = \cosh{r_{p}}$ and $d_{-}(p) = - e^{- i \theta_{p}} \, \sinh{r_{p}}$. 

\subsection{The cavity mode}
The explicit result of Eq. (\ref{SH4}), once inserted into Eq. (\ref{HG4}), determines  the evolution of the cavity mode. In this respect it is  practical to introduce the operator $\widehat{C}(\tau) = e^{i\, \omega_{c}\tau} \, \widehat{c}(\tau)$; in terms of $\widehat{C}(\tau)$  Eq. (\ref{HG4}) becomes:
\begin{equation}
\widehat{C}^{\,\prime} = \frac{i}{2} \biggl\{\int d^{3} k\,\, k\, \lambda_{k}\,\, \biggl[ 
\widehat{{\mathcal P}}_{\vec{k},\gamma} \, e^{- i k \Delta \tau} + \widehat{{\mathcal P}}_{\vec{k},\gamma}^{\dagger}\, e^{i k\, \Delta \tau} + 
4 \, \lambda_{k} \, \widehat{C}^{\dagger} \widehat{C} ( 1 - \cos{k\,\Delta \tau})\delta_{\gamma\beta}\biggr] \biggr\} \widehat{C}, 
\label{CH1}
\end{equation}
where, as before, the prime denotes a derivation with respect to $\tau$. 
Equation (\ref{CH1}) can be solved in different ways but the simplest 
strategy is to introduce the unitary operator $\widehat{U}(\tau, \tau_{\ast})$ 
that relates $\widehat{C}(\tau)$ to $\widehat{C}(\tau_{\ast})$ where 
$\tau_{\ast} \geq \tau_{re}$. The range of $\tau_{\ast}$ and $\tau$  coincides in practice with the present time: $\tau_{\ast}$ is the moment at which we start assessing the photon correlations and $\tau -\tau_{\ast}$ is the range of the measurement itself. We then write $\widehat{C}(\tau) = \widehat{U}(\tau, \tau_{\ast}) \widehat{C}(\tau_{\ast})$ with the boundary conditions $\widehat{U}(\tau_{\ast},\tau_{\ast}) = \mathds{1}$. From Eq. (\ref{CH1}) we deduce an integral equation (of Volterra type) for $\widehat{U}(\tau, \tau_{\ast})$ 
\begin{equation}
\widehat{U}(\tau,\tau_{\ast}) = \mathds{1} + i \int_{\tau_{\ast}}^{\tau} \widehat{{\mathcal F}}(\tau^{\,\prime},\tau_{\ast}) \, \widehat{U}(\tau^{\prime}, \tau_{\ast}) \, d\tau^{\prime},
\label{CH2}
\end{equation}
where $\widehat{{\mathcal F}}(\tau,\tau_{\ast})$ can be read off from the general expression of Eq. (\ref{CH1}) and it is given by:
\begin{equation}
\widehat{{\mathcal F}}(\tau,\tau_{\ast})=\frac{1}{2} \int d^{3} k\,\, k\, \lambda_{k}\,\, \biggl[ 
\widehat{{\mathcal P}}_{\vec{k},\gamma} \, e^{- i k \Delta \tau} + \widehat{{\mathcal P}}_{\vec{k},\gamma}^{\dagger}\, e^{i k\, \Delta \tau} + 
4 \, \lambda_{k} \, \widehat{n}_{c} ( 1 - \cos{k\,\Delta \tau}) \delta_{\gamma\beta}\biggr].
\label{CH3}
\end{equation}
Note that, by definition,  $\widehat{n}_{c} =\widehat{C}^{\dagger} \widehat{C} = \widehat{c}^{\dagger} \, \widehat{c}$ is a constant of motion since 
$\widehat{n}_{c}$ commutes with the total Hamiltonian of Eq. (\ref{HG1a}). Equation (\ref{CH3}) is then solved by iteration and the final result is
\begin{equation}
\widehat{U}(\tau, \tau_{\ast}) = e^{\int d^{3} k [ \alpha_{k}(x) \widehat{{\mathcal P}}_{\vec{k},\gamma}^{\dagger} - \alpha_{k}^{\ast}(x) \, \widehat{{\mathcal P}}_{\vec{k},\gamma}]}\, \,e^{2 \,i \, \widehat{n}_{c} \int d^{3} k \beta_{k}(x)},
\label{CH4}
\end{equation}
where $x = k (\tau -\tau_{re})$ while $\alpha_{k}(x)$ and $\beta_{k}(x)$ are given by
\begin{equation}
\alpha_{k}(x) = \lambda_{k}(e^{i x} -1)/2, \qquad \beta_{k}(x) = \lambda_{k}^2 \,\,(x - \sin{x}).
\label{CH5}
\end{equation}
While $\alpha(x) = {\mathcal O}(\lambda_{k})$, the values of $\beta$ are systematically smaller, i.e.
$ \beta_{k}(x) = {\mathcal O}(\lambda_{k}^2)$.
We may now use the observation that the Baker-Campbell-Hausdorff theorem \cite{QQ3} also holds when a 
certain operator $\widehat{B}$ is replaced by a continuous and differentiable function of $\widehat{B}$ (be it 
$Q(\widehat{B})$); in this case the  Baker-Campbell-Hausdorff theorem implies\footnote{In the case of the operator $Q(\widehat{B}) = e^{\widehat{B}}$ we can formally expand the exponential 
and apply the Baker-Campbell-Hausdorff theorem order by order. After all the contributions are resummed the result of Eq. (\ref{CH5a}) is obtained in the form  $e^{\widehat{A}} \, e^{\widehat{B}} \, e^{- \widehat{A}} = \exp{\bigl[e^{\widehat{A}} \, \widehat{B} \, e^{- \widehat{A}}\bigr]}$.}:
\begin{equation}
Q \biggl[ e^{\widehat{A}} \, \widehat{B} \, e^{- \widehat{A}} \biggr]= e^{\widehat{A}} \, Q(\widehat{B}) \, e^{- \widehat{A}}.
\label{CH5a}
\end{equation}
Thus using Eq. (\ref{CH5a}) (and recalling that $\widehat{\Sigma}$ commutes with $\widehat{{\mathcal R}}$), the expression of 
 $\widehat{c}(\tau)$ can be explicitly written as
 \begin{equation}
 \widehat{c}(\tau) = e^{- i\, \omega_{c}(\tau- \tau_{\ast})}\,\,\,\widehat{\Sigma}^{\dagger}[z] \,\,\,
 \widehat{{\mathcal D}}[\alpha(x)]\,\,\, \widehat{{\mathcal R}}[\beta(x)] \,\,\, \widehat{\Sigma}[z] \,\,\,\widehat{c}_{\ast}.
 \label{CH6}
 \end{equation}
The two operators $\widehat{{\mathcal D}}[\alpha(x)]$ and $\widehat{{\mathcal R}}[\beta(x)]$ can be understood, respectively, as generalized displacement and rotation operators:
\begin{equation}
\widehat{{\mathcal D}}[\alpha(x)] = e^{\int d^3\,k \bigl[ \alpha_{k}(x) (\widehat{b}_{\vec{k}}^{\,\,\dagger} + \widehat{b}_{-\vec{k}}^{\,\,\dagger}) - \alpha_{k}^{\ast}(x) (\widehat{b}_{\vec{k}} +  \widehat{b}_{-\vec{k}})\bigr]}, \qquad \widehat{{\mathcal R}}[\beta(x)] = e^{ 2 \, i \, \widehat{n}_{c} \int d^{3} k \,\,\beta_{k}(x)}.
\label{CH7}
\end{equation}
In the following two sections we are going to show that, in spite of the effective coupling, the degree of second-order coherence associated with the cavity mode does not depend on the correlation properties of the gravitons.

\renewcommand{\theequation}{6.\arabic{equation}}
\setcounter{equation}{0}
\section{Photon optics and second-order correlations}
\label{sec6}
\subsection{Degrees of coherence of the photons}
\subsubsection{The degree of first-order coherence of the cavity mode}
In the case of the photons the general expression of the degrees of first- and second-order coherence follows from Eqs. (\ref{HBT1})--(\ref{HBT2}) and (\ref{HBT5})--(\ref{HBT6}) 
(see also Eq. (\ref{HBT9})). We stress that in section \ref{sec2} $x = (\vec{x},\tau)$ but we find more convenient to separate the temporal from the spatial coordinates. In the case of an optical resonator in one spatial dimension  the degree of first-order 
coherence for the cavity mode is:
\begin{equation}
g_{c}^{(1)}(x_{1}, \tau_{1}; x_{2}, \tau_{2}) = \frac{\langle \widehat{E}^{(-)}(x_{1}, \tau_{1}) \, \widehat{E}^{(+)}(x_{2}, \tau_{2}) \rangle}{\sqrt{\langle \widehat{E}^{(-)}(x_{1}, \tau_{1}) \, \widehat{E}^{(+)}(x_{1}, \tau_{1}) \rangle \,\,\, \langle \widehat{E}^{(-)}(x_{2}, \tau_{2}) \, \widehat{E}^{(+)}(x_{2}, \tau_{2})\rangle}},
\label{FOC1}
\end{equation}
where we prefer to use $g_{c}^{(1)}(x_{1}, \tau_{1}; x_{2}, \tau_{2})$ [rather than $g_{\gamma}^{(1)}(x_{1}, \tau_{1}; x_{2}, \tau_{2})]$ to stress that we here dealing with a cavity mode.
Recalling the cartoon of Fig. \ref{Figure1}, the components of the cavity electric field 
are polarized in the $\hat{y}$ direction, i.e. $\widehat{E}(x,\tau) \equiv \widehat{E}_{y}(x,\tau)$. 
At the right hand side of Eq. (\ref{FOC1}) the spatial dependence of the cavity field is the same in the numerator and in the denominator;  thus the degree of first-ordercoherence can also be written more explicitly as: 
\begin{equation}
g_{c}^{(1)}( \tau; \tau + \Delta \tau) = \frac{\langle \widehat{c}^{\,\,\dagger}(\tau) \, \widehat{c}(\tau + \Delta\tau) \rangle}{\sqrt{\langle \widehat{c}^{\,\,\dagger}(\tau) \, \widehat{c}(\tau) \rangle \,\,\, \langle \widehat{c}^{\,\,\dagger}( \tau+ \Delta\tau) \, \widehat{c}(\tau+\Delta\tau)\rangle}},
\label{FOC1a}
\end{equation}
According to Eqs. (\ref{CH6})--(\ref{CH7}) the evolution of  $\widehat{c}(\tau)$ is governed by the product of 
three unitary operators so that the expectation values entering the denominator of Eq. (\ref{FOC1a}) always coincide with their respective expressions at $\tau_{\ast}$
\begin{equation}
\langle \widehat{c}^{\,\,\dagger}(\tau) \,\,\widehat{c}(\tau)\rangle = \langle \widehat{c}^{\,\,\dagger}(\tau+\Delta\tau) \,\,\widehat{c}(\tau+\Delta\tau) \rangle = \langle \widehat{c}_{\ast}^{\,\,\dagger} \,\, \widehat{c}_{\ast}\rangle.
\label{FOC1b}
\end{equation}
The expectation value in Eq. (\ref{FOC1b}) is then computed from 
the density matrix at $\tau_{\ast}$ 
\begin{equation}
\langle \widehat{c}_{\ast}^{\,\,\dagger} \, \widehat{c}_{\ast}\rangle = 
 \mathrm{Tr} \biggl[ \widehat{\rho}_{\ast} \, \widehat{c}_{\ast}^{\,\dagger} \, \widehat{c}_{\ast}\biggr] =  \mathrm{Tr} \biggl[ \widehat{\rho}_{c} \, \widehat{c}_{\ast}^{\,\dagger} \, \widehat{c}_{\ast}\biggr].
 \label{FOC1ba}
 \end{equation}
Since the density 
matrices of the cavity mode and of the gravitons span two different Hilbert spaces 
at $\tau_{\ast}$ we have that 
$\widehat{\rho}_{\ast} = \widehat{\rho}_{g} \otimes \widehat{\rho}_{c}$ 
(with $\mathrm{Tr} \widehat{\rho}_{c}= \mathrm{Tr} \widehat{\rho}_{g} =1$): this is the rationale for the last equality of  Eq. (\ref{FOC1ba}). Both $\widehat{\rho}_{c}$ and $\widehat{\rho}_{g}$ may describe either pure or mixed states.  After these necessary specifications, the numerator of Eq. (\ref{FOC1a}) 
is given by
\begin{eqnarray}
 \langle \widehat{c}^{\dagger}(\tau) \,\, \widehat{c}(\tau+ \Delta\tau) \rangle = \bigg\langle  \widehat{c}_{\ast}^{\,\,\dagger} \,\,\widehat{\Sigma}[z] \,\, \widehat{{\mathcal R}}^{\dagger}[\beta(x)]
 \widehat{{\mathcal D}}^{\dagger}[\alpha(x)] \widehat{{\mathcal D}}[\alpha(x + y)] 
\widehat{{\mathcal R}}[\beta(x + y)] \,\, \widehat{\Sigma}^{\dagger}[z] \,\, \widehat{c}_{\ast} \bigg\rangle. 
\label{CUNO}
\end{eqnarray}
Equation (\ref{CUNO}) requires the evaluation of the operators $\widehat{{\mathcal D}}[\alpha(x)]$ and $\widehat{{\mathcal R}}[\beta(x)]$ at two different times\footnote{As in Eqs. (\ref{CH6})--(\ref{CH7}) $x = k(\tau - \tau_{\ast})$ while, in what follows, $y = k \Delta\tau$}, i.e. $x+ y$ and $x$; the relevant products of operators appearing in Eq. (\ref{CUNO}) are then given by:
\begin{eqnarray}
\widehat{{\mathcal R}}^{\,\,\dagger}[\beta(x)] \, \widehat{{\mathcal R}}[\beta(x+y)] = e^{ - 2 \, i\,  \widehat{n}_{c} \Delta(x, y)}, \qquad
\widehat{{\mathcal D}}^{\,\,\dagger}[\alpha(x)] \, \widehat{{\mathcal D}}[\alpha(x+y)] = e^{- \widehat{\Delta}_{{\mathcal D}}(x, y)} \, e^{- i \Delta(x,y)},
\label{FOC1d}
\end{eqnarray}
where, as usual, $ \widehat{n}_{c}= \widehat{c}_{\ast}^{\,\,\dagger} \, \widehat{c}_{\ast}$.
In Eq. (\ref{FOC1d}) 
 the function $\Delta(x, y)$ and the operator $\widehat{\Delta}_{{\mathcal D}}(x, y)$ have been introduced for convenience and they are defined as:
\begin{eqnarray}
&& \Delta(x, y) = \int d^{3} k \,\, \lambda_{k}^2 \biggl[ \sin{(x+ y)} - \sin{x} - \sin{y}\biggr],
\nonumber\\
&& \widehat{\Delta}_{{\mathcal D}}(x, y) = \int d^{3}k \biggl[\alpha_{k}^{\ast}(x, y) (\widehat{b}_{\vec{k}} + \widehat{b}_{- \vec{k}}) - \alpha_{k}(x, y) (\widehat{b}_{\vec{k}}^{\dagger} + \widehat{b}_{- \vec{k}}^{\dagger})\biggr],
\label{FOC1e}
\end{eqnarray}
where $\alpha_{k}(x, y) = \lambda_{k}^2 e^{i \, x} [ e^{i \, y} -1]/2$. It is relevant to note that 
when $y \to 0$ (i.e. $\Delta \tau \to 0$) also $\alpha(x,y) \to 0$ and $\Delta(x, y) \to 0$. In this 
limit, as expected, $\widehat{{\mathcal R}}^{\dagger}[\beta] \, \widehat{{\mathcal R}}[\beta] \to \mathds{1}$ and also $\widehat{{\mathcal D}}^{\dagger}[\alpha] \, \widehat{{\mathcal D}}[\alpha] \to \mathds{1}$. From Eqs. (\ref{FOC1d})--(\ref{FOC1e}) the degree of first-order coherence reads 
\begin{equation}
g_{c}^{(1)}( \tau; \tau + \Delta \tau) = e^{i \, \gamma(x,y)} \mathrm{Tr}\biggl[ \widehat{\rho}_{\ast} \, \, e^{- 2\, i \Delta(x,y) \,\widehat{c}_{\ast}^{\,\dagger} \widehat{c}_{\ast} } \,\,\widehat{c}_{\ast}^{\,\dagger} \widehat{c}_{\ast}  \,\,\widehat{\Sigma}^{\dagger}[z] \,\, e^{-\widehat{\Delta}_{{\mathcal D}}(x, y) }\widehat{\Sigma}[z]\biggr]/ \mathrm{Tr}\biggl\{ \widehat{\rho}_{\ast} \, \, \widehat{c}_{\ast}^{\dagger} \widehat{c}_{\ast}\biggr\}.
\label{FOC1g}
\end{equation}
In Eq. (\ref{FOC1g}) $\gamma(x,y) = (\omega_{c}/k) y + \Delta(x,y)$ is an overall phase and the combinations appearing inside the traces are a direct consequence of Eq. (\ref{CUNO}) after 
noticing that $\widehat{{\mathcal R}}[\beta(x)]$ commutes with $\widehat{n}_{c}$ (but not with $\widehat{c}_{\ast}$ and $\widehat{c}^{\,\,\dagger}_{\ast}$ alone). Recalling that 
$\widehat{\rho}_{\ast} = \widehat{\rho}_{c} \otimes \widehat{\rho}_{g}$, Eq. (\ref{FOC1g}) 
becomes:
\begin{equation}
g_{c}^{(1)}( \tau; \tau + \Delta \tau) = e^{i \, \gamma(x,y)}  \mathrm{Tr}\biggl\{ \widehat{\rho}_{g} \widehat{\Sigma}^{\dagger}[z] \,\, e^{-\widehat{\Delta}_{{\mathcal D}}(x, y) }\widehat{\Sigma}[z]\biggr\} \mathrm{Tr}\biggl[ \widehat{\rho}_{c} \, \, e^{- 2\, i \Delta(x,y) \,\widehat{c}_{\ast}^{\,\dagger} \widehat{c}_{\ast} } \,\,\widehat{c}_{\ast}^{\,\dagger} \widehat{c}_{\ast}  \biggr]/\mathrm{Tr}\biggl[ \widehat{\rho}_{c}  \,\,\widehat{c}_{\ast}^{\,\dagger} \widehat{c}_{\ast} \biggr].
\label{FOChh}
\end{equation}
In the zero-delay limit $y \to 0$ (i.e. $\Delta\tau\to 0$) both $\gamma(x,y)$ and $\Delta(x,y)$ vanish and  the degree of first-order coherence 
of Eq. (\ref{FOChh}) goes to $1$, i.e.
\begin{equation}
\lim_{\Delta\tau\to 0}\,\,g_{c}^{(1)}( \tau; \tau + \Delta \tau) = 1.
\label{FOCea}
\end{equation}
In quantum optical applications (see e.g. Refs. \cite{QQ3} and
\cite{loudon}) the result of Eq. (\ref{FOCea}) holds for different states with opposite physical properties and it is the main motivation for the analysis of intensity correlations: quantum states describing very different physical situations all lead to the same degree of first-order coherence in the zero-delay limit. Thus $g_{c}^{(1)}( \tau; \tau + \Delta \tau)$ cannot be used (alone) to infer the properties of the underlying quantum state; this is why, as we are going to see in the following paragraph, the degrees of second-order coherence are, in practice, more essential.  Let us consider, as an example the case where the total state of the system is pure, i.e. $| \{s\}\rangle = | s_{c} \rangle \otimes |\{s_{g}\}\rangle$ and the gravitons are produced from the vacuum\footnote{ In this case the expectation value $\langle \{ s_{g} \} | \widehat{\Sigma}^{\dagger}[z] \,\, e^{-\widehat{\Delta}_{{\mathcal D}}(x, y) }\widehat{\Sigma}[z] |\{ s_{g} \}\rangle $ can be reduced, for a single mode of the field, to the average over the vacuum of a generalized 
displacement operator, i.e. $\langle 0| e^{\overline{\alpha} \,\,\widehat{a}^{\dagger} - \overline{\alpha}^{\ast}\,\, \widehat{a}}\, |0\rangle \to 1$.
We omit the details that are anyway similar to the ones discussed in the last part of this section.}. In this situation  $\langle \{s_{g}\} \widehat{\Sigma}^{\dagger}[z] \,\, e^{-\widehat{\Delta}_{{\mathcal D}}(x, y) }\widehat{\Sigma}[z]|\{s_{g}\}\rangle \to 1$ while the cavity contribution 
evaluates to a phase. Indeed in the case of a coherent or Fock state the expectation value involving $|s_{c} \rangle$ 
is given by 
\begin{equation}
\bigg\langle e^{- 2\, i \Delta(x,y) \,\widehat{n}_{c}} \,\, \widehat{n}_{c}\bigg\rangle/\langle \widehat{n}_{c}\rangle = 
e^{ - 2 i \Delta(x,y) \overline{n}_{c}},
\label{EX1}
\end{equation}
where $\overline{n}_{c}$ is the averaged multiplicity of the state, i.e. $\overline{n}_{c} = |\alpha|^2$ for a coherent state (i.e. $\widehat{c}_{\ast} | \alpha \rangle = \alpha |\alpha\rangle$) and $\overline{n}_{c} = n$ for a Fock state (i.e. $\overline{n}_{c}| n \rangle = n |\rangle $).
In this situation, even for $\Delta \tau \neq 0$, we have that $|g_{c}^{(1)}( \tau; \tau + \Delta \tau)| \to 1$. 
We may also evaluate the expectation value at the left hand side of Eq. (\ref{EX1}) for a chaotic mixture; in this 
case the density operator reads $\widehat{\rho}_{g}= \sum_{n =0}^{\infty} p_{n} \, |n\rangle \langle n|$ where the statistical
weights $p_{n} = \overline{n}^{n}/(\overline{n} +1)^{n+1}$ correspond to the ones of a Bose-Einstein distribution with averaged multiplicity $\overline{n}$:
\begin{equation}
\bigg\langle e^{- 2\, i \Delta(x,y) \,\widehat{n}_{c}} \,\, \widehat{n}_{c}\bigg\rangle/\langle \widehat{n}_{c}\rangle = e^{- 2 \, i\,  \Delta(x,y)}/[ \overline{n} ( 1 - e^{- 2 \, i\,  \Delta(x,y)})+1].
\label{EX2}
\end{equation}
It follows from Eq. (\ref{EX2})  that $|g_{c}^{(1)}( \tau; \tau + \Delta \tau)| = 1/[ \cos^2{\Delta(x,y)} + ( 1 + 2 \overline{n}) \sin^2{\Delta(x,y)}]$; thus in the case of a thermal mixture $0 <  |g_{c}^{(1)}( \tau; \tau + \Delta \tau)| \leq 1$; once more $|g_{c}^{(1)}( \tau; \tau + \Delta \tau)| \to 1$ 
when $y \to 0$ and $\Delta(x,y) \to 0$. As in the conventional quantum optical situation the degrees of first-order coherence give
inconclusive indications on the nature of the underlying quantum states.

\subsubsection{The degree of second-order coherence of the cavity mode}
The analysis of the correlation of the intensities 
of the photons (originally suggested by Hanbury-Bown and Twiss \cite{QQ1,QQ2}) 
is generally suitable for the determination of the statistical properties of the photons.   
The HBT correlations are encoded in the degree of second-order coherence
\begin{eqnarray}
g_{c}^{(2)}(x_{1}, \tau_{1}; x_{2}, \tau_{2}) = \frac{\langle \widehat{E}^{(-)}(x_{1}, \tau_{1}) \,\,\,\widehat{E}^{(-)}(x_{2}, \tau_{1})\,\,\, \widehat{E}^{(+)}(x_{2}, \tau_{2}) \,\,\, \widehat{E}^{(+)}(x_{1}, \tau_{1}) \rangle}{\langle \widehat{E}^{(-)}(x_{1}, \tau_{1}) \,\, \widehat{E}^{(+)}(x_{1}, \tau_{1})\rangle \,\,\, \langle \widehat{E}^{(-)}(x_{2}, \tau_{2}) \,\,\, \widehat{E}^{(+)}(x_{2}, \tau_{2})\rangle},
\label{FOC2}
\end{eqnarray}
where, as already recalled, the electric field operators are polarized along  $\hat{y}$  since 
the cavity is directed along $\hat{x}$ (see Fig. \ref{Figure1} and discussion thereafter). Equation (\ref{FOC2}) follows from Eqs. (\ref{HBT5})--(\ref{HBT6}) written with a non-covariant notation.
As in the case of the degrees of first-order coherence the spatial dependence is the same in the numerator and in the denominator; thus the degree of second-order 
coherence defined in Eq. (\ref{FOC2}) becomes: 
\begin{equation}
g_{c}^{(2)}(\tau, \tau + \Delta \tau) = \frac{\langle \widehat{c}^{\dagger}(\tau)\,\, \widehat{c}^{\dagger}(\tau + \Delta \tau)\,\, \widehat{c}(\tau + \Delta \tau)\,\, \widehat{c}(\tau)\rangle }{\langle \widehat{c}^{\dagger}(\tau)\,\widehat{c}(\tau)\rangle \,\, \langle \widehat{c}^{\dagger}(\tau+ \Delta\tau)\,\widehat{c}(\tau+ \Delta\tau)\rangle}.
\label{FOC2a}
\end{equation}
The structure of Eq. (\ref{FOC2a}) already shows that the degree of second-order 
coherence of the photons cannot be employed to infer the statistical properties 
of the gravitons. To prove this statement we consider preliminarily the case 
of a pure state given in the form $| \{s\}\rangle = | s_{c} \rangle \otimes |\{s_{g}\}\rangle$; in this situation
it is immediate to realize that the denominator of Eq. (\ref{FOC2a}) is 
simply given by $\langle s_{c} | \widehat{c}_{\ast}^{\dagger} \,\, \widehat{c}_{\ast} | s_{c}\rangle^2$. 
With the same conventions the numerator appearing at the right-hand side of Eq. (\ref{FOC2a}) 
equals $\langle s_{c} | \widehat{c}_{\ast}^{\dagger} \,\,\widehat{c}_{\ast}^{\dagger}\,\, \widehat{c}_{\ast}\,\,\widehat{c}_{\ast} | s_{c}\rangle$. This result is immediate if we observe 
 that the product of operators $\widehat{c}^{\dagger}(\tau + \Delta \tau)\,\, \widehat{c}(\tau + \Delta \tau)$ is explicitly given by:
  \begin{equation}
 \widehat{c}_{\ast}^{\dagger} \,\,\widehat{\mathcal R}^{\dagger}[\beta(x+y)]\,\,\widehat{\Sigma}^{\dagger}[z] \,\,\widehat{{\mathcal D}}^{\dagger}[\alpha(x+y)]\, \widehat{\Sigma}[z]\, \, \widehat{\Sigma}^{\dagger}[z] \,\, \widehat{{\mathcal D}}[\alpha(x+y)]\,\,\widehat{\Sigma}[z] \,\,\widehat{\mathcal R}[\beta(x+y)] \widehat{c}_{\ast}.
  \label{FOC2b}
  \end{equation}
Since the operators $\widehat{\mathcal R}[\beta]$, $\widehat{\mathcal D}[\alpha]$ and $\widehat{\Sigma}[z]$ are all unitary the result of Eq. (\ref{FOC2b}) is just 
  $\widehat{c}_{\ast}^{\dagger} \widehat{c}_{\ast}$. This conclusion also implies that 
\begin{equation}
\langle \{s\} |\widehat{c}^{\dagger}(\tau)\,\, \widehat{c}^{\dagger}(\tau + \Delta \tau)\,\, \widehat{c}(\tau + \Delta \tau)\,\, \widehat{c}(\tau)|\{s\}\rangle = \langle \{s\} |\widehat{c}^{\dagger}(\tau)\,\, \widehat{n}_{c}\,\, \widehat{c}(\tau)|\{s\}\rangle,
\label{FOC2c}
\end{equation}
where we introduced, as usual, $\widehat{n}_{c} = \widehat{c}_{\ast}^{\dagger}\,\, \widehat{c}_{\ast}$ (i.e. 
the number operator associated with the cavity mode); the result of Eq. (\ref{FOC2c}) is then justified since 
$[\widehat{H}, \widehat{n}_{c}]=0$ and this implies $\widehat{n}_{c}(\tau+ \Delta\tau) = \widehat{n}_{c}(\tau) = \widehat{c}_{\ast}^{\dagger}\,\, \widehat{c}_{\ast}$. For the same reason we also have that 
\begin{equation}
\biggl[ \widehat{{\mathcal R}}[\beta(x)], \widehat{n}_{c}(x)\biggr] =  \biggl[ \widehat{{\mathcal R}}[\beta(x+y)], \widehat{n}_{c}(x+y)\biggr] =0.
\label{FOC2d}
\end{equation}
Equation (\ref{FOC2d}) follows  because the operator $\widehat{{\mathcal R}}[\beta(x)]$ 
depends on $\widehat{n}_{c}$ and therefore commutes with the total Hamiltonian.
Also $[ \widehat{{\mathcal D}}, \widehat{n}_{c}] =0$ and $[ \widehat{\Sigma}, \widehat{n}_{c}] =0$:
this happens since neither $\widehat{{\mathcal D}}[\beta(x)]$ nor $\widehat{\Sigma}$ 
depend the cavity mode. Thanks to these three observations we then obtain 
\begin{equation}
\langle \{s\} |\widehat{c}^{\dagger}(\tau)\,\, \widehat{n}_{c}\,\, \widehat{c}(\tau)|\{s\}\rangle 
= \langle s_{c}| \widehat{c}^{\dagger}_{\ast} \,\,\widehat{c}^{\dagger}_{\ast}\,\,\widehat{c}_{\ast}\,\,
\widehat{c}_{\ast} |s_{c} \rangle.
\label{FOC2e}
\end{equation}
This means that the degree of second-order coherence of Eq. (\ref{FOC2a}) is generally given by
\begin{equation}
g_{c}^{(2)}(\tau, \tau + \Delta \tau) = \frac{\langle s_{c}| \widehat{c}^{\dagger}_{\ast} \,\,\widehat{c}^{\dagger}_{\ast}\,\,\widehat{c}_{\ast}\,\,
\widehat{c}_{\ast} |s_{c} \rangle}{\langle s_{c}| \widehat{c}^{\dagger}_{\ast} \,\,
\widehat{c}_{\ast} |s_{c} \rangle^2},
\label{FOC2f}
\end{equation}
irrespective the time delay and {\em of the quantum state of the gravitons}. Equation (\ref{FOC2f}) has been derived for a pure state but it also holds for a statistical mixture characterized by a total density matrix $\widehat{\rho}_{\ast} = \widehat{\rho}_{g} \otimes \widehat{\rho}_{c}$:
\begin{equation}
g_{c}^{(2)}(\tau, \tau + \Delta \tau) = \frac{\mathrm{Tr} [\widehat{\rho}_{c} \,\,\widehat{c}^{\dagger}_{\ast} \,\,\widehat{c}^{\dagger}_{\ast}\,\,\widehat{c}_{\ast}\,\,
\widehat{c}_{\ast}]}{\mathrm{Tr} [\widehat{\rho}_{c}\,\, \widehat{c}^{\dagger}_{\ast} \,\,
\widehat{c}_{\ast}]^2} = 1+ \frac{\sigma_{c}^2 - \overline{n}_{c} }{ \overline{n}_{c}^2},
\label{FOC2g}
\end{equation}
where $\overline{n}_{c} = \langle \widehat{n}_{c} \rangle = \mathrm{Tr}[ \widehat{\rho}_{c} \, \widehat{n}_{c} ]$ is the averaged multiplicity of the photons and  $\sigma_{c}^2 = \langle \widehat{n}_{c}^2 \rangle - \langle \widehat{n}_{c} \rangle^2$ the total dispersion.
Equation (\ref{FOC2g}) means that, in spite of the state of the gravitons, the degree of the second-order coherence of the photons is only determined by $| s_{c} \rangle$. So if we imagine, for instance,  that a photon is created inside our cavity  in resonance with the laser drive we always have that $g_{c}^{(2)}(\tau, \tau+ \Delta\tau) =1$; this is because for a coherent state $\sigma_{c}^2 = \overline{n}_{c}$. The Poissonian limit 
holds for a single-mode coherent state (i.e. $\widehat{c}_{\ast} |\alpha \rangle = \alpha  |\alpha \rangle$) and Eq. (\ref{FOC2g}) implies
$g_{c}^{(2)}\to 1$. If the state of the cavity photons is different we may have $g_{c}^{(2)} > 1$ but this has nothing to do with 
contribution of the gravitons. For instance, for a chaotic state charaterized by a Bose-Einstein distribution Eq. (\ref{FOC2g}) 
would imply, as usual, $g_{c}^{(2)} \to 2$ (see e.g. \cite{QQ3} and also \cite{loudon}). Finally for a single Fock state (with 
$\widehat{c}_{\ast} |\, n\,\rangle = n  |\, n\,\rangle$) we would have $g_{c}^{(2)} \to (1 - 1/n )< 1$ implying a sub-Poissonian 
statistics. This conclusion tells that the degree of second-order coherence 
of the photons in a cavity are always insensitive to the statistical properties 
of the gravitons. This is not only true in practice (because of the smallness of the coupling) 
but also in principle. It is also interesting that the results deduced for the degree of second-order coherence 
also hold for all the degrees of higher-order coherence. As we saw in section \ref{sec2}
the degree of third-order coherence $g_{c}^{(3)}(x_{1}, \tau_{1}; x_{2}, \tau_{2}; x_{3}, \tau_{3})$ is given by:
\begin{equation}
\frac{\langle \widehat{E}^{(-)}(x_{1}, \tau_{1}) \,\,\,\widehat{E}^{(-)}(x_{2}, \tau_{1})\,\,\, \widehat{E}^{(-)}(x_{3}, \tau_{3})\,\,\,\widehat{E}^{(+)}(x_{3}, \tau_{3})\,\,\widehat{E}^{(+)}(x_{2}, \tau_{2}) \,\,\, \widehat{E}^{(+)}(x_{1}, \tau_{1}) \rangle}{\sqrt{\langle \widehat{E}^{(-)}(x_{1}, \tau_{1}) \,\, \widehat{E}^{(+)}(x_{1}, \tau_{1})\rangle \, \langle \widehat{E}^{(-)}(x_{2}, \tau_{2}) \,\,\, \widehat{E}^{(+)}(x_{2}, \tau_{2})\rangle \langle \widehat{E}^{(-)}(x_{3}, \tau_{3})\, \,\,\, \widehat{E}^{(+)}(x_{3}, \tau_{3})\rangle}}.
\label{FOC3}
\end{equation}
In the case of the cavity mode we then have that the spatial dependence disappears and the result 
for the degree of third-order coherence is 
\begin{equation}
g_{c}^{(3)}(\tau_{1}; \tau_{2}; \tau_{3}) = \frac{\langle \widehat{c}^{\,\,\dagger}(\tau_{1}) \,\,\widehat{c}^{\,\,\dagger}(\tau_{2})\,\, \widehat{c}^{\,\,\dagger}(\tau_{3})\,\, \widehat{c}(\tau_{3})\,\, \widehat{c}(\tau_{2}) \widehat{c}(\tau_{1})\rangle }{\sqrt{\langle \widehat{c}^{\,\,\dagger}(\tau_{1})\,\, \widehat{c}(\tau_{1}) \rangle \,\,\langle \widehat{c}^{\,\,\dagger}(\tau_{2})\,\, \widehat{c}(\tau_{2}) \rangle\,\, \langle \widehat{c}^{\,\,\dagger}(\tau_{3})\,\, \widehat{c}(\tau_{3}) \rangle}}.
\label{FOC4}
\end{equation}
Because of the unitarity of the evolution (and using the same steps discussed in the case of $g^{(2)}_{c}(\tau_{1},\tau_{2})$)  the degree of third-order coherence reduces to
\begin{equation}
g_{c}^{(3)}(\tau_{1}; \tau_{2}; \tau_{3}) = \frac{\langle \widehat{c}^{\,\,\dagger}_{\ast}\,\, \widehat{c}^{\,\,\dagger}_{\ast}\,\,\widehat{c}^{\,\,\dagger}_{\ast}\,\, \widehat{c}_{\ast}\,\, \widehat{c}_{\ast} \widehat{c}_{\ast} \rangle }{[\langle \widehat{c}^{\,\,\dagger}_{\ast}\,\, \widehat{c}_{\ast}\rangle ]^{3/2}}.
\label{FOC5}
\end{equation}
Equations (\ref{FOC4})--(\ref{FOC5}) can also be generalized even further to the degree of $n$-th order coherence. We therefore conclude that the degree of higher-order coherence 
of the photons is not sensitive to the quantum state of the gravitons. 

\subsection{Degrees of coherence of the gravitons}
To complete the discussion it is now relevant to determine independently the degree of coherence 
of the gravitons. The operators entering the Glauber correlators describing the degrees of first- and second- order coherence of the gravitons must always be Hermitian and since the gravitons are produced in pairs of opposite three momenta the field operators (for a single tensor polarization) can be 
expressed as:
\begin{equation}
\widehat{\mu}(\vec{x}, \tau) = \widehat{\mu}^{(+)}(\vec{x}, \tau) + \widehat{\mu}^{(-)}(\vec{x}, \tau),
\label{muop1}
\end{equation}
where $\widehat{\mu}^{(+)\,\dagger}= \widehat{\mu}^{(-)}$ and the explicit form of 
$\mu^{(\pm)}(\vec{x},\tau)$ is
\begin{eqnarray}
\widehat{\mu}^{(-)}(\vec{x}, \tau) = \frac{\sqrt{2} \ell_{P}}{ 2\, (2\pi)^{3/2}} \int \frac{d^{3} k}{\sqrt{2 k}} \,\, 
\widehat{V}_{\vec{k}}^{\dagger}(\vec{x}, \tau), \qquad \widehat{\mu}^{(+)}(\vec{x}, \tau) = \frac{\sqrt{2} \ell_{P}}{ 2\, (2\pi)^{3/2}} \int \frac{d^{3} k}{\sqrt{2 k}}\,\, \widehat{V}_{\vec{k}}(\vec{x}, \tau).
\label{muop3}
\end{eqnarray}
The two operators $\widehat{V}_{\vec{k}}(\vec{x}, \tau)$ and $\widehat{V}_{\vec{k}}^{\dagger}(\vec{x}, \tau)$ depend both on $\vec{k}$ and $\vec{x}$:
\begin{equation}
\widehat{V}_{\vec{k}}(\vec{x}, \tau) = \widehat{a}_{\vec{k}}(\tau) \,\, e^{- i \vec{k}\cdot\vec{x}} + \widehat{a}_{-\vec{k}}(\tau) \,\,e^{ i \vec{k}\cdot\vec{x}}, \qquad \widehat{V}_{\vec{k}}^{\dagger}(\vec{x}, \tau) = \widehat{a}_{\vec{k}}^{\dagger}(\tau)\,\, e^{i \vec{k}\cdot\vec{x}} + \widehat{a}_{-\vec{k}}^{\dagger}(\tau) \,\,e^{- i \vec{k}\cdot\vec{x}}.
\label{muop4}
\end{equation}
In what follows we shall primarily consider the case of a single tensor polarization since 
this is the interaction appearing in the Hamiltonian of Eq. (\ref{HG1}). This is 
also the exclusive perspective customarily adopted in quantum optical applications \cite{QQ3} (see 
aslo \cite{loudon}). It is however true that the degree of second-order coherence of the gravitons, originally introduced in Ref. \cite{MG1}, can be discussed within different perspectives \cite{MG2,MG3,MG4,MG5}. It is possible to consider more inclusive descriptions where all the polarizations are treated on equal footing but this is not essential in the present discussion, as we shall 
briefly comment at the end of the analysis.

\subsubsection{The degree of first-order coherence of the gravitons}
From Eqs. (\ref{HBT3})--(\ref{HBT4}) and (\ref{HBT4a}) the degree of first-order coherence of the gravitons is
\begin{equation}
g_{g}^{(1)}(\vec{x}_{1}, \tau_{1}; \vec{x}_{2}, \tau_{2}) = \frac{\langle \widehat{\mu}^{(-)}(\vec{x}_{1},\tau_{1})\,\, \widehat{\mu}^{(+)}(\vec{x}_{2}, \tau_{2}) \rangle }{\sqrt{\langle \widehat{\mu}^{(-)}(\vec{x}_{1},\tau_{1})\,\, \widehat{\mu}^{(+)}(, \vec{x}_{1},\tau_{1}) \rangle\,\,\,\langle \widehat{\mu}^{(-)}(\vec{x}_{2}, \tau_{2})\,\, \widehat{\mu}^{(+)}(\vec{x}_{2}, \tau_{2}) \rangle }}.
\label{GC1}
\end{equation}
We remind that in Eq. (\ref{HBT4a}) a four-dimensional notation has been employed for the 
space-time coordnates (i.e. $x = (\vec{x}, \tau)$); in Eq. (\ref{GC1}) and in the following discussion
we prefer to separate spatial and temporal coordinates. 
As in the case of the photons it is practical to analyze preliminarily the numerator of Eq. (\ref{GC1}) 
and then assess the degree of first-order coherence. In this perspective, recalling Eqs. (\ref{muop3})--(\ref{muop4}) the numerator of Eq. (\ref{GC1}) is:
\begin{eqnarray}
\langle \widehat{\mu}^{(-)}(\vec{x}_{1},\tau_{1})\,\, \widehat{\mu}^{(+)}(\vec{x}_{2}, \tau_{2}) \rangle &=& \frac{\ell_{p}^2}{4 (2\pi)^{3}} 
\int \frac{d^3 k_{1}}{\sqrt{k_{1}}} \int \frac{d^{3} k_{2}}{\sqrt{k_{2}}}\biggl[ \biggl\langle \widehat{a}_{\vec{k}_{1}}^{\dagger}(\tau_{1}) \,\, \widehat{a}_{\vec{k}_{2}}(\tau_{2}) \biggr\rangle\,\, e^{i \, (\vec{k}_{1}\cdot\vec{x}_{1} -  \vec{k}_{2}\cdot\vec{x}_{2})}
\nonumber\\
&+& \biggl\langle \widehat{a}_{\vec{k}_{1}}^{\dagger}(\tau_{1}) \,\, \widehat{a}_{- \vec{k}_{2}}(\tau_{2}) \biggr\rangle
e^{i \, (\vec{k}_{1}\cdot\vec{x}_{1} +  \vec{k}_{2}\cdot\vec{x}_{2})} 
\nonumber\\
&+& \biggl\langle \widehat{a}_{-\vec{k}_{1}}^{\dagger}(\tau_{1}) \,\, \widehat{a}_{\vec{k}_{2}}(\tau_{2}) \biggr\rangle
e^{- i \, (\vec{k}_{1}\cdot\vec{x}_{1} +  \vec{k}_{2}\cdot\vec{x}_{2})}
\nonumber\\
&+&  \biggl\langle \widehat{a}_{-\vec{k}_{1}}^{\dagger}(\tau_{1}) \,\, \widehat{a}_{- \vec{k}_{2}}(\tau_{2}) \biggr\rangle
e^{-i \, (\vec{k}_{1}\cdot\vec{x}_{1} - \vec{k}_{2}\cdot\vec{x}_{2})}\biggr].
\label{GC1a}
\end{eqnarray}
Thanks to the results of Eq. (\ref{SH2})  
the relevant field operators entering Eq. (\ref{GC1a}) must be evaluated when the (physical) wavelengths of the gravitons are all shorter than the Hubble radius (i.e. $k > a \, H$): 
this is the relevant physical regime where the degrees of coherence 
of the photons and of the gravitons are comparable. In the language of Fig. \ref{Figure3} 
we must then consider wavenumbers $k > a_{re} \, H_{re}$ and time-scales $\tau \geq \tau_{\ast}$. In this concrete situation\footnote{To avoid confusions we stress that in Eqs. (\ref{GCt1})--(\ref{GCt2}) $x = k (\tau - \tau_{\ast})$.} the explicit form of the field operators of Eq. (\ref{GC1a}) is:
\begin{eqnarray}
\widehat{a}_{\vec{k}}(\tau) &=& \biggl[ d_{+}(k) \,\,\widehat{b}_{\vec{k}} + d_{-}^{\ast}(k)\,\,\widehat{b}_{-\vec{k}}^{\,\dagger} \biggr] \,\, e^{- i \,x} 
+ \lambda_{k} \, k\, \widehat{c}_{\ast}^{\dagger} \, \widehat{c}_{\ast} \,\,\biggl( 1 - e^{-i \, x}\biggr),
\label{GCt1}\\
\widehat{a}_{-\vec{k}}^{\dagger}(\tau) &=& \biggl[ d_{+}^{\ast}(k) \,\,\widehat{b}_{-\vec{k}}^{\,\dagger} + 
d_{-}(k) \,\,\widehat{b}_{\vec{k}} \biggr] \,\,e^{ i \, x} + \lambda_{k} \, k\, \widehat{c}_{\ast}^{\dagger} \, \widehat{c}_{\ast}\,\,\biggl( 1 - e^{i\, x}\biggr).
\label{GCt2}
\end{eqnarray}
In Eqs. (\ref{GCt1})--(\ref{GCt2}) the first term describes the direct production of the gravitons 
thanks to the evolution of the space-time curvature. The second (subleading) term ${\mathcal O}(\lambda_{k})$ accounts for the contribution of the photons to the degree of coherence 
of the gravitons. If we would be able to measure directly the degrees of coherence of the gravitons
the terms ${\mathcal O}(\lambda_{k})$ could be used (at least in principle) to infer 
the properties of the cavity photons. We are unfortunately in the opposite situation since 
from the properties of the cavity photons we would like to obtain some clue about the gravitons
This is why, in our context, the terms ${\mathcal O}(\lambda_{k})$ appearing in Eqs. 
(\ref{GCt1})--(\ref{GCt2}) are conceptually less relevant since they give the correlation properties of the photons once the degrees of coherence of the gravitons are measured. In our discussion we rather focus on the leading-order result and only keep track, for the sake of accuracy, of the corrections.
With these caveats we then insert Eqs. (\ref{GCt1})--(\ref{GCt2}) into  Eq. (\ref{GC1a}) and obtain, as expected, a result that only depends upon  $r = |\vec{x}_{2} - \vec{x_{1}}|$:
\begin{equation}
\langle \{s\} |\widehat{\mu}^{(-)}(\vec{x}_{1},\tau)\,\, \widehat{\mu}^{(+)}(\vec{x}_{2}, \tau)| \{\ s\} \rangle  = \int_{k_{min}}^{k_{max}} k \, d k \,\,e^{i k (\tau_{1} - \tau_{2})}\, j_{0}(k\, r) A(k) + 4\pi \lambda^2 \langle n_{c}^2 \rangle G^{\ast}(\tau_{1}) G(\tau_{2}),
\label{GCt3}
\end{equation}
where $j_{0}(k\,r)$ is the zeroth-order spherical Bessel function  \cite{SP1,SP2} and the integrals over the momenta extend over the whole spectrum of the gravitons ranging between $k_{min} = 2 \pi \nu_{min}$ and $k_{max} = 2 \pi \nu_{max}$. The lowest frequency range of the gravitons 
coincides with the aHz region where the maximum of the spectral energy density is expected 
in the context of the concordance scenario; in this region the direct limits on the 
tensor to scalar ratio imply $r_{T} < {\mathcal O}(0.03)$ \cite{SF6,SF7}. The maximal wavenumber 
of the gravitons coincides with the region where the expansion rate peaks (see Fig. \ref{Figure3} and discussion therein). In a quantum mechanical perspective the maximal frequency coincides 
with the region where only few gravitons are produced and this is why $\nu_{max}$ cannot 
exceed the THz region \cite{numax} (see also \cite{numax2}). What matters the most 
for the present considerations is the actual existence of $k_{min}$ and $k_{max}$. This is why 
the assorted integrals appearing in the degrees of coherence do not lead to divergent results and 
make the various limits well defined, as we are going to see in a moment.
To maintain a concise notation in Eq. (\ref{GCt3}) the following two functions have been introduced:
\begin{eqnarray}
A(k) = \bigl| d_{+}(k) \bigr|^2 \overline{n}_{k} + \bigl| d_{-}(k) \bigr|^2\, (\overline{n}_{k}+1),
\qquad G(\tau) = \int_{k_{min}}^{k_{max}} \,k \, [ 1 - e^{-i k (\tau- \tau_{\ast})}] \,\, d k.
\label{GCt4}
\end{eqnarray}
In Eq. (\ref{GCt4}) $\overline{n}_{k}$ denotes the averaged multiplicity of the initial state of the gravitons  and it appears when evaluating the expectation values of the field operators; so for instance we would have that 
\begin{equation}
\langle \widehat{b}_{\vec{k}}^{\,\dagger} \, \widehat{b}_{\vec{p}} \rangle = \mathrm{Tr} \biggl[\widehat{\rho}_{\ast}\widehat{b}_{\vec{k}}^{\,\dagger} \,\,\widehat{b}_{\vec{p}} \biggr] = \mathrm{Tr} \biggl[\widehat{\rho}_{g} \,\,\widehat{b}_{\vec{k}}^{\,\dagger} \,\, \widehat{b}_{\vec{p}}\biggr] 
= \overline{n}_{k} \,\,\delta^{(3)}(\vec{k} - \vec{p}).
\label{GCt5}
\end{equation}
When the initial state is the vacuum $\overline{n}_{k} \to 0$ and from Eq. (\ref{GCt4}) we have that  $A(k) \to  \bigl| d_{-}(k) \bigr|^2$. If we now drop for conciseness the extrema of integration (with 
the proviso that they actually exist), the degree of first-order coherence defined in Eq. (\ref{GC1}) can 
be rewritten as:
\begin{equation}
g_{g}^{(1)}(r,\,\tau_{1},\tau_{2}) = \frac{\int k \, d k \,\,e^{i k (\tau_{1} - \tau_{2})}\, j_{0}(k\, r) A(k) + 4\pi \lambda^2 \langle \widehat{n}_{c}^2 \rangle G^{\ast}(\tau_{1}) G(\tau_{2})}{\sqrt{[\int k \, d k \, A(k) + 4\pi \lambda^2 \langle \widehat{n}_{c}^2 \rangle |G(\tau_{1}) |^2]\, [\int k \, d k \, A(k) + 4\pi \lambda^2 \langle \widehat{n}_{c}^2 \rangle |G(\tau_{2}) |^2]}}.
\label{GCt5a}
\end{equation}
In the zero-delay limit $\tau_{1}\to \tau_{2} = \tau$ and Eq. (\ref{GCt5a}) gets even simpler:
\begin{equation}
g_{g}^{(1)}(r,\,\tau) = \frac{\int k\, d k\, j_{0}(k\, r)  \, [ \bigl| d_{+}(k) \bigr|^2 \overline{n}_{k} + 
\bigl| d_{-}(k) \bigr|^2\, (\overline{n}_{k}+1)]}{\int k\, d k \, [ \bigl| d_{+}(k) \bigr|^2 \overline{n}_{k} + 
\bigl| d_{-}(k) \bigr|^2\, (\overline{n}_{k}+1)]}  +  {\mathcal O}[ \lambda^2 \, \langle \widehat{n}_{c}^2 \rangle ].
\label{GC3}
\end{equation}
If the frequency of the gravitons is large in comparison with the inverse length of the cavity (i.e. $|k \, L |\ll 1$) the spherical Bessel function can always be evaluated 
in its small argument limit, i.e. $j_{0}(k\, r) \to 1$; thus, as expected the degree of first-order coherence goes always to $1$ when the corresponding wavelengths are shorter than the Hubble radius after 
reentry (i.e. in the rightmost region of Fig. \ref{Figure3}): 
\begin{equation}
\lim_{k r \to 0} g_{g}^{(1)}(r,\,\tau) = 1 + {\mathcal O}[ \lambda^2 \, \langle \widehat{n}_{c}^2 \rangle ],
\label{GC4}
\end{equation}
where we also indicated a series of corrections whose first term is of the order $\lambda^2 \langle \widehat{n}_{c}^2\rangle$. Equation (\ref{GC4}) is closely related Eq. (\ref{FOCea}: In both cases the photons and the gravitons are first-order coherent but this result does not clarify their statistical properties and their quantum mechanical origin.

\subsubsection{The degree of second-order coherence of the gravitons}
We now go back to Eqs. (\ref{HBT7}) and (\ref{HBT8}) with the purpose 
of evaluating the degree of second-order coherence of the gravitons. As before 
we consider a single polarization so that the relevant expression is given by:
\begin{equation}
g_{g}^{(2)}(\vec{x}_{1}, \tau_{1}; \vec{x}_{2}, \tau_{2}) = \frac{\langle \widehat{\mu}^{(-)}(\vec{x}_{1}, \tau_{1})\,\, \widehat{\mu}^{(-)}(\vec{x}_{2}, \tau_{2}) \,\, \widehat{\mu}^{(+)}(\vec{x}_{2}, \tau_{2}) \,\, \widehat{\mu}^{(+)}(\vec{x}_{1}, \tau_{1})\rangle }{\langle \widehat{\mu}^{(-)}(\vec{x}_{1},\tau_{1})\,\, \widehat{\mu}^{(+)}(\vec{x}_{1},\tau_{1}) \rangle\,\,\langle \widehat{\mu}^{(-)}(\vec{x}_{2}, \tau_{2})\,\, \widehat{\mu}^{(+)}(\vec{x}_{2}, \tau_{2}) \rangle }.
\label{GD1}
\end{equation}
The denominator appearing at the right hand side of Eq. (\ref{GD1}) follows 
from Eq. (\ref{GC1a}) and from the explicit expression of the first Glauber correlator 
already analyzed above. The numerator of Eq. (\ref{GD1}) is apparently 
more cumbersome than the ones evaluated so far and it formally corresponds to:
\begin{eqnarray}
&&\biggl\langle \widehat{\mu}^{(-)}(\vec{x}_{1}, \tau_{1})\,\, \widehat{\mu}^{(-)}(\vec{x}_{2}, \tau_{2}) \,\, \widehat{\mu}^{(+)}(\vec{x}_{2}, \tau_{2}) \,\, \widehat{\mu}^{(+)}(\vec{x}_{1}, \tau_{1})\biggr\rangle = \frac{\ell_{P}^4}{(2\pi)^{6}} \int \frac{d^{3} k_{1}}{k_{1}} \int \frac{d^{3} k_{2}}{k_{2}}
\int \frac{d^{3} k_{3}}{k_{3}}  \int \frac{d^{3} k_{4}}{k_{4}} 
\nonumber\\
&& \times  \biggl\langle \widehat{V}_{\vec{k}_{1}}^{\dagger}(\vec{x}_{1},\tau_{1}) \,\, \widehat{V}^{\dagger}_{\vec{k}_2}(\vec{x}_{2}, \tau_{2}) \,\,\widehat{V}_{ \vec{k}_3}(\vec{x}_{2},\tau_{2})  \,\,\widehat{V}_{ \vec{k}_4}(\vec{x}_{1},\tau_{1})  \biggr\rangle.
\label{GD1a}
\end{eqnarray}
The same steps leading to Eqs. (\ref{GCt1})--(\ref{GCt2}) will now be followed with the 
difference that now the relevant integrals involve $4$ three-momenta (i.e. $\vec{k}_{1}$, $\vec{k}_{2}$, $\vec{k}_{3}$ and $\vec{k}_{4}$) and $16$ different expectation values; each of these expectation values involves four operators. We do now write them all since they are just repetitive but an example is given by 
\begin{eqnarray}
&&\biggl\langle \hat{a}^{\dagger}_{- \vec{k}_1}(\tau_{1}) \,\, \hat{a}^{\dagger}_{- \vec{k}_2}(\tau_{2}) \,\,\hat{a}_{ \vec{k}_3}(\tau_{2})  \,\,\hat{a}_{ \vec{k}_4}(\tau_{1})  \biggr\rangle=\biggl\langle \hat{B}^{\dagger}_{- \vec{k}_1}(\tau_{1}) \,\, \hat{B}^{\dagger}_{- \vec{k}_2}(\tau_{2}) \,\,\hat{B}_{ \vec{k}_3}(\tau_{2})  \,\,\hat{B}_{ \vec{k}_4}(\tau_{1})  \biggr\rangle
\nonumber\\
&&+ \langle \widehat{n}_{c}^4\rangle D_{k_{1}}(\tau_{1})D_{k_{2}}(\tau_{2})D_{k_{3}}^{\ast}(\tau_{2})D_{k_{4}}^{\ast}(\tau_{1})
 + \biggl\langle \hat{B}^{\dagger}_{- \vec{k}_1}(\tau_{1}) \,\,\hat{B}_{ \vec{k}_3}(\tau_{2})  \biggr\rangle\,\,\langle \widehat{n}_{c}^2\rangle \,\,D_{k_{2}}(\tau_{2}) D_{k_{4}}^{\ast}(\tau_{1})
\nonumber\\
&& + \biggl\langle \hat{B}^{\dagger}_{- \vec{k}_2}(\tau_{2}) \,\,\hat{B}_{ \vec{k}_4}(\tau_{1})  \biggr\rangle\,\,\langle \widehat{n}_{c}^2\rangle \,\,D_{k_{1}}(\tau_{1}) D_{k_{3}}^{\ast}(\tau_{2}) + \biggl\langle \hat{B}^{\dagger}_{- \vec{k}_1}(\tau_{1}) \,\,\hat{B}_{ \vec{k}_4}(\tau_{1})  \biggr\rangle\,\,\langle \widehat{n}_{c}^2\rangle \,\,D_{k_{3}}(\tau_{1}) D_{k_{2}}^{\ast}(\tau_{2})
\nonumber\\
&&+ \biggl\langle \hat{B}^{\dagger}_{- \vec{k}_2}(\tau_{2}) \,\,\hat{B}_{ \vec{k}_3}(\tau_{2})  \biggr\rangle\,\,\langle \widehat{n}_{c}^2\rangle \,\,D_{k_{1}}(\tau_{2}) D_{k_{4}}^{\ast}(\tau_{1}) +\biggl\langle \hat{B}^{\dagger}_{- \vec{k}_1}(\tau_{1}) \,\,\hat{B}_{ -\vec{k}_2}^{\dagger}(\tau_{2})  \biggr\rangle\,\,\langle \widehat{n}_{c}^2\rangle \,\,D_{k_{3}}(\tau_{2}) D_{k_{4}}^{\ast}(\tau_{1})
\nonumber\\
&& + \biggl\langle \hat{B}^{\dagger}_{\vec{k}_3}(\tau_{2}) \,\,\hat{B}_{ \vec{k}_4}(\tau_{1})  \biggr\rangle\,\,\langle \widehat{n}_{c}^2\rangle \,\,D_{k_{1}}(\tau_{1}) D_{k_{2}}^{\ast}(\tau_{2}).
\label{GD2}
\end{eqnarray}
In Eq. (\ref{GD2}), always for conciseness, the following shorthand notation has been employed: 
\begin{equation}
\widehat{B}_{\vec{k}}(\tau) = \biggl[ d_{+}(k) \widehat{b}_{\vec{k}} + d_{-}^{\ast}(k)\, \widehat{b}_{-\vec{k}}^{\dagger} \biggr] \,\, e^{- i \,k\,\tau(-\tau_{\ast})}, \qquad D_{k}(\tau)= \biggl( 1 - e^{-i \, k(\tau- \tau_{\ast})}\biggr).
\label{GD3}
\end{equation}
Each of the $16$ terms of the type illustrated in Eq. (\ref{GD2}) lead to two Dirac delta functions over the three-momenta by so eliminating two of the four integrals appearing in the original expression.
The final result for the degree of second-order coherence is 
\begin{equation}
g_{g}^{(2)}(\vec{r}, \tau_{1}, \tau_{2}) = \overline{g}_{g}^{(2)}(\vec{r}, \tau_{1}, \tau_{2}) + {\mathcal O}(\lambda^2) \langle \widehat{n}_{c}^2 \rangle + {\mathcal O}(\lambda^4) \langle \widehat{n}_{c}^4 \rangle, 
\label{GD4}
\end{equation}
where $\overline{g}_{g}^{(2)}(\vec{r}, \tau_{1}, \tau_{2})$ denotes the leading contribution that 
we write in the zero time-delay limit (i.e. $\tau_{1}- \tau_{2}\to 0$) and for a vacuum initial state 
(i.e. $\overline{n}_{k} \to 0$):
\begin{eqnarray}
\overline{g}^{(2)}_{g}(\vec{r},\tau) &=&  1 + \frac{\int k_{1} d k_{1} |d_{-}(k_{1})|^2 \, j_{0}(k_{1} r) \,\,\int k_{2} d k_{2} |d_{-}(k_{2})|^2\, j_{0}(k_{2} r) }{\int k_{1} \,d k_{1} |d_{-}(k_{1})|^2\,\int k_{2} \,d k_{2} |d_{-}(k_{2})|^2}
\nonumber\\
&+&  \frac{\int k_{1} d k_{1} \,d_{+}^{*}(k_{1}) d_{-}^{*}(k_{1})\, j_{0}(k_{1} r) \,\,\int k_{2} d k_{2} \, d_{+}(k_{2}) d_{-}(k_{2})\,j_{0}(k_{2} r)}{\int k_{1} d k_{1} |d_{-}(k_{1})|^2 \,\,\int k_{2} d k_{2} |d_{-}(k_{2})|^2},
\label{GD5}
\end{eqnarray}
where, as before, $j_{0}(k_{1} r)$ and  $j_{0}(k_{2} r)$ denote the spherical Bessel function of zeroth order \cite{SP1,SP2}. In the  limit $k_{1} r \ll 1$ and $k_{2} r\ll1$, Eq. (\ref{GD5}) first implies 
that 
\begin{eqnarray}
\overline{g}^{(2)}_{g}(\vec{r},\tau) &=&  2 +  \frac{\int k_{1} d k_{1} \,d_{+}^{*}(k_{1}) d_{-}^{*}(k_{1})\,  \,\,\int k_{2} d k_{2} \, d_{+}(k_{2}) d_{-}(k_{2})}{\int k_{1} d k_{1} |d_{-}(k_{1})|^2 \,\,\int k_{2} d k_{2} |d_{-}(k_{2})|^2}
\nonumber\\
&=& 2  + \frac{\int k_{1} d k_{1}  |d_{-}(k_{1})|^2 \epsilon(k_{1})\,  \,\,\int k_{2} d k_{2} \, |d_{-}(k_{2})|^2\epsilon^{\ast}(k_{2})}{\int k_{1} d k_{1} |d_{-}(k_{1})|^2 \,\,\int k_{2} d k_{2} |d_{-}(k_{2})|^2},
\label{GD5a}
\end{eqnarray}
where $\epsilon(k_{1}) =\,d_{+}^{\ast}(k_{1})/d_{-}(k_{1})$ and $\epsilon^{\ast}(k_{2}) =d_{+}(k_{2})/d_{-}^{\ast}(k_{2})$. We now recall that the explicit expressions for $d_{\pm}(k)$ have been 
introduced in Eqs. (\ref{SH6}) (see also Eqs. (\ref{WKB6})--(\ref{WKB8})). In that context 
we stressed that the relevant physical limit coincides with $a_{re}\gg a_{ex}$ 
implying that the Universe expands between $\tau_{ex}$ and $\tau_{re}$ (see also Fig. \ref{Figure3}).
When $a_{re} \gg a_{ex}$ we then have that
\begin{equation}
|\epsilon(k) |^2 =\,\frac{|d_{+}(k)|^2}{|d_{-}(k)|^2} = 1 + \frac{1}{|d_{-}(k)|^2} \to 1 + {\mathcal O}(a_{ex}^2/a_{re}^2)
\label{GD5b}
\end{equation}
since $|d_{-}(k)|^2 = {\mathcal O}(a_{re}^2/a_{ex}^2)\gg 1$. If we then consider together Eqs. (\ref{GD5a}) and (\ref{GD5b}) we conclude that $\epsilon(k_1) \simeq  |\epsilon(k_{2})\to 1$ and  $\overline{g}^{(2)}(\vec{r},\tau) \to 3$.
The degree of second-order coherence of the gravitons is therefore always larger than the 
Poisson limit (i.e. $\overline{g}^{(2)}_{g} \to 1$) and also generally larger than the chaotic limit 
(i.e. $\overline{g}^{(2)}_{g} \to 2$). The most important aspect of the present analysis is that 
the degree of second-order coherence of the cavity mode is not sensitive to the degree 
of coherence of the gravitons. This is not only true in practice (because of the smallness of the coupling) but also in principle because of the symmetries of the total Hamiltonian.

\newpage 
\renewcommand{\theequation}{7.\arabic{equation}}
\setcounter{equation}{0}
\section{Concluding considerations}
\label{sec7}
If the adiabatic paradigm is supplemented by an early inflationary stage the entangled pairs of gravitons with opposite (comoving) three-momenta are produced thanks to the evolution of the space-time curvature. While the statistical properties of these squeezed states are encoded in the degrees of coherence (and in the related Hanbury-Brown Twiss correlations), their actual  existence is consistent with the success of the current lore for structure formation where the Universe expands both during inflation and 
in the subsequent decelerated phase. As a consequence, the degree of second-order coherence of the gravitons is super-Poissonian (i.e. larger than in the case of a coherent 
 state) and also super-chaotic (i.e. larger than in the case of a mixed quantum state with Bose-Einstein statistical weights). The spectra of squeezed gravitons generally depend upon three related physical aspects, namely the curvature scale reached at the end of inflation, the timeline of the post-inflationary expansion rate and the low-frequency bounds on the tensor-to-scalar ratio. Together with these three key aspects the Hanbury-Brown Twiss correlations can be used to infer the statistical properties of the squeezed gravitons 
 and to distinguish them from other diffuse sources of gravitational radiation. 
 
 From the physical viewpoint it is conceivable that the intensity correlations 
 of the gravitons could be potentially deduced, at least in principle, from 
 the ones of the photons. To scrutinize this possibility the simplest 
 strategy is to analyze the interaction of the  gravitons with the fundamental mode of a quantized electromagnetic field confined inside an optical resonator with perfectly-reflecting walls. The first step along this direction is a proper definition of the various degrees of coherence in the scalar, vector and tensor case 
 according to the tenets of the Glauber-Sudarshan approach. After 
 deducing the total Hamiltonian of the problem (including the interaction between the gravitons and the cavity mode) the degrees of coherence associated with the photons can be determined. In our context the transverse intensity profiles are not important for the degrees of coherence and it is the sufficient to consider the light beams as exciting a single mode of the field, as it also happens in conventional optomechanical applications.

To close the circle the degrees of coherence of the gravitons have been separately discussed and compared with the ones associated with the cavity modes. The conclusion of this analysis is  
 that the Hanbury-Brown Twiss correlations of the gravitons cannot be deduced 
 from the ones of the photons not only in practice (because of the minuteness 
 of the couplings) but also in principle. The physical reason for this 
 result is ultimately related to the symmetries of the total Hamiltonian 
 that commutes with the number of photons of the cavity. This 
 occurrence implies that the degree of second order coherence 
 of the cavity mode is insensitive to the gravitons and only depend 
 on the statistical properties of the driving source of photons inside the cavity. 
We may imagine, for instance,  that a laser drive term describes a photon inside the cavity with an approximate  frequency coinciding with the one of the laser, then the degrees of second-order coherence of the photons generally reflect the ones of the drive term. Consequently the statistical properties of the gravitons (and their super-Poissonian statistics) cannot be inferred, even in principle, from the intensity correlations of the optical photons contained in a cavity with perfectly reflecting walls.
 
\section*{Acknowledgements}
It is a pleasure to thank A. Gentil-Beccot, L. Pieper, S. Rohr, J. Vigen of the CERN Scientific Information Service for their kind help.

\newpage
\begin{appendix}
\renewcommand{\theequation}{A.\arabic{equation}}
\setcounter{equation}{0}
\section{Relic gravitons and photons in curved backgrounds}
\label{APPA}
\subsection{General considerations}
In this appendix we are going to derive the specific form of Eq. (\ref{AC1}) with the purpose of describing 
the interactions  of the relic gravitons with the cavity modes. We then start from the general form of the four-dimensional action that we write in the following manner:
\begin{equation}
S = - \frac{1}{2 \ell_{P}^2} \int d^{4} x \sqrt{-g} \,\,R - \frac{1}{4} \int d^{4} x \sqrt{-g} \, F_{\mu\nu}\, F^{\mu\nu} + S_{m},
\label{AP1}
\end{equation} 
where $F_{\mu\nu} = \nabla_{[\mu} \, A_{\nu]}$ denotes the Maxwell field strength defined in terms of the vector potential and of the covariant derivative; $S_{m}$ indicates the action 
of the matter sources that are characterized, for simplicity,  in terms of a total pressure $p_{t}$ and total energy density $\rho_{t}$. Furthermore, Eq. (\ref{AP1}) $g$ denotes the determinant of the four-dimensional metric whose general expression is separated into a background value $\overline{g}_{\mu\nu}$ 
supplemented by a perturbation $|f_{\mu\nu}|<1$:
\begin{equation}
g_{\mu\nu}(x) = \overline{g}_{\mu\nu}(x) + f_{\mu\nu},\qquad g^{\mu\nu} = \overline{g}^{\mu\nu} - 
f^{\mu\nu} + f^{\nu\alpha} \, f_{\alpha}^{\,\,\,\mu}.
\label{AP1a}
\end{equation}
The second-order correction to $\overline{g}^{\mu\nu}$ has been also included in Eq. (\ref{AP1a}) to stress that Eq. (\ref{AP1}) must be eventually expanded to second-order in the amplitude of the tensor modes of the geometry, i.e. the transverse and solenoidal excitations of the background metric. Although some of the 
results can be deduced in more general terms we are now going to focus on a conformally flat background 
metric and on its traceless and transverse inhomogeneities:
\begin{equation}
\overline{g}_{\mu\nu}(\tau) = a^2(\tau)\, \eta_{\mu\nu}, \quad f_{i\,j} = - a^2 h_{i\, j}, \quad \partial_{i} h^{i}_{\,\,\,j} = h^{i}_{\,\,\,i}=0.
\label{AP1b}
\end{equation}
As usual, in Eq. (\ref{AP1b}) $a(\tau)$ denotes the scale factor, $\tau$ indicates the conformal time coordinate and $\eta_{\mu\nu} = \mathrm{diag}(1, \, -1,\, -1,\, -1)$ is the Minkowski metric. 
The gravity part of the action appearing in Eq. (\ref{AP1}) can be rewritten in a way that excludes automatically a total derivative \cite{LAND}:
\begin{eqnarray}
S &=& S_{m} + \frac{1}{2 \ell_{P}^2} \int d^{4} x \sqrt{-g} \, g^{\alpha\beta} {\mathcal Q}_{\alpha\beta} - \frac{1}{4} \int d^{4} x \sqrt{-g} \, F_{\mu\nu}\, F^{\mu\nu} 
\nonumber\\
&+& \frac{1}{2\ell_{P}^2} \int d^{4} x \sqrt{-g}\, \, g^{\alpha\beta} \biggl(\nabla_{\beta} \,\Gamma_{\alpha\lambda}^{\,\,\,\,\,\lambda} - \nabla_{\lambda}  \,\Gamma_{\alpha\beta}^{\,\,\,\,\,\lambda}\biggr), 
\label{AP2}
\end{eqnarray}
where $\Gamma_{\alpha\beta}^{\,\,\,\,\,\lambda}$ are, as usual, the Christoffel symbols while
$\nabla_{\alpha}$ indicates the covariant derivatives with respect to the full curved metric $g_{\mu\nu}$. 
As anticipated, the second line of Eq. (\ref{AP2}) corresponds to a total derivative that does not contribute to the second-order action while ${\mathcal Q}_{\alpha\beta}$ is now
\begin{equation}
{\mathcal Q}_{\alpha\beta} =\biggl[ \Gamma_{\alpha\beta}^{\,\,\,\,\,\mu} \Gamma_{\mu\nu}^{\,\,\,\,\,\nu} - \Gamma_{\alpha\nu}^{\,\,\,\,\,\mu} \Gamma_{\mu\beta}^{\,\,\,\,\,\nu} \biggr].
\label{AP3}
\end{equation}
From Eq. (\ref{AP3}) we have that 
 $\overline{{\mathcal Q}}_{\alpha\beta}$ indicates the background value of ${\mathcal Q}_{\alpha\beta}$;  in particular $\overline{{\mathcal Q}}_{00} = 0$ and $\overline{{\mathcal Q}}_{i\, j} = 2 {\mathcal H}^2 \delta_{i\, j}$ where, as in the bulk of the paper, ${\mathcal H} = a^{\prime}/a$ and the prime indicates a derivation with respect to the conformal time coordinate. We remind that, according to the standard notations, ${\mathcal H} = a \, H$ where $H = \dot{a}/a$ is the Hubble rate and the overdot stands for a derivation with respect to the cosmic time coordinate. The first-order fluctuations of ${\mathcal Q}_{\alpha\beta} $ are given, respectively, by 
$\delta_{t}^{(1)} {\mathcal Q}_{00} =0$ and $\delta_{t}^{(1)} {\mathcal Q}_{i\,j} = 2 {\mathcal H}^2\, h_{i\,j}$; finally the only non-vanishing components of $\delta_{t}^{(2)} {\mathcal Q}_{\mu\nu}$ read
\begin{eqnarray}
\delta_{t}^{(2)} {\mathcal Q}_{00} &=& - \frac{1}{4} h_{k\ell}^{\,\prime} \,  h^{k\ell\,\,\prime} + \frac{{\mathcal H}}{2} h_{k\ell}^{\prime}\, h^{k\ell},
\nonumber\\
\delta_{t}^{(2)} {\mathcal Q}_{ij} &=& - \frac{{\mathcal H}}{2} h_{k\ell}^{\prime}\, h^{k\ell} \,\,\delta_{ij} - \frac{1}{4} \biggl[ h_{i}^{\,\,k\,\,\prime}\,\, h_{k\, j}^{\prime} + h_{j}^{\,\,k\,\,\prime}\,\, h_{k\, i}^{\prime}\biggr] 
\nonumber\\
&-& \frac{1}{4} \biggl[ \partial_{\ell} \,h_{i}^{\,\, k} + \partial_{i} h_{\ell}^{\,\, k} - \partial^{k} \,h_{\ell \,i} \biggr] \biggl[ \partial_{k} \,\,h_{j}^{\,\, \ell} + \partial_{j} \,\,h_{k}^{\,\, \ell} - \partial^{\ell} \,\,h_{k \,j} \biggr].
\label{AP4c}
\end{eqnarray}
The second-order variation of the first line of Eq. (\ref{AP1}) gives the contribution of the gravitons 
while the remaining terms account for the contributions of the cavity modes and of the mutual interactions between the two components.

\subsection{The contribution of the gravitons}
The action of the gravitons is obtained by 
considering the second-order variation of the gravity and matter 
parts of $S$ given in Eq. (\ref{AP1}):
\begin{equation}
S_{g} = \delta_{t}^{(2)} S_{m}  + \frac{1}{2 \ell_{P}^2} \int d^{4} x \biggl[ \delta_{t}^{(2)} \sqrt{-g} \,\,\overline{g}^{\mu\nu} \,\, \overline{{\mathcal Q}}_{\mu\nu} + \sqrt{- \overline{g}} \, \,\delta_{t}^{(2)} \biggl( g^{\mu\nu} \, {\mathcal Q}_{\mu\nu} \biggr)\biggr].
\label{APsec1}
\end{equation}
Recalling Eqs. (\ref{AP1b}) we have that $\delta^{(2)} \sqrt{- g} = - a^4 h_{i\,j}\,h^{i\,j}/4$ and therefore 
the first term of Eq. (\ref{APsec1}) is easily evaluated; the remaining part of Eq. (\ref{APsec1}) 
is slightly more cumbersome; for this purpose we recall that 
\begin{equation}
\delta_{t}^{(2)} \biggl( g^{\mu\nu} \, {\mathcal Q}_{\mu\nu} \biggr)= \overline{{\mathcal Q}}_{\mu\nu} \,\delta_{t}^{(2)} g^{\mu\nu}\,\,  + \overline{g}^{\mu\nu}  \,\, \delta_{t}^{(2)} {\mathcal Q}_{\mu\nu} +  \delta_{t}^{(1)}\, g^{\mu\nu} \,\,\delta_{t}^{(1)} {\mathcal Q}_{\mu\nu}.
\label{APsec2}
\end{equation}
Equation (\ref{APsec2}) is explicitly evaluated thanks to Eq. (\ref{AP4c}) together with the expressions of 
$\overline{Q}_{\alpha\beta}$ 
and $\delta_{t}^{(1)}{\mathcal Q}_{\alpha\beta}$; the final form of $S_{g}$ becomes then  
\begin{eqnarray}
&&S_{g} = \frac{1}{8 \ell_{P}^2} \int d^{4} x \,\, a^2\, \, \biggl[ 
\partial_{\tau} h_{ij} \,\partial_{\tau} h^{\,\,i\,j} - \partial_{k} h_{i\,j} \,\partial^{k} h^{\,\,i\,j}\biggr]
- \frac{1}{2 \ell_{P}^2} \int d^4 x \,\,a^2 h_{ij}\, h^{\,\,ij} \biggl[ \biggl( {\mathcal H}^{\prime} + 
\frac{{\mathcal H}^2}{2} \biggr) + \frac{\ell_{P}^2}{2} a^2 p_{t}\biggr]
\nonumber\\
&& +   \frac{1}{2 \ell_{P}^2} \int d^{4} x \, \biggl\{ \partial_{\tau} \biggl[ a^2 {\mathcal H} h_{ij} \, h^{\,ij} \biggr] - \partial_{k} \biggl[a^2 \,h_{\,i\,\ell} \,\,\partial^{\ell} h^{\,i k} \biggr] \biggr\}.
\label{APsec3}
\end{eqnarray}
The first term in Eq. (\ref{APsec3}) is the physically relevant contribution and it coincides 
with the action originally derived by Ford and Parker \cite{IOTA2} (see also \cite{ACF2,ACF3}). The second term   of Eq. (\ref{APsec3}) cancels on-shell since it is proportional to the background Friedmann equation $(2 {\mathcal H}^{\prime} + {\mathcal H}^2 + \ell_{P}^2 a^2 p_{t}) =0$. Note that the same happens if the sources are different and this will just affect the explicit form of $p_{t}$; note that the third term of Eq. (\ref{APsec3}) is a total derivative that does not affect the equations of motion.

\subsection{The contribution of the gauge fields}
We now consider the action of the cavity modes and since the electromagnetic background is effectively absent the corresponding action is swiftly deduced after recalling that, in the conformally flat background of Eq. (\ref{AP1b}), the explicit expressions of the 
field strengths is:
\begin{equation}
F^{i0} =e^{i}(\vec{x},\tau)/a^2(\tau), \qquad F^{ij} = - \epsilon^{i\,j\,k} b_{k}(\vec{x},\tau)/a^2(\tau).
\label{APG1}
\end{equation}
In Eq. (\ref{APG1}) $e^{i}(\vec{x},\tau)$ and $b_{k}(\vec{x},\tau)$ are the (physical) electric and magnetic fields; note that $F_{i0} = - a^2(\tau)\, e_{i}(\vec{x},\tau)$ and $F_{ij} = - a^2(\tau)\, \epsilon^{i\,j\,k}\,b_{k}(\vec{x},\tau)$. In terms of the physical fields the action becomes therefore 
\begin{eqnarray}
S_{\gamma} + S_{\gamma\,g} &=& \frac{1}{2} \int d^{4} x\, a^4 ( e^2 - b^2) - 
\frac{1}{4}\int d^{4} x a^4 [(e_{i} e_{j} + b_{i} b_{j} ) - (e^2 + b^2) \delta_{i\,j}/3 ]\, h^{i\, j} 
\nonumber\\
&-& \frac{1}{4}\int d^{4} x a^4 [(e^{i} e^{j} + b^{i} b^{j} ) - (e^2 + b^2) \delta^{i\,j}/3 ]\, h_{i\, j} 
\label{APG2}
\end{eqnarray}
Equation (\ref{APG2}) can be further simplified by introducing the physical electric and magnetic 
fields $\vec{E}(\vec{x},\tau) = a^2(\tau) \, \vec{e}(\vec{x},\tau)$ and $\vec{B}(\vec{x},\tau) = a^2(\tau)  \, \vec{b}(\vec{x},\tau)$: 
\begin{equation}
S_{\gamma} + S_{\gamma\,g} = \frac{1}{2} \int d^{4} x\, (E^2 - B^2) - \frac{1}{4}\int d^{4} x \Pi_{i\,j} \, \,h^{i\, j} - \frac{1}{4}\int d^{4} x\, \Pi^{i\,j} \,\, h_{i\, j},
\label{APG3}
\end{equation}
Where $\Pi_{i\,j} = [E_{i} E_{j} + B_{i} B_{j} - (B^2 +E^2) \delta_{i\,j}/3]$ indicates the anisotropic stress. Note that the term containing the Kroeneker delta is contracted with $h^{i\,j}$ (which is traceless); this means that, for the present ends, only the first two terms count in the expression of $\Pi_{i\,j}$ [see, however, Eq. (\ref{AC5a}) and discussions thereafter].

\subsection{The full action} 
From the results of the two previous subsection the total action can be written in the following manner
\begin{eqnarray}
S&=&  \frac{1}{8 \ell_{P}^2} \int d^{4} x \,\, a^2(\tau)\, \, \biggl[ 
\partial_{\tau} h_{i\,j} \,\partial_{\tau} h^{\,\,i\,j} - \partial_{k} h_{i\,j} \,\partial^{k} h^{\,\,i\,j}\biggr]
\nonumber\\
&+& \frac{1}{2} \int d^{4} x\,\, (E^2 - B^2) -  \frac{1}{4}\int d^{4} x \,\,(E_{i} E_{j} + B_{i} B_{j}) \,\,h^{i\, j} 
\nonumber\\
&-& \frac{1}{4}\int d^{4} x\,\,(E^{i} E^{j} + B^{i} B^{j})\,\, h_{i\, j},
\label{APG4}
\end{eqnarray}
where, by definition, $S = S_{g}+ S_{\gamma} + S_{\gamma g} $. It may superficially seem that the action (\ref{APG4}) 
does not differ much from its flat space-time analog; to clarify this point we can introduce the rescaled 
tensor amplitude $ a(\tau) h_{i\,j}(\vec{x},\tau) = \mu_{i\, j}(\vec{x},\tau)$ and obtain 
\begin{eqnarray}
S&=&  \frac{1}{8 \ell_{P}^2} \int d^{4} x \,\,  \biggl[ 
\partial_{\tau} \mu_{i\,j} \,\partial_{\tau} \mu^{\,\,i\,j} - \partial_{k} \mu_{i\,j} \,\partial^{k} \mu^{\,\,i\,j} + {\mathcal H}^2 \mu_{i\,j}\,\, \mu^{i\,j}
- {\mathcal H} \,\,\mu_{i\,j}\, \partial \mu^{i\,j} - {\mathcal H} \partial_{\tau} \mu_{i\,j} \, \mu^{i\, j} \biggr]
\nonumber\\
&+& \frac{1}{2} \int d^{4} x\, (E^2 - B^2) -  \frac{1}{4}\int d^{4} x  (E_{i} E_{j} + B_{i} B_{j}) \, \,\mu^{i\, j}/a 
\nonumber\\
&-& \frac{1}{4}\int d^{4} x\,(E^{i} E^{j} + B^{i} B^{j})\,\, \mu_{i\, j}/a.
\label{APG5}
\end{eqnarray}
As already mentioned in the bulk of the paper it is practical to employ the Coulomb gauge since it is invariant 
under a conformal rescaling; in this gauge $\vec{E} = - \vec{A}^{\,\prime}$ and $\vec{B} = \vec{\nabla} \times \vec{A}$ and the 
second part of the action (\ref{APG4}) can be rewritten accordingly:
\begin{eqnarray}
S_{\gamma} + S_{\gamma g} &=& \frac{1}{2} \int d^{4} x\, (\partial_{\tau}
A_{i} \partial_{\tau} A^{i}- \partial_{k} A_{i} \partial^{k} A^{i}) 
\nonumber\\
&-&  \frac{1}{4}\int d^{4} x  [\partial_{\tau} A_{i} \partial_{\tau} A_{j} + (\vec{\nabla}\times \vec{A})_{i}(\vec{\nabla}\times \vec{A})_{j} ] \, \,\mu^{i\, j}/a 
\nonumber\\
&-& \frac{1}{4}\int d^{4} x\,[\partial_{\tau} A^{i} \partial_{\tau} A^{j} + (\vec{\nabla}\times \vec{A})^{i}(\vec{\nabla}\times \vec{A})^{j} ] \, \mu_{i\, j}/a.
\label{APG6}
\end{eqnarray}
\end{appendix}
The expressions in Eq. (\ref{APG6}) can also be made more explicit by using the properties of antisymmetric tensor fields in 3-dimensions;
so for instance we have that 
\begin{eqnarray}
&& (\vec{\nabla}\times \vec{A})_{i} (\vec{\nabla}\times \vec{A})_{j} = \biggl[ (\partial^{m} A^{j}) (\partial_{i} A_{m}) - (\partial^{m} A^{j}) (\partial_{m} A_{i}) + (\partial^{j} A^{m}) (\partial_{m} A_{i}) 
\nonumber\\
&& -( \partial^{j} A^{m}) (\partial_{i} A_{m}) \biggr]
+ \biggr[ (\partial^{m} A^{n}) (\partial_{m} A_{n}) - (\partial^{m} A^{n}) (\partial_{n} A_{m})\biggr] \delta_{i\,j},
\label{APG7}
\end{eqnarray}
and similarly for the analog expressions with two contravariant indices.

\renewcommand{\theequation}{B.\arabic{equation}}
\setcounter{equation}{0}
\section{Reduced Hamiltonians with two and three modes}
\label{APPB}
At the beginning of section \ref{sec5} we stressed that the multimode Hamiltonian discussed 
here has many analogies with different (reduced) Hamiltonians describing 
the interaction of two or three modes. In this appendix we shall clarify more accurately 
this analogy.
\subsection{The conventional optomechanical Hamiltonian}
We showed that the degrees of second-order coherence of the gravitons are not reflected into the corresponding degrees 
 of coherence associated with the photons. One of the reasons why this happens is that the Hamiltonian commutes with the number operator of the cavity photons. This property holds also in the case of Eq. (\ref{HAMF3}) and 
recalling that $\widehat{n}_{c} = \widehat{c}^{\dagger} \widehat{c}$
is the number operator of the cavity photons, 
$[\widehat{H}, \widehat{n}_{c}] =0$, where $\widehat{H}$ now refers to Eq. (\ref{HAMF3}). On the contrary the number of gravitational excitations  
$\widehat{n}_{b} = \widehat{b}^{\dagger} \widehat{b}$ does not commute with 
the Hamiltonian, i.e. $[\widehat{H}, \widehat{n}_{b}] \neq 0$. From Eq. (\ref{HAMF3}) we have 
\begin{eqnarray}
\widehat{c}^{\,\prime} =  - i \,\omega_{c}\, \widehat{c} + i \,g_{\omega} \,\widehat{c}\, (\widehat{b} +  \widehat{b}^{\dagger}), \qquad
\widehat{b}^{\,\prime} =  - i \,\omega \,\widehat{b} + i\,g_{\omega} \,\widehat{n}_{c},
\label{HEIS2}
\end{eqnarray}
where the prime denotes, as usual, a derivation with respect to $\tau$.
Since $[\widehat{H}, \widehat{n}_{c}] =0$, the evolution of $\widehat{b}(\tau)$ (second equation in Eq. (\ref{HEIS2})) can be directly solved: 
\begin{equation}
\widehat{b}(\tau) = \widehat{b}_{\ast} \,e^{- i x} + \overline{g}_{\omega} \,\widehat{n}_{c} \,\bigl( 1 - e^{- i x}\bigr),
\label{HEIS3}
\end{equation}
where the variables $ x = \omega( \tau - \tau_{\ast})$ and $\overline{g}_{\omega}= g_{\omega}/\omega$ have been introduced for convenience; by definition $\widehat{b}_{\ast} = \widehat{b}(\tau_{\ast})$.  From Eq. (\ref{HEIS3}) the Hermitian combination $\widehat{b}(\tau) + \widehat{b}^{\dagger}(\tau)$ can be explicitly constructed 
\begin{equation}
\widehat{b}(\tau) + \widehat{b}^{\dagger}(\tau) = \widehat{b}_{\ast} \,\, e^{- i x} + \widehat{b}_{\ast}^{\dagger} \, e^{i x} + 2 \, \overline{g} \,\, \widehat{n}_{c}\,\, (1 - \cos{x}),
\label{HEIS4}
\end{equation}
and eventually inserted back into the evolution of $\widehat{c}(\tau)$ (first equation in Eq. (\ref{HEIS2})); the resulting 
expression takes the following form: 
\begin{equation}
\widehat{C}^{\,\,\prime} = i g_{\omega} \,\, [ \widehat{b}(\tau) + \widehat{b}^{\dagger}(\tau)] \,\, \widehat{C}, \qquad \widehat{C}(\tau) = e^{i \omega_{c} \tau} \, \widehat{c}(\tau).
\label{HEIS5}
\end{equation}
The solution of Eq. (\ref{HEIS5}) 
involves  the operator $\widehat{Z}(\tau, \tau_{\ast})$  obeying 
\begin{eqnarray}
&&\partial_{\tau} \widehat{Z}(\tau, \tau_{\ast}) = i g_{\omega} \widehat{F}(\tau, \tau_{\ast})\, \widehat{Z}(\tau, \tau_{\ast}),
\label{HEIS6}\\
&& \widehat{F}(\tau, \tau_{\ast}) = \widehat{M}_{\ast} \,\cos{x}  - i \widehat{N}_{\ast} \, \sin{x} + \widehat{Q}_{\ast} \, (1 - \cos{x}).
\label{HEIS7}
\end{eqnarray}
Equations (\ref{HEIS6})--(\ref{HEIS7}) ultimately follow from Eq. (\ref{HEIS5}) since $\widehat{C}(\tau) = \widehat{Z}(\tau, \tau_{\ast}) \widehat{C}(\tau_{\ast})$.  For immediate convenience in Eq. (\ref{HEIS7}) the auxiliary operators $\widehat{M}_{\ast} = (\widehat{b}_{\ast} + \widehat{b}_{\ast}^{\dagger})$, $\widehat{N}_{\ast} =  (\widehat{b}_{\ast} - \widehat{b}_{\ast}^{\dagger})$ and 
$\widehat{Q}_{\ast} = 2 \,\overline{g}_{\omega}\,\widehat{c}_{\ast}^{\dagger} \widehat{c}_{\ast}$ have been introduced. If the time derivative in Eq. (\ref{HEIS7}) is traded for a derivative with respect to $x$ we obtain: 
\begin{equation}
\partial_{x} \widehat{Z}(x,\, x_{\ast}) = i\,\overline{g}_{\omega}\,  \widehat{F}(x, x_{\ast})
Z(x,\, x_{\ast}),\qquad\widehat{Z}(x_{\ast},\, x_{\ast}) = \mathds{1}.
\label{HEIS7a}
\end{equation}
Equation (\ref{HEIS7a}) can be further transformed into an integral equation
written in the form:
\begin{equation}
Z(x, \,x_{\ast}) = \mathds{1} + i \overline{g}_{\omega} \int_{x_{\ast}}^{x} \widehat{F}(y, x_{\ast}) \widehat{Z}(y, x_{\ast}) \, d y.
\label{HEIS8}
\end{equation}
Equation (\ref{HEIS8}) can be finally solved by iteration:
\begin{eqnarray}
Z(x, x_{\ast}) &=& \mathds{1} + i \, \overline{g}_{\omega} \int_{x_{\ast}}^{x} \widehat{F}(x_{1}, x_{\ast}) \, d x_{1}
+( i \overline{g}_{\omega})^2 \int_{x_{\ast}}^{x} \widehat{F}(x_{1}, x_{\ast}) \, d x_{1}
\int_{x_{\ast}}^{x_{1}} \widehat{F}(x_{2}, x_{\ast}) \, d x_{2} + \,.\,.\,.\,.
\nonumber\\
&+& ( i \overline{g}_{\omega})^n \int_{x_{\ast}}^{x} \widehat{F}(x_{1}, x_{\ast}) \, d x_{1} \,.\,.\,.\,.\int_{x_{\ast}}^{x_{n-1}} \widehat{F}(x_{n}, x_{\ast}) \, d x_{n}.
\label{HEIS8a}
\end{eqnarray}
We may compute the various corrections order by order and then realize that 
the final result can be 
expressed in the following manner 
\begin{equation}
\widehat{c}(\tau) = e^{- i \, z} \, \widehat{{\mathcal R}}[\beta(x)] \,\widehat{{\mathcal D}}[\alpha(x)] \,
\widehat{c}_{\ast}, 
\label{HEIS9}
\end{equation}
where $z = \omega_{c}(\tau- \tau_{\ast})$.
The two operators $\widehat{{\mathcal R}}[\beta(x)]$ and 
$\widehat{{\mathcal D}}[\alpha(x)]$ introduced in Eq. (\ref{HEIS9}) are: 
\begin{equation}
\widehat{{\mathcal R}}[\beta(x)] = e^{ 2 i \, \beta(x)\, \widehat{c}_{\ast} \widehat{c}_{\ast}}, \qquad 
\widehat{{\mathcal D}}[\alpha(x)] = e^{\alpha(x) \widehat{b}_{\ast}^{\dagger} - 
\alpha^{\ast}(x) \widehat{b}_{\ast}},
\label{HEIS10}
\end{equation}
where $\alpha(x) = \overline{g}_{\omega} (e^{i x} -1)$ and $\beta(x) = \overline{g}_{\omega}^2(x - \sin{x})$.

\subsection{Reduced Hamiltonians with three modes}
We now consider the case of Eq. (\ref{HAMF2}); the evolution in the 
Heisenberg description is given by:
\begin{eqnarray}
&&  \widehat{a}^{\,\,\prime} = 
- i \omega\, \widehat{a} + i g_{\omega} \,\widehat{c}^{\dagger}\, \widehat{c} + {\mathcal H} \,\widehat{b}^{\dagger},
\label{HEIP1}\\
&&  \widehat{b}^{\,\, \prime} =  
- i \omega\, \widehat{b} + i g_{\omega} \widehat{c}^{\dagger}\, \widehat{c} + {\mathcal H} \,\widehat{a}^{\dagger},
\label{HEIP2}\\
&& \widehat{c}^{\,\,\prime} =  - i \omega_{c}\, \widehat{c}
+ i \, g_{\omega} \,(\widehat{a} + \widehat{b}^{\dagger} + \widehat{a}^{\dagger} + \widehat{b}) \,\widehat{c}.
\label{HEIP3}
\end{eqnarray}
We can now sum and subtract Eqs. (\ref{HEIP2})--(\ref{HEIP3}); the result 
of these manipulations can be expressed in terms of the operators $\widehat{{\mathcal Z}}_{\pm} = ( \widehat{a} \pm \widehat{b}^{\dagger})$:
\begin{eqnarray}
&& \widehat{{\mathcal Z}}_{+}^{\,\,\prime} - {\mathcal H} \,\,\widehat{{\mathcal Z}}_{+} = - i \, \omega\, \widehat{{\mathcal Z}}_{-},
\label{HEIP6}\\
&& \widehat{{\mathcal Z}}_{-}^{\,\,\prime} + {\mathcal H} \,\, \widehat{{\mathcal Z}}_{-} = - i \,\omega\,\widehat{{\mathcal Z}}_{+} + 2 \,i \,g_{\omega} \,\widehat{n}_{c}.
\label{HEIP7}
\end{eqnarray}
If we now deduce $\widehat{{\mathcal Z}}_{-}$ from Eq. (\ref{HEIP6})
we obtain $\widehat{{\mathcal Z}}_{-} = i\, [\widehat{{\mathcal Z}}_{+}^{\prime} - {\mathcal H} \widehat{{\mathcal Z}}_{+}]/\omega$; this 
expression can then be inserted into Eq. (\ref{HEIP7}) to obtain a decoupled equation for $\widehat{{\mathcal Z}}_{+}$:
\begin{equation}
\widehat{{\mathcal Z}}_{+}^{\,\prime\prime} + \biggl[ \omega^2 - \frac{a^{\prime\prime}}{a} \biggr] \widehat{{\mathcal Z}}_{+} = 2 \, g_{\omega} \,\omega \,\widehat{n}_{c}.
\label{HEIP8}
\end{equation}
From Eq. (\ref{HEIP6}) we can then deduce $\widehat{{\mathcal Z}}_{-}$ and eventually construct 
$(\widehat{{\mathcal Z}}_{+} + \widehat{{\mathcal Z}}_{-})$, i.e. is the source term of Eq. (\ref{HEIP3}) 
that can be eventually solved for different sets of initial data.

\end{document}